\documentclass[fleqn,usenatbib]{mnras}

\usepackage[T1]{fontenc}

\DeclareRobustCommand{\VAN}[3]{#2}
\let\VANthebibliography\thebibliography
\def\thebibliography{\DeclareRobustCommand{\VAN}[3]{##3}\VANthebibliography}

\usepackage{graphicx}	
\usepackage{amsmath}	
\usepackage{amssymb}	
\usepackage{pdflscape}
\usepackage{soul}
\usepackage{pdflscape}

\newcommand{\hml}{H$_2\:2.1218\,\mu$m}
\newcommand{\brg}{Br$\gamma$}

\usepackage{lscape}

\title[The AGNIFS survey: gas kinematics]{The AGNIFS survey: spatially resolved observations of hot molecular and ionised outflows in nearby active galaxies}

\author[R. A. Riffel et al.]{R. A. Riffel,$^{1}$\thanks{E-mail: rogemar@ufsm.br (RAR)} 
T. Storchi-Bergmann,$^{2}$
R. Riffel,$^{2}$ 
M. Bianchin,$^{1}$
N. L. Zakamska,$^{3,4}$
\newauthor
D. Ruschel-Dutra,$^{5}$
M. C. Bentz,$^{6}$
L. Burtscher,$^{7}$
D. M. Crenshaw,$^{6}$
L. G. Dahmer-Hahn,$^{8}$
\newauthor
N. Z. Dametto,$^{9,10}$
R. I. Davies,$^{11}$
M. R. Diniz,$^{1}$
T. C. Fischer,$^{12}$
C.~M. Harrison,$^{13}$
\newauthor
V. Mainieri,$^{14}$
M. Revalski,$^{15}$
A. Rodriguez-Ardila,$^{16,17}$
D. J. Rosario,$^{12}$
A. J. Sch\"onell$^{18}$
\\
$^{1}$Departamento de F\'isica, Centro de Ci\^encias Naturais e Exatas, Universidade Federal de Santa Maria, 97105-900, Santa Maria, RS, Brazil\\
$^{2}$Instituto de F\'isica, Universidade Federal do Rio Grande do Sul, Av. Bento Gon\c calves 9500, 91501-970 Porto Alegre, RS, Brazil \\
$^{3}$Department of Physics \& Astronomy, Johns Hopkins University, Bloomberg Center, 3400 N. Charles St, Baltimore, MD 21218, USA\\
$^{4}$Institute for Advanced Study, Einstein Dr., Princeton NJ 08540\\
$^{5}$Departamento de F\'isica, Universidade Federal de Santa Catarina, P.O. Box 476, 88040-900, Florian\'opolis, SC, Brazil \\
$^{6}$Department of Physics and Astronomy, Georgia State University, 25 Park Place, Suite 605, Atlanta, GA 30303, USA\\
$^{7}$Leiden Observatory, PO Box 9513, 2300 RA, Leiden, The Netherlands\\
$^{8}$Shanghai Astronomical Observatory, Chinese Academy of Sciences, 80 Nandan Road, Shanghai 200030, China\\
$^{9}$Instituto de Astrof\'isica de Canarias, E-38205 La Laguna, Tenerife, Spain\\
$^{10}$ Universidad de La Laguna, Dpto. Astrof\'isica, E-38206 La Laguna, Tenerife, Spain\\
$^{11}$Max-Planck-Institut f\"ur Extraterrestrische Physik, Postfach 1312, 85741, Garching, Germany\\
$^{12}$AURA for ESA, Space Telescope Science Institute, 3700 San Martin Drive, Baltimore, MD 21218\\
$^{13}$School of Mathematics, Statistics and Physics, Newcastle University, NE1 7RU, UK\\
$^{14}$European Southern Observatory, Karl-Schwarzschild-Strasse 2, Garching bei M\"unchen, Germany\\
$^{15}$Space Telescope Science Institute, 3700 San Martin Drive, Baltimore, MD 21218, USA\\
$^{16}$Laborat\'orio Nacional de Astrof\'isica. Rua dos Estados Unidos, 154, 37504-364 Itajub\'a, MG, Brazil \\
$^{17}$Instituto Nacional de Pesquisas Espaciais, Av. dos Astronautas, 1758 - Jardim da Granja S\~ao Jos\'e dos Campos/SP - 12227-010, Brazil.\\
$^{18}$Instituto Federal de Educa\c c\~ao, Ci\^encia e Tecnologia Farroupilha, BR287, km 360, Estrada do Chapad\~ao, 97760-000, Jaguari - RS, Brazil\\
}

\date{Accepted XXX. Received YYY; in original form ZZZ}

\pubyear{2020}

\begin{document}
\label{firstpage}
\pagerange{\pageref{firstpage}--\pageref{lastpage}}
\maketitle

\begin{abstract}
We present the hot molecular and warm ionised gas kinematics for 33 nearby ($0.001\lesssim z\lesssim0.056$) X-ray selected active galaxies using the H$_2\,2.1218\,\mu$m and Br$\gamma$  emission lines observed in the K-band with the Gemini Near-Infrared Field Spectrograph (NIFS). The observations cover the inner 0.04--2\,kpc of each AGN at spatial resolutions of 4-250\,pc with a velocity resolution of $\sigma_{\rm inst}\approx$20\,${\rm km\,s^{-1}}$. We find that 31 objects (94 per cent) present a kinematically disturbed region (KDR) seen in ionised gas, while such regions are observed in hot molecular gas for 25 galaxies (76 per cent). We interpret the KDR as being due to outflows with masses of 10$^2$--10$^7$ M$_\odot$ and 10$^0$--10$^4$ M$_\odot$ for the ionised and hot molecular gas, respectively. The ranges of mass-outflow rates ($\dot{M}_{\rm out}$) and kinetic power ($\dot{E}_{\rm K}$) of the outflows are 10$^{-3}$--10$^{1}$  M$_\odot$\,yr$^{-1}$ and $\sim$10$^{37}$--10$^{43}$ erg\,s$^{-1}$ for the ionised gas outflows, and 10$^{-5}$--10$^{-2}$  M$_\odot$\,yr$^{-1}$ and 10$^{35}$--10$^{39}$ erg\,s$^{-1}$ for the hot molecular gas outflows. The median  coupling efficiency in our sample is $\dot{E}_{K}/L_{\rm bol}\approx1.8\times10^{-3}$ and the estimated momentum fluxes of the outflows suggest they are produced by radiation-pressure in low-density environment, with possible contribution from shocks.

\end{abstract}



\section{Introduction}

The co-evolution of galaxies and their super massive black holes (SMBHs) is supported by a large number of recent observational and theoretical studies \citep[e.g.][]{magorrian98,ferrarese00,gebhardt00,murr05,dimatteo05,gulterkin09,heckman11,kormendy13,harrison17,harrison18,costa18,sb19,caglar20}. This co-evolution is due to both feeding and feedback processes in Active Galactic Nuclei (AGNs). The feedback processes comprise jets  of relativistic particles emitted from the inner rim of the accretion disc,  winds emanating from outer regions of the disc and radiation emitted by the hot gas in the disc or by its corona \citep[e.g.][]{elvis00,frank02,ciotti10}, which is believed to play an important role in shaping galaxies in all mass ranges by quenching star formation in the hosts during cycles of nuclear activity \citep[e.g.][]{dimatteo05,hopkins_elvis10,Schaye_eagle_15,weinberger17,silk17,penny18,xu22}.

AGN feedback is strongly dependent on luminosity. For instance,  quasars may inject enough energy into the galactic medium so that the wind can overcome the inertia of the gas in the galactic potential. In low luminosity AGNs (LLAGN), on the other hand, the outflows may not be powerful enough to affect the large scale recent star formation in their hosts, in spite of some simulations predict that LLAGN can produce significantly feedback \citep{ward22}. In these LLAGN, the connection seems to be rather related to the feeding process of the AGN --  inflow of gas to the inner region -- in the sense that recent studies have revealed an excess of intermediate age stellar components, that can be interpreted as due to a delay between the onset of star formation and triggering of the AGN \citep[e.g.][]{rogemar_m1066_SP,rogerio_mrk1157,rogerio22,sb12,diniz17,mallmann18,burtscher21}. Although AGN feedback may had a more profound impact on galaxy evolution at the cosmic noon ($z\approx2-3$), AGN winds extending from hundred of parsecs to a few kiloparsecs \citep[e.g.][]{fischer18,forster_schreiber19,mingozzi19,santoro20,avery21,trindade-falcao21,lamperti21,vayner21,Luo21,Speranza21,deconto-machado22,kakkad20,kakkad22,Singha22} are hardly spatially resolved at these distances. Thus, it is nearby galaxies that offer the only opportunity to test in detail the prescriptions used in models of galaxy and SMBH co-evolution.

Near-infrared (hereafter, near-IR) integral field spectroscopy (IFS) observations -- and in particular with adaptive optics -- provide resolutions of a few tens of parsecs in nearby AGN hosts, allowing to spatially resolve the gas emission structure and kinematics. Near-IR observations are less affected by dust extinction, probing more obscured regions than observations in optical bands. In addition, the near-IR spectra of AGN hosts typically present emission lines, both from hot molecular ($\sim$2000\,K) and ionised gas \citep[e.g.][]{ardila04,rogerio06,rogemar21_extended,lamperti17,U19,denbrok22}, allowing observations of multi-phase AGN winds \citep[e.g.][]{santoro18,Shimizu19,ramos-almeida19,rogemar_cygA,bianchin22}, fundamental to understand the role of AGN feedback in galaxy evolution. 

The physical properties of outflows (e.g. mass-outflow rate and kinetic power) have been estimated using distinct methods and assumptions. These include using: (i) single component fits of the line profiles and comparison with the rest-frame stellar velocity in single aperture spectra \citep[e.g.][]{kova22} and IFS \citep[e.g.][]{ilha19,deconto-machado22}; (ii) decomposition of the emission lines in multiple-kinematic components using nuclear spectra \citep[e.g.][]{perrota19} and Hubble Space Telescope long slit data \citep[e.g.][]{revalski21}, and IFS observations \citep[e.g.][]{fischer19,bianchin22,kakkad22,speranza22}; (iii) and non-parametric measurements of the emission lines using both single aperture \citep[e.g.][]{zakamska14} and IFS \citep[e.g.][]{wylezalek20,ruschel-dutra21} data. A precise determination of outflow properties requires high quality data to spatially and spectrally resolve the outflow component, as well as detailed photoionisation models to calculate the gas masses and a correct determination of the gas density \citep[e.g.][]{baron19,davies20,revalski22}. However, this procedure is time-demanding and hard to be applied for large samples.  On the other hand, non-parametric measurements do not depend on details of the line-profile fitting procedure (e.g. choice of the number of components and their physical interpretations), can be applied to large samples and result in estimates of outflow properties consistent with those obtained with other methods \citep[e.g.][]{ruschel-dutra21}. 

Here, we use use non-parametric measurements to map the hot molecular and ionised gas kinematics in a sample of 33 X-ray selected AGN of the local Universe, observed with the Gemini Near-Infrared Integral Field Spectrograph (NIFS).  Our sample is drawn from \citet{rogemar21_extended}, who presented the NIFS data of 36 objects, 34 of them with extended emission observed in  H$_2\,2.1218\,\mu$m and Br$\gamma$ emission lines. In this previous work, we found that the  H$_2$ emission is mainly due to thermal processes -- X-ray heating and shocks -- and its flux distribution is more extended than that of Br$\gamma$. In addition, regions of H$_2$ emission due to shocks are observed in about 40 per cent of the sample. 
The estimated masses of hot molecular and ionised gas in the inner 250\,pc diameter are in the ranges $10^1-10^4$ M$_\odot$ and  $10^4-10^6$ M$_\odot$, respectively. Finally, the only difference found between type 1 and type 2 AGN is that the nuclear emission-line equivalent widths of type 1 objects are smaller than in type 2, attributed to a larger contribution of hot dust emission to the galaxy continuum in the former. In the present paper, we analyse the molecular and ionised gas kinematics using non-parametric measurements of the  H$_2\,2.1218\,\mu$m and Br$\gamma$ emission lines, define the kinematically disturbed region (hereafter KDR) as the region where the AGN significantly affects the gas kinematics (e.g. through AGN winds). The identification of KDRs allow to spot locations where the gas is strongly impacted by outflows, and estimate the outflow properties for both gas phases. 

This work is organized as follows: Section 2 summarizes the sample properties, observations and measurement procedures. Sec.~\ref{sec:kdr} presents the selection criteria to identify KDRs and regions where the gas motions are dominated by the gravitational potential of the galaxies. In Sec.~\ref{sec:outflow}, we estimate the outflow properties, which are discussed in Sec.~\ref{sec:discussion}. Our conclusions are listed in Sec.~\ref{sec:conc}. Additional maps of the gas kinematics for individual objects are included as Supplementary Materials.

\section{Data and Measurements}\label{sec:data} 

\subsection{The sample and data}
The sample used in this work is the same from \citet{rogemar21_extended}, which is composed of 36 AGN observed with Gemini Near-Infrared Integral Field Spectrograph (NIFS) in the K band. In short, the sample was defined by cross-correlating the list of objects included in the 105 month catalogue of the Swift Burst Alert Telescope (BAT) survey \citep{BAT105} at redshifts $z<0.12$, with the objects from the Gemini Science Archive with K-band NIFS data available. As the main aim of this paper is to identify KDRs by the AGN, we have excluded the advanced stage merger NGC\,6240 from the analysis performed in this study, as the disturbed gas may be mainly due to shocks from the interaction process. This object has been  extensively studied, including by near-IR IFS \citep{Ilha16,sanchez18}. In addition, no extended H$_2$ or Br$\gamma$ emission is detected with the NIFS data for two galaxies in the sample of \citet{rogemar21_extended} --  NGC\,3393 and Mrk\,352. Thus, in this paper we present the molecular and ionised gas kinematics for 33 AGN hosts, 16 classified as type 2 AGN and 17 as type 1 \citep{BAT105}. 

\begin{figure}
    \centering
    \includegraphics[width=0.44\textwidth]{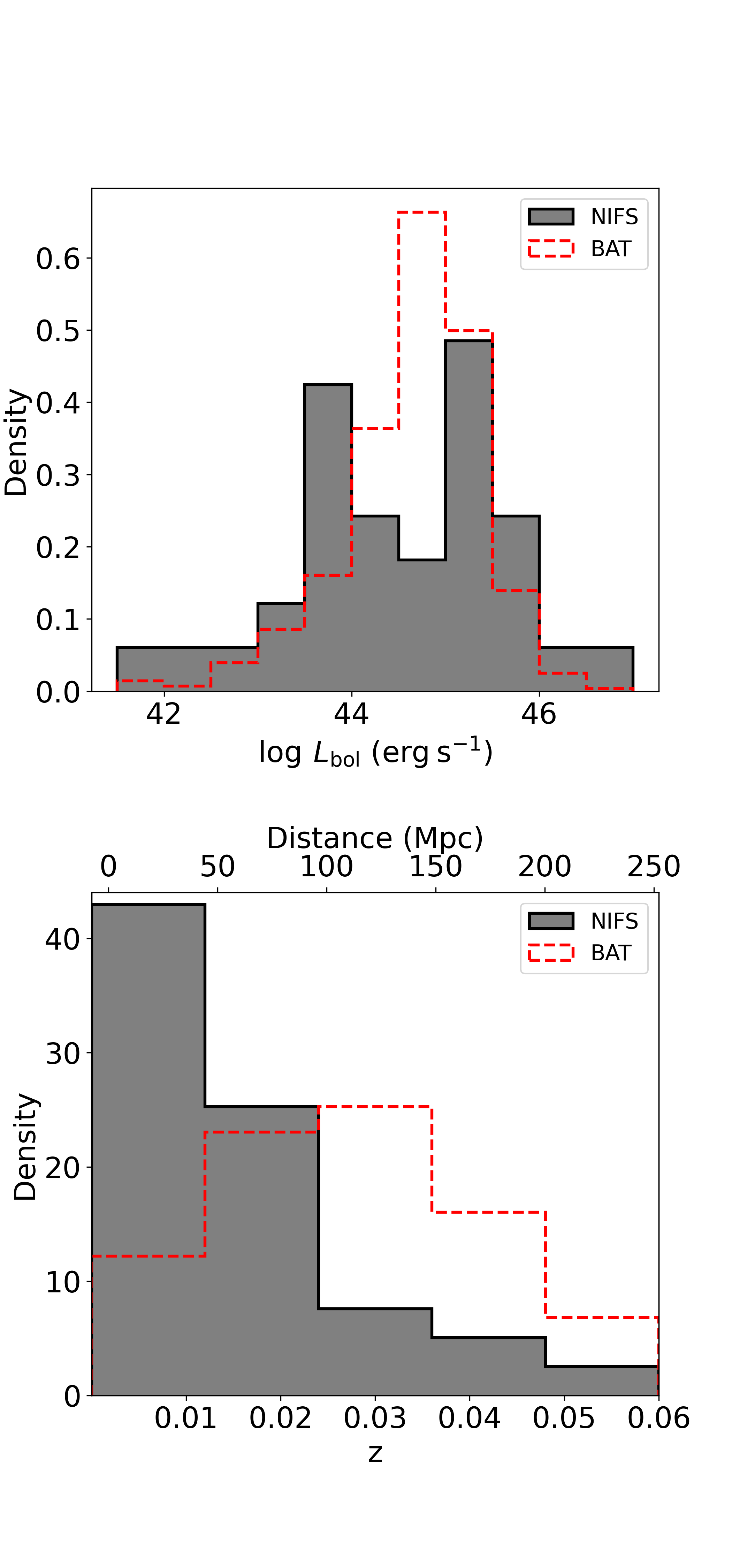} 
\caption{\small AGN bolometric (top) and redshift (bottom) distribution of our sample (gray) and BAT survey \citep[red;][]{BAT105} for the same redshift range.} 
    \label{fig:sample}
\end{figure}

In Figure~\ref{fig:sample} we present the AGN bolometric luminosity (top panel) and redshift (bottom panel) distributions of our sample. The AGN bolometric luminosities are obtained from the hard X-ray (14-195 keV) intrinsic luminosities presented in \citet{ricci17} using the relation  $\log{L_{\rm bol}} = 0.0378(\log{L_{\rm X}})^2 - 2.03\log{L_{\rm X}}+61.6$ from \citep{ichikawa17}. For Mrk607 and Mrk1066, which are not in the sample of \citet{ricci17}, we use the observed X-ray luminosities from \citet{BAT105}. For most galaxies we adopt distances based on their redshifts, except for those with accurate distance determinations: NGC\,3227  \citep[20.5 Mpc; ][]{tonry01} , NGC\,4051 \citep[16.6 Mpc; ][]{yuan21},  NGC\,4151 \citep[15.8 Mpc; ][]{yuan20},  NGC\,4258  \citep[7.6 Mpc; ][]{reid19}, NGC\,4395  \citep[4.0 Mpc; ][]{thim04} and NGC\,6814  \citep[21.65 Mpc; ][]{bentz19}. 

The comparison of the luminosity and redshift distributions of our sample with those from the whole 105 month BAT catalogue \citep{BAT105} for the same redshift range (Fig.~\ref{fig:sample}) shows that these distributions are distinct. In comparison to the BAT catalogue, our sample is biased to lower redshifts and distinct luminosity distribution. The results presented in this paper should not be considered as statisticaly significant  for a complete, volume limited sample
of nearby AGN. The different luminosity and redshift distributions between the BAT catalogue and our sample is due to the fact that we have used archival data with observations obtained to address distinct scientific goals. However, it should be mentioned that the sample used in this work provides one of the largest comparisons of hot molecular and ionized gas kinematics available in the literature, which can provide important information about the emission structure and gas dynamics in these phases in the central region of AGN hosts.

The data were obtained with the Gemini NIFS \citep{mcgregor03}, which has a square field of view of 3$\times$3 arcsec$^2$. The angular resolution of the observations is in the range 0\farcs11--0\farcs44 and velocity resolution of $\sigma_{\rm inst}\sim$20\,km\,s$^{-1}$ \citep[see ][]{rogemar_sample,rogemar21_extended}. The data reduction followed the standard procedures, as described in \citet{rogemar_stellar}, resulting in a single datacube for each galaxy at angular sampling of 0\farcs05$\times$0\farcs05. More details about the sample, observational strategy and data reduction can be found in \citet{rogemar_stellar,rogemar_sample,rogemar21_extended}.

\subsection{Measurements}
We characterise the hot molecular and ionised gas kinematics by measuring the $W_{\rm 80}$, $V_{\rm peak}$ and $V_{\rm cen}$ parameters for the \hml\ and \brg\ emission lines, respectively. The $W_{\rm 80}$ is defined as the smalest width of the line that contains 80\,per cent of its total flux and has been used to look for signatures of ionised gas outflows in AGN hosts \citep{zakamska14,mcelroy15,wylezalek17,wylezalek20,rogemar_N1275,kakkad20}. $V_{\rm peak}$ is the velocity corresponding to the peak of the emission line, which is expected to trace emission from gas in the galaxy disc, while  $V_{\rm cen}$ is the centroid velocity, which is expected to be different from $V_{\rm peak}$ for asymmetric profiles. These properties are computed by using the fits of the spectra performed by \citet{rogemar21_extended}  using the {\sc ifscube} code \citep{ifscube,ruschel-dutra21}, where the \hml\ and \brg\ emission lines are represented by up to three Gaussian components and the underlying continuum is reproduced by a first order polynomial function.  The measurements of $W_{\rm 80}$ and $V_{\rm cen}$ are also obtained with the {\sc ifscube} code, while  $V_{\rm peak}$ is obtained directly from the modelled spectra by computing the velocity corresponding to the maximum flux value within a spectral window of 1500 km\,s$^{-1}$ centred at the peak of each emission line. 

\begin{figure*}
    \centering
     \begin{tabular}{c}
    \includegraphics[width=0.98\textwidth]{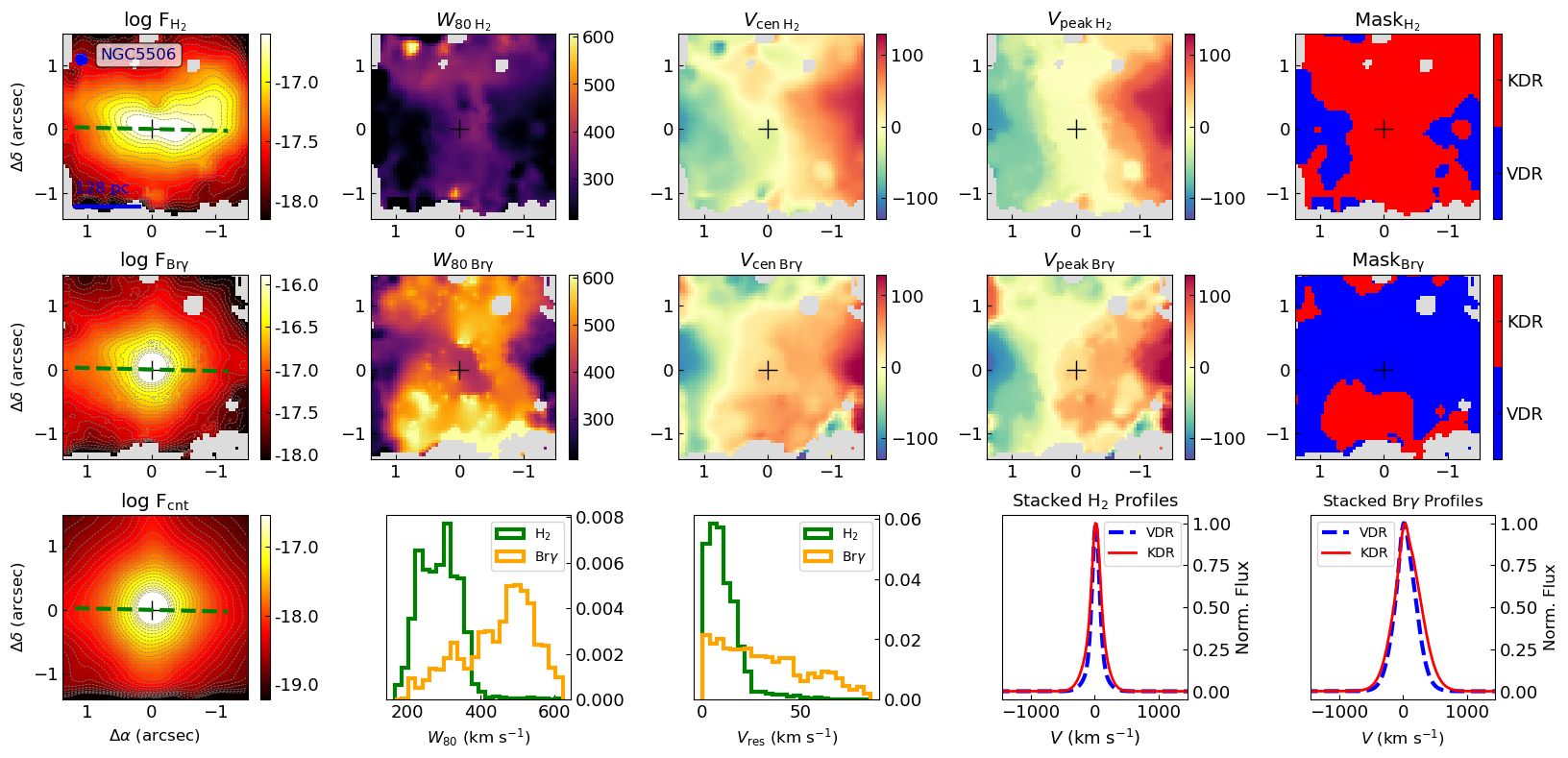} \\
        \hline
      \includegraphics[width=0.98\textwidth]{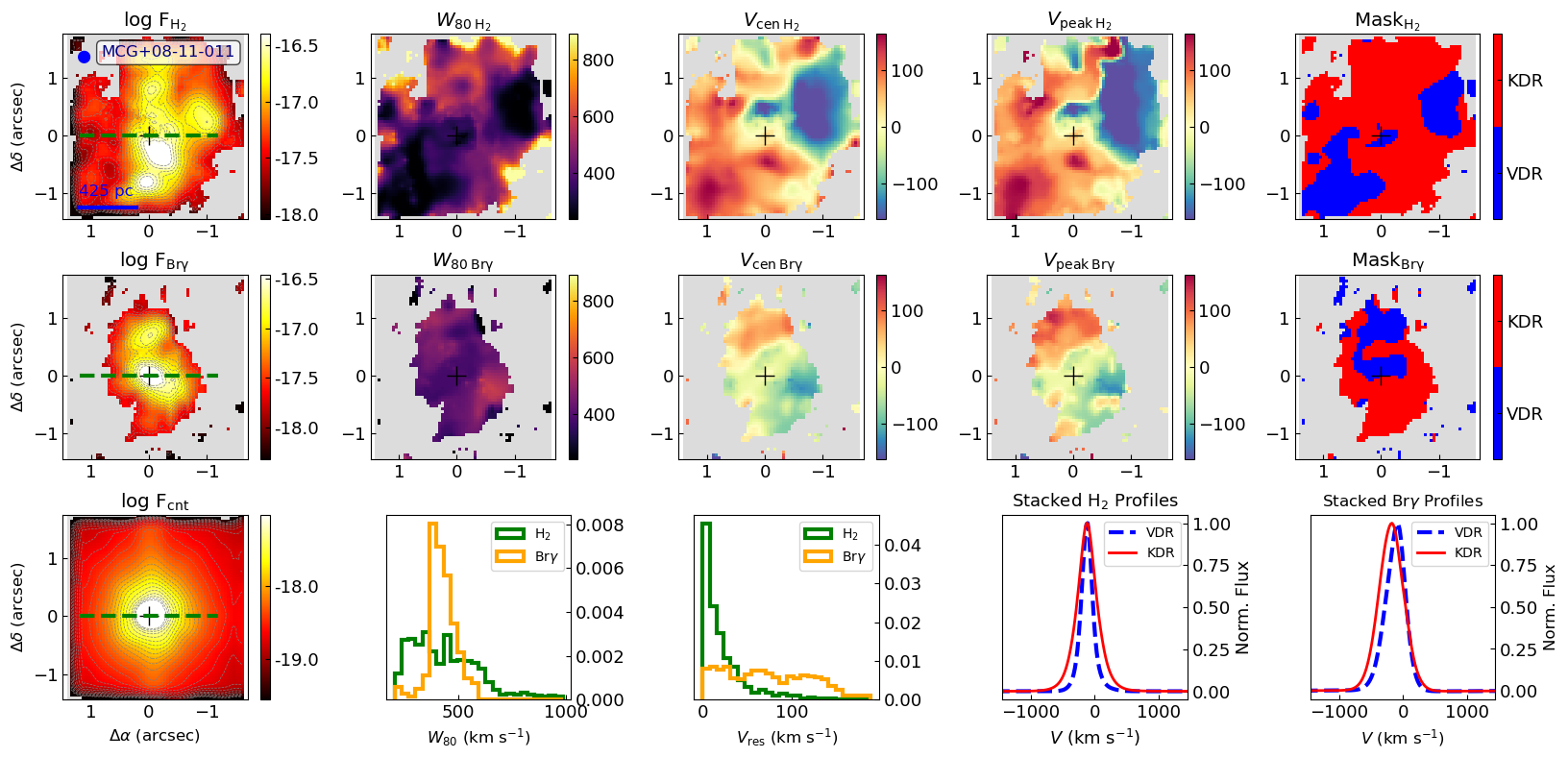} 
\end{tabular}    
\caption{\small Examples of maps for NGC\,5506 (top panels) and MCG+08-11-011 (bottom panels). For each galaxy, the first row  shows the results for the \hml\, and second row  show results for the the \brg\ emission line. From left to right: emission line flux distribution, $W_{\rm 80}$, $V_{\rm cen}$,  $V_{\rm peak}$, and a ``kinematic map'' identifying the kinematically disturbed region (KDR) in red and the virially-dominated region (VDR) in blue. The colour bars show fluxes in erg\,s$^{-1}$\,cm$^{-2}$\,spaxel$^{-1}$ and velocities in km\,s$^{-1}$.  The grey areas identify locations where the emission-line amplitude is below 3 times the continuum noise amplitude (3$\sigma$). The bottom rows show a K-band continuum image in erg\,s$^{-1}$\,cm$^{-2}$\,\AA$^{-1}$\,spaxel$^{-1}$, the density distributions of $W_{\rm 80}$ and $V_{\rm res}=|V_{\rm cen}-V_{\rm peak}|$ and stacked profiles of the H$_2$ and Br$\gamma$ emission lines from the VDR and KDR. Stacked profiles for the KDR are presented only if it corresponds to at least 10 per cent of the spaxels with detected emission. The green dashed lines in the leftmost panels show the orientation of the major axis of the large-scale disk, as presented in \citet{rogemar21_extended}. In all maps, North is up and East is to the left. }
    \label{fig:examplemaps}
\end{figure*}

In Fig.~\ref{fig:examplemaps} we present examples of the  resulting line flux and kinematic maps.  Besides the maps for $W_{\rm 80}$, $V_{\rm peak}$ and $V_{\rm cen}$ for \hml\ and \brg, we also present the flux maps for these emission lines, obtained by integrating the line profiles within a spectral window of 1500 km\,s$^{-1}$ centred at each emission line and a continuum image in the K-band -- already presented in \citet{rogemar21_extended}, as well as maps identifying the KDRs and virially-dominated region (VDRs; see Sec.~\ref{sec:kdr}).   In the bottom row of Fig.\,\ref{fig:examplemaps} we also present histograms for the distributions of $W_{\rm 80}$ values and residual velocities, $V_{\rm res}$, defined as $V_{\rm res}=V_{\rm cen}$-$V_{\rm peak}$ for both emission lines, as well as representative line profiles that will be discussed in the forthcoming sections. In all maps, we masked out regions where the peak of the line profile is not above 3 times the noise in the neighbouring continuum. These regions are shown as grey areas in the maps.

\section{The kinematically disturbed region -- KDR}\label{sec:kdr}

Ionised outflows have been extensively studied in AGN hosts, mostly by using the [O\,{\sc iii}]$\lambda$5007  emission line as its tracer \citep[e.g.][]{zakamska14,wylezalek17,wylezalek20,rogemar_N1275,kakkad20,ruschel-dutra21}. The $W_{\rm 80}$ parameter can be used to identify the KDR, defined as the region where the AGN significantly affects the gas kinematics (e.g through AGN winds or in situ acceleration of the clouds via radiation pressure).  Usually  $W_{\rm 80}>$600\,km\,s$^{-1}$, observed in the [O\,{\sc iii}]$\lambda$5007 line is associated with ionised outflows in quasars \citep[e.g.][]{kakkad20}, while in lower luminosity AGN hosts  $W_{\rm 80}>$500 km\,s$^{-1}$ may already be tracing the KDR emission and consistent with radiatively or mechanically driven AGN outflows  \citep[e.g.][]{wylezalek20}.  The justification for the choice of this threshold is that even for the deepest galaxy gravitational potential of the most massive galaxies, normal orbital velocities and velocity dispersions correspond to lower $W_{\rm 80}$ values.

\begin{figure*}
    \centering
    \includegraphics[width=0.45\textwidth]{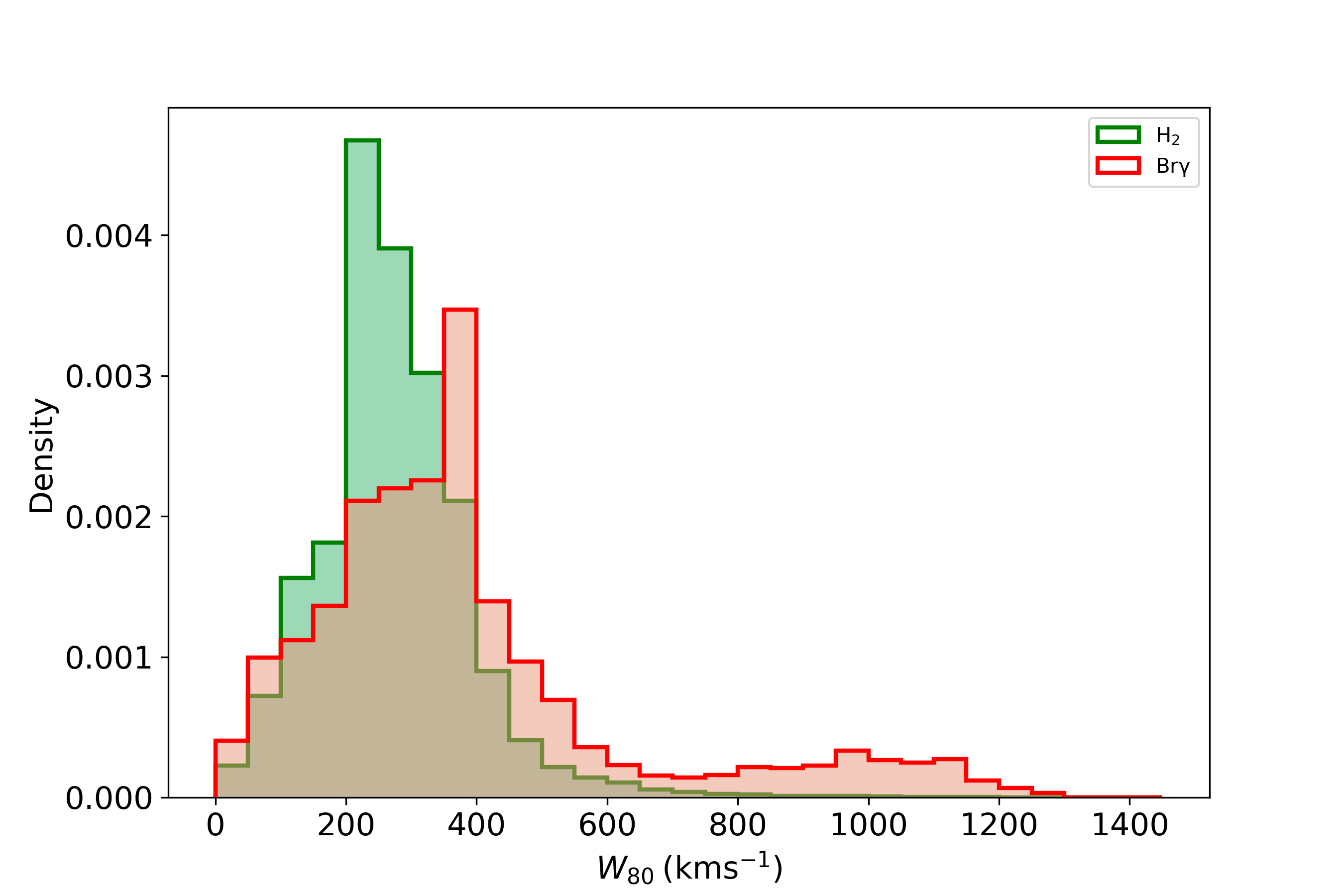}
    \includegraphics[width=0.45\textwidth]{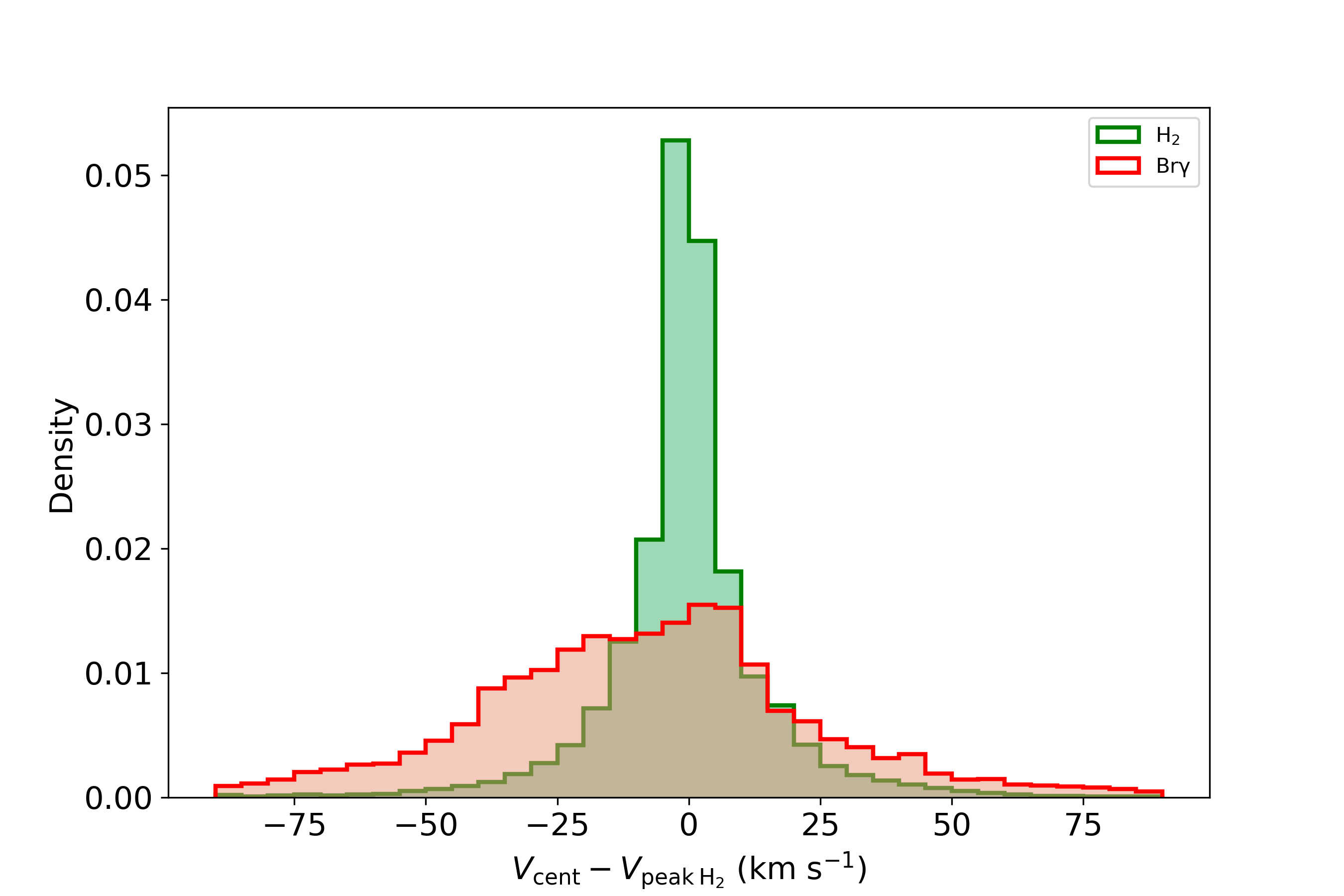}
\caption{\small Left: \brg\ (in red) and \hml\ (in green) $W_{\rm 80}$ distributions for our sample in bins of 50 km\,s$^{-1}$ using measurements in all spaxels.  Right: \brg\ (in red) and \hml\ (in green) residual centroid velocity distributions for our sample in bins of 5 km\,s$^{-1}$. }
    \label{fig:w80nir}
\end{figure*}

The $W_{\rm 80}$ cuts mentioned above are determined using the [O\,{\sc iii}]$\lambda$5007 emission as a tracer of the KDR, and different cuts may be used for distinct tracers, considering the multiple gas phases in the KDR. 
 The Br$\gamma$ emission line is more sensitive to star-formation for which narrower profiles are usually observed, compared to [O\,{\sc iii}] which is a better tracer of the highest ionization gas. Thus, if an outflow component is superimposed to a disc component, the $W_{\rm 80}$ cutoff value for  Br$\gamma$ is expected to be smaller than that for [O\,{\sc iii}]. Similarly, the H$_2$ near-IR emission from the inner region of nearby Seyfert galaxies originates mostly from gas rotating in the disc, also resulting in narrower line profiles compared to those from ionised gas emission lines \citep[e.g.][]{rogemar_sample}.   Figure~\ref{fig:w80nir} presents the $W_{\rm 80}$ distributions of the H$_2$\,2.1218\,$\mu$m (in green) and Br$\gamma$ (in red) emission lines for our sample. Overall, higher $W_{\rm 80}$ values are observed for Br$\gamma$ than for \hml, confirming previous results. In addition, the Br$\gamma$ $W_{\rm 80}$ distribution clearly presents a tail of high values. A less prominent tail is also observed in the H$_2$ distribution, but with smaller values.  Furthermore, a more accurate way of identifying the KDR requires defining different $W_{\rm 80}$ thresholds not only for each line but also for each object.

For the definition of the KDR we also use the galaxies nuclear spectra to measure the emission-line and stellar kinematics. We extract an spectrum within a circular aperture of 0\farcs25 radius centred at the peak of the continuum emission. The size of the aperture is comparable to the angular resolution of the data \citep{rogemar21_extended} and so, the measured kinematics is representative of the nucleus of each galaxy.  We measure the stellar line-of-sight velocity distribution (LOSVD) of each galaxy by fitting the CO absorption band-heads ($\sim$2.29--2.40\,$\mu$m -- rest wavelengths) with the penalized Pixel-Fitting {\sc ppxf} method \citep{cappellari04,cappellari17} using the Gemini library of late spectral type stars observed with the Gemini Near-Infrared Spectrograph (GNIRS) IFU and  NIFS \citep{winge09}. We were able to obtain measurements of the stellar kinematics for 14 objects (Mrks 1066, 348, 607 and NGCs 1052, 1125, 1241, 2110, 3227, 3516, 4051, 4258, 4388, 5899 and 788) in our sample. For the other objects the CO absorption bands are not detected (or are too weak), mainly due to the dilution of the bands by dust emission \citep{burtscher15,rogemar_stellar}.  We measure the emission-line properties using the same procedure described in the previous section, using the {\sc ifscube} code.

The near-IR H$_2$ emission in nearby AGN hosts is usually dominated by emission of gas in rotation in the plane of the galaxy \citep[e.g.][]{hicks13,mazzalay14,rogemar_sample}, thus the peak velocity measured from the \hml\ is expected to be similar to that of the stars. Indeed, the comparison between the nuclear H$_2$ peak velocities and the stellar velocities show that they are consistent with an average difference of $\langle V_{\rm peak\,H_2}-V_{\rm stars}\rangle=-6$\,km\,s$^{-1}$, which is within NIFS velocity resolution of $\sim$50 km\,s$^{-1}$,  and a standard deviation of 34\,km\,s$^{-1}$. The velocity differences are in the range from $-45\pm13$\,km\,s$^{-1}$ (for NGC\,1241) and $70\pm9$\,km\,s$^{-1}$ (for NGC\,1052). This indicates that both the peak \hml\ velocity and stellar velocity trace the systemic velocities of the galaxies and that $V_{\rm peak\,H_2}$  can be used as a proxy of the bulk velocity of the virially-dominated region.

\begin{figure}
    \centering
    \includegraphics[width=0.48\textwidth]{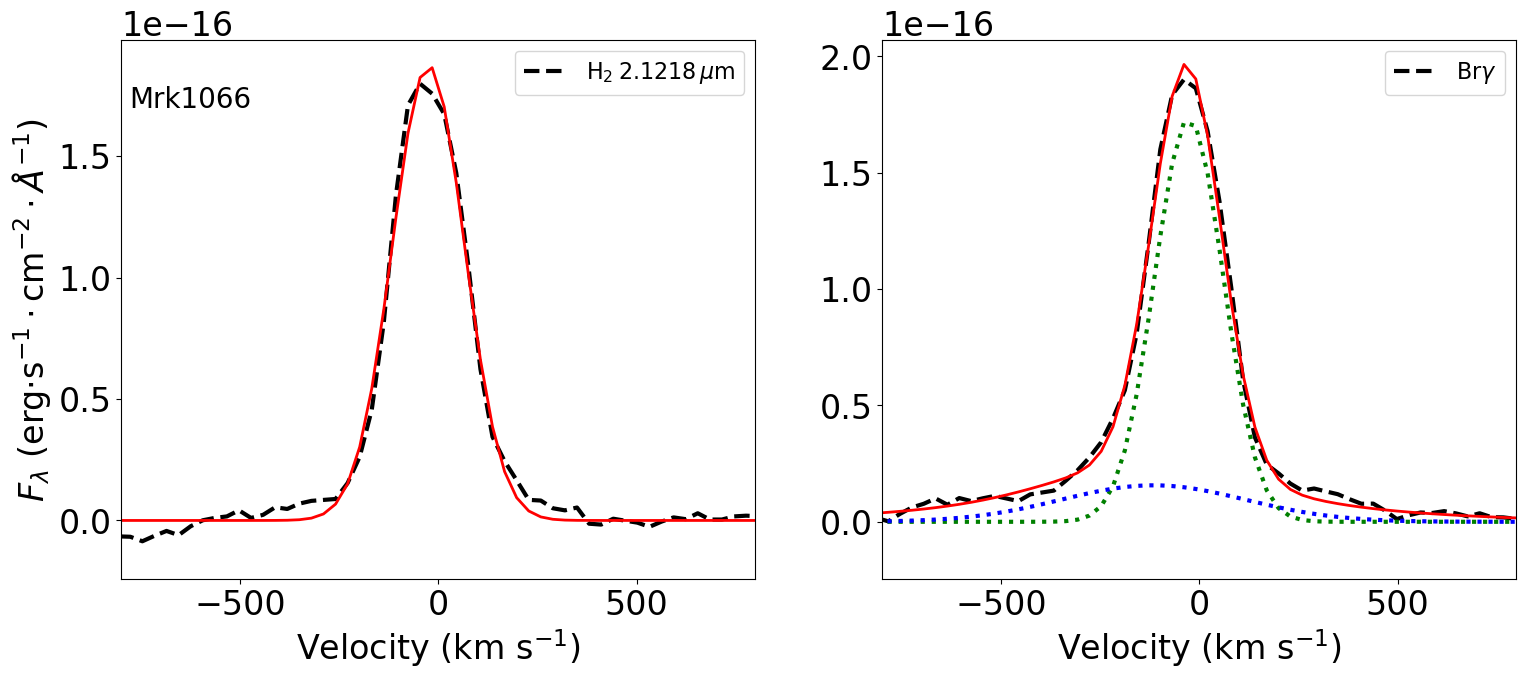}
        \includegraphics[width=0.48\textwidth]{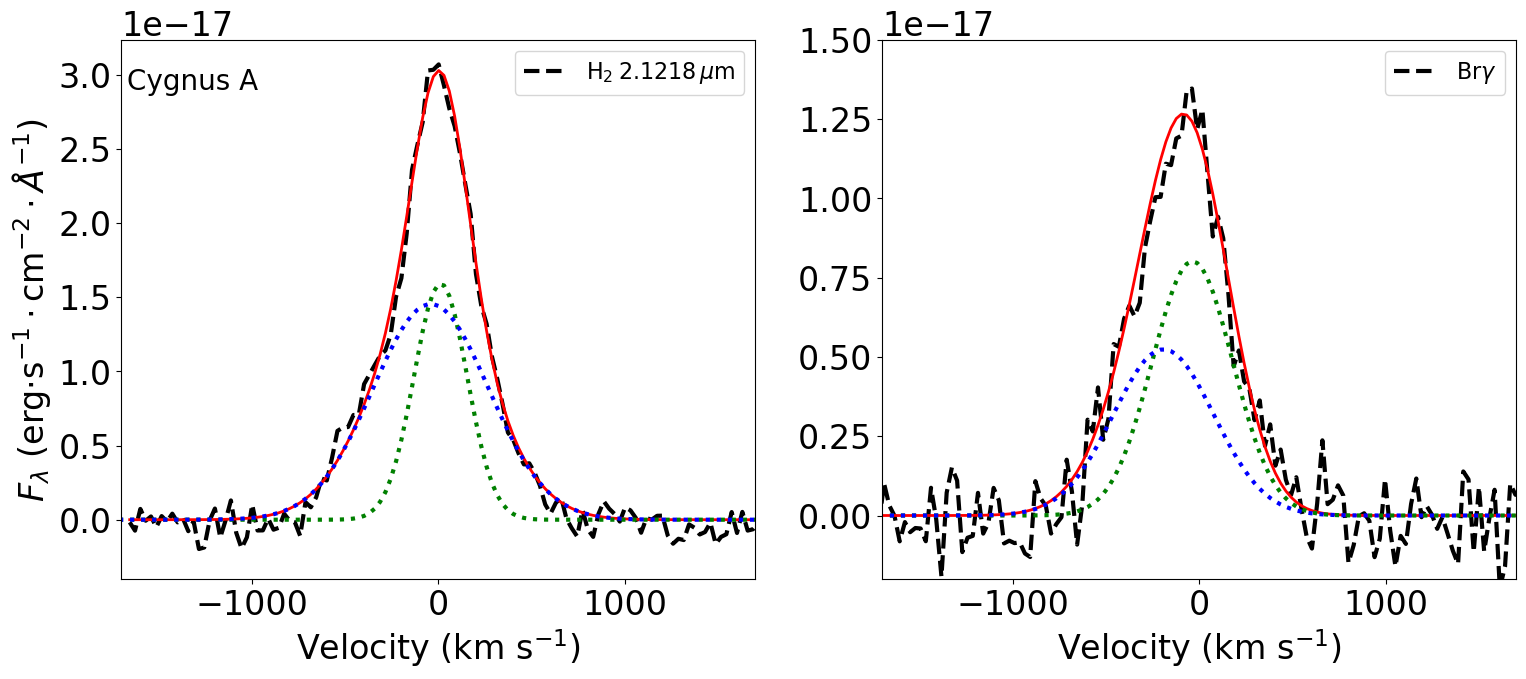}
            \includegraphics[width=0.48\textwidth]{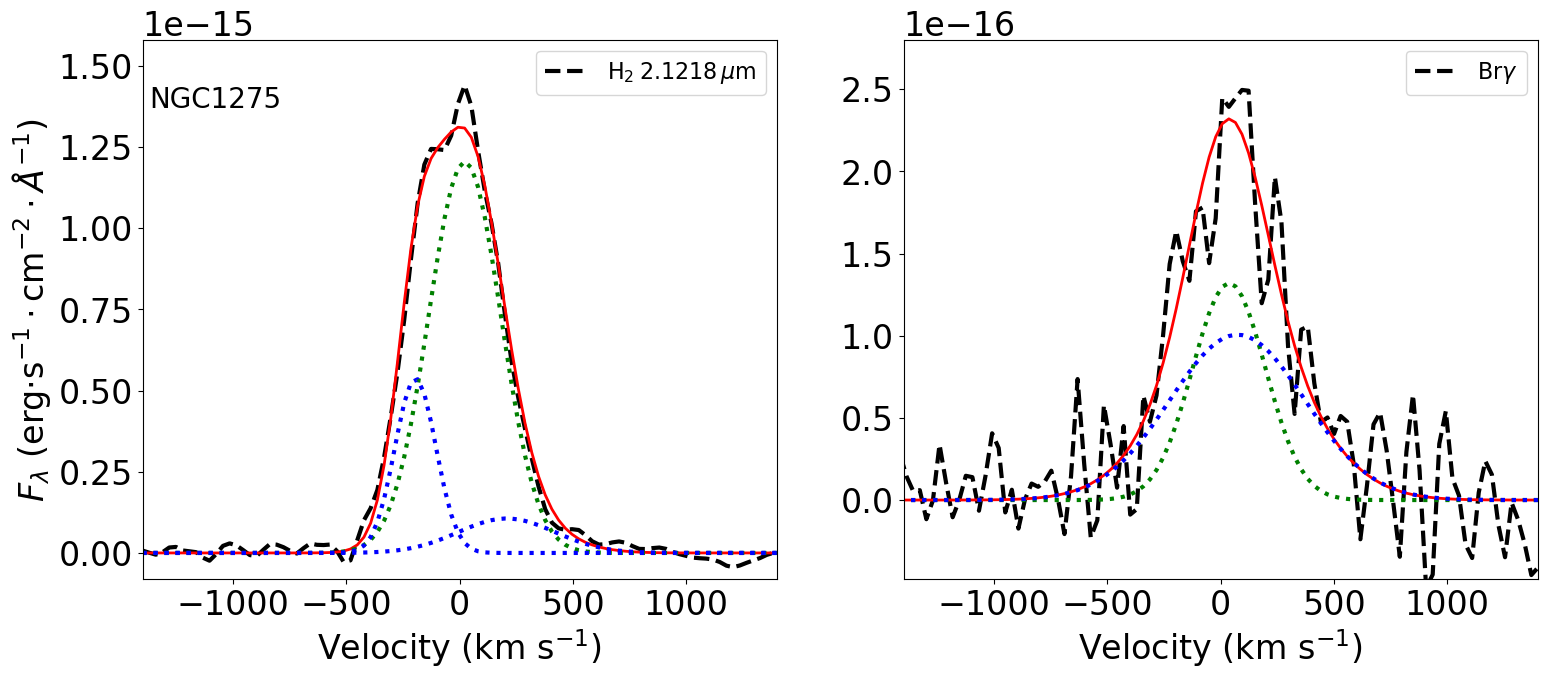}
\caption{\small Examples of fits of the nuclear emission-line profiles, within an aperture of 0\farcs25 radius. Left panels show the fits of the \hml\ and right panels show results for the Br$\gamma$ for Mrk\,1066 (top), Cygnus A (middle) and NGC\,1275 (bottom). The observed profiles are shown in black and the best-fit model in red. If more than one Gaussian function is used to represent the line profile, the dotted green line shows the component attributed to VDR, while the blue dotted lines represents the KDR emission. } 
    \label{fig:fit_line}
\end{figure}

If the gas motions are dominated by the gravitational potential, it is expected that the velocity dispersion (measured here by the $W_{\rm 80}$ parameter) decreases with the distance to the nucleus, so that the nuclear value can be used as the maximum velocity dispersion that can be attributed to the gravitational potential. 
We fit the emission line profiles in the nuclear spectra by a single Gaussian curve. In order to minimize the inclusion of outflows in this nuclear spectrum, we restrict the centroid velocity of the Gaussian to differ by at most 50 km\,s$^{-1}$  from the stellar velocity (50 km\,s$^{-1}$ is roughly the NIFS velocity resolution -- FWHM). If this condition does not apply, we include another Gaussian component in the fit and adopt as representative of the orbital motion the one with centroid velocity closer to the stellar one. For galaxies with no stellar  kinematics measurements, we use the peak \hml\ velocity as reference. If the line profile is well reproduced by a single Gaussian function, we use its $W_{\rm 80}$ plus the corresponding uncertainty as a threshold to define the KDR. Spaxels with $W_{\rm 80}$ values larger than this threshold are associated to the KDR, while spaxels with smaller values are attributed to virialized gas motions, corresponding to the VDR. 
In Fig.~\ref{fig:fit_line} we present examples of the line profile fits and in Table~\ref{tab:w80cut} we show the maximum $W_{\rm 80}$ values attributed to motions under the gravitational potentials for each galaxy.

\begin{table}
    \caption{$W_{\rm 80}$ values for the disc component in our sample (see text) for an aperture of 0\farcs25 radius centred at the peak of the continuum emission. (1) Object, (2) Number of Gaussian functions used to represent the \hml\ emission line, (3) $W_{\rm 80}$ measured for the H$_2$ from the Gaussian component that represents the disc and (4) its uncertainty. (5)--(7) Same as (2)--(4), but for the Br$\gamma$ emission line. We consider spaxels with  $W_{\rm 80}$ values larger than the nuclear values plus their uncertainties as a signature of kinematically disturbed gas. For objects with no nuclear emission, we adopt a $W_{\rm 80}$ threshold of 500 km\,s$^{-1}$ as a lower limit to identify the KDR, following \citet{wylezalek20}. We identify these objects with the superscript $^*$.}
    \begin{tabular}{lcccccc}
    \hline
 (1) & (2) & (3) & (4) & (5) &(6) & (7) \\
 \hline
 	 & \multicolumn{3}{c}{\hml} & \multicolumn{3}{c}{Br$\gamma$} \\
	     	 \hline
	     	 \multicolumn{7}{c}{Type 2}\\
	     	 \hline 
 NGC788  &  1  &  197  &  30  &  2  &  227  &  29  \\    
NGC1052  &  1  &  345  &  36  &  1  &  329  &  65  \\    
NGC1068  &  1  &  200  &  50  &  1  &  444  &  68 \\    
NGC1125  &  2  &  212  &  37  &  2  &  228  &  30 \\    
NGC1241  &  1  &  262  &  29  &  1  &  305  &  41 \\    
NGC2110  &  2  &  347  &  47  &  2  &  230  &  61  \\    
NGC4258  &  1  &  294  &  47  &  1  &  521  &  52    \\    
NGC4388  &  1  &  226  &  28  &  1  &  220  &  29  \\    
NGC5506  &  1  &  197  &  42  &  2  &  580  &  33 \\    
NGC5899  &  1  &  239  &  26  &  2  &  366  &  43 \\    
   Mrk3  &  2  &  275  &  55  &  2  &  330  &  47  \\    
 Mrk348   &  2  &  156  &  54  &  2  &  167  &  69   \\    
 Mrk607   &  3  &  237  &  43  &  3  &  350  &  43   \\    
Mrk1066    &  1  &  232  &  26  &  2  &  222  &  27 \\    
ESO578-G009 &  --  & 500$^*$  &  --  &  --  &  500$^*$  &  -- \\   
Cygnus\,A  &  2  &  365  &  35  &  2  &  551  &  42 \\
	     	 \hline
	     	 \multicolumn{7}{c}{Type 1}\\
	     	 \hline 
NGC1275  &  3  &  408  &  26  &  2  &  412  &  45 \\
NGC3227   &  2  &  224  &  34  &  2  &  356  &  42 \\    
NGC3516   &  2  &  193  &  54  &  1  &  220  &  53 \\    
NGC4051   & 1  &  158  &  32  &  2  &  211  &  38 \\    
NGC4151  &  1  &  355  &  91  &  1  &  393  &  41 \\    
NGC4235   &  1  &  371  &  52  &  --  &  500$^*$  &  -- \\    
NGC4395   &  1  &  84  &  24  &  1  &  101  &  25 \\    
NGC5548    &  1  &  311  &  46  &  2  &  384  &  64 \\    
NGC6814  &  1  &  161  &  30  &  2  &  291  &  43    \\    
  Mrk79   &  1  &  315  &  48  &  1  &  250  &  53 \\    
Mrk509 &  --  & 500$^*$  &  --  &  --  &  500$^*$  &   --\\   
 Mrk618   &  1  &  199  &  46  &  1  &  360  &  43 \\    
 Mrk766  &  1  &  155  &  38  &  2  &  169  &  32    \\    
 Mrk926   &  2  &  201  &  52  &  2  &  382  &  52 \\    
Mrk1044 &  --  & 500$^*$  &  --  &  --  &  500$^*$  &  -- \\   
Mrk1048   &  2  &  191  &  44  &  1  &  148  &  47\\    
MCG+08-11-011  &  1  &  279  &  43  &  1  &  445  &  43 \\    
  \hline

    \end{tabular}
    \label{tab:w80cut}
\end{table}

In some cases, instead of a significant enhancement of the gas velocity dispersion, an outflow produces only a deviation of the centroid velocity of the gas with respect to that corresponding to the galaxy rest frame. This may occur, for instance, in a bipolar outflow launched from the galaxy nucleus at an angle almost perpendicular to the galaxy disc, so that the outflow just weakly interacts with the gas in the disc \citep[e.g.][]{rogemarM79,bianchin22}. In order to account for this possibility, we consider that the $V_{\rm peak}$ parameter traces  
virially-dominated motion and compute the residual velocity  $V_{\rm res}=|V_{\rm cen} - V_{\rm peak}|$.  As the H$_2$ is a better tracer of the emission of the disc, we use the  $V_{\rm peak}$ measured for the \hml\ to compute the $V_{\rm res}$ for both lines. If $V_{\rm res}>{\rm  50\:km\,s^{-1}}$, we assume that the gas motions are not dominated by the gravitational potential. We point out that only a few spaxels are selected using this criteria. The fraction of spaxels with  $V_{\rm res}>{\rm  50\:km\,s^{-1}}$ correspond to only 1 percent of the total number of spaxels for the H$_2$ and about 4 percent for the Br$\gamma$.

In summary, throughout this paper, locations where $W_{\rm 80}$ are larger than the values listed in Table~\ref{tab:w80cut} plus their uncertainties or $V_{\rm res}>{\rm 50\:km\,s^{-1}}$ are identified as the gas KDR. The KDR is assumed to be produced by outflows.  Other regions are identified as the VDRs. In the next section we derive the outflow properties and discuss their uncertainties. 

\section{Properties of the outflows}\label{sec:outflow}

The origin of the KDR in central region of AGN hosts may be due to gas outflows \citep[e.g.][]{wylezalek20,ruschel-dutra21,deconto-machado22} and thus, we can use the flux and kinematic measurements to determine the properties of the hot molecular and ionised gas outflows, traced by the H$_2$ and Br$\gamma$ emission lines, respectively.  In the two-bottom right panels for each galaxy in Fig.~\ref{fig:examplemaps}  we present stacked emission line profiles for the KDR (in red) and VDR (blue). These profiles were constructed by summing up all spaxels of each region using the peak velocity of the line as reference. Such profiles are shown for all objects in the supplementary material. In order to avoid  possible spurious measurements, we only plot the stacked profile for the KDR if this region corresponds to at least 10 per cent of the spaxels with detection of the corresponding emission line. As expected (by definition) the profiles from the KDR are broader and in several cases present distinctly different peak velocities than those from the VDR. 

We find that 31 galaxies (94 per cent)  present at least 10 per cent of the spaxels in the KDR considering only spaxels with detected Br$\gamma$ emission. For the H$_2$, the number of objects with more than 10 per cent of the spaxels in the KDR is 25 (76 per cent). We estimate the outflow properties only for these objects, in each gas phase. 

\subsection{Estimates of the outflows properties}

We estimated the mass of hot H$_2$ and H\,{\sc ii} using the fluxes of the \hml\ and \brg\ emission lines, respectively. The mass of hot molecular gas can be derived by

\begin{equation}
 \left(\frac{M_{\rm H_2}}{M_\odot}\right)=5.0776\times10^{13}\left(\frac{F_{\rm H_{2}\,2.1218}}{\rm erg\,s^{-1}\,cm^{-2}}\right)\left(\frac{D}{\rm Mpc}\right)^2,
\label{mh2}
\end{equation}
where $F_{\rm H_{2}\,2.1218}$ is the H$_2\,2.1218\mu$m emission-line flux and $D$ is the distance to the galaxy. Local thermal equilibrium is assumed with an excitation temperature of 2000\,K \citep[e.g.][]{scoville82,rogemarN1068}. 

The mass of ionised gas is obtained by 
\begin{equation}
 \left(\frac{M_{\rm H\,II}}{M_\odot}\right)=3\times10^{19}\left(\frac{F_{\rm Br\gamma}}{\rm erg\,cm^{-2}\,s^{-1}}\right)\left(\frac{D}{\rm Mpc}\right)^2\left(\frac{N_e}{\rm cm^{-3}}\right)^{-1},
\label{mhii}
\end{equation}
where $F_{\rm Br\gamma}$ is the Br$\gamma$ flux and $N_e$ is the electron density \citep{Osterbrock06,sbN4151Exc}. We adopt an electron density of $N_e=1000\,{\rm cm^{-3}}$, which is a typical value measured in AGN hosts from the [S\,{\sc ii}]$\lambda\lambda$6717,6730 lines \citep[e.g.][]{dors14,dors20,brum17,Freitas18,kakkad18}.

\begin{figure*}
    \centering
    \includegraphics[width=1\textwidth]{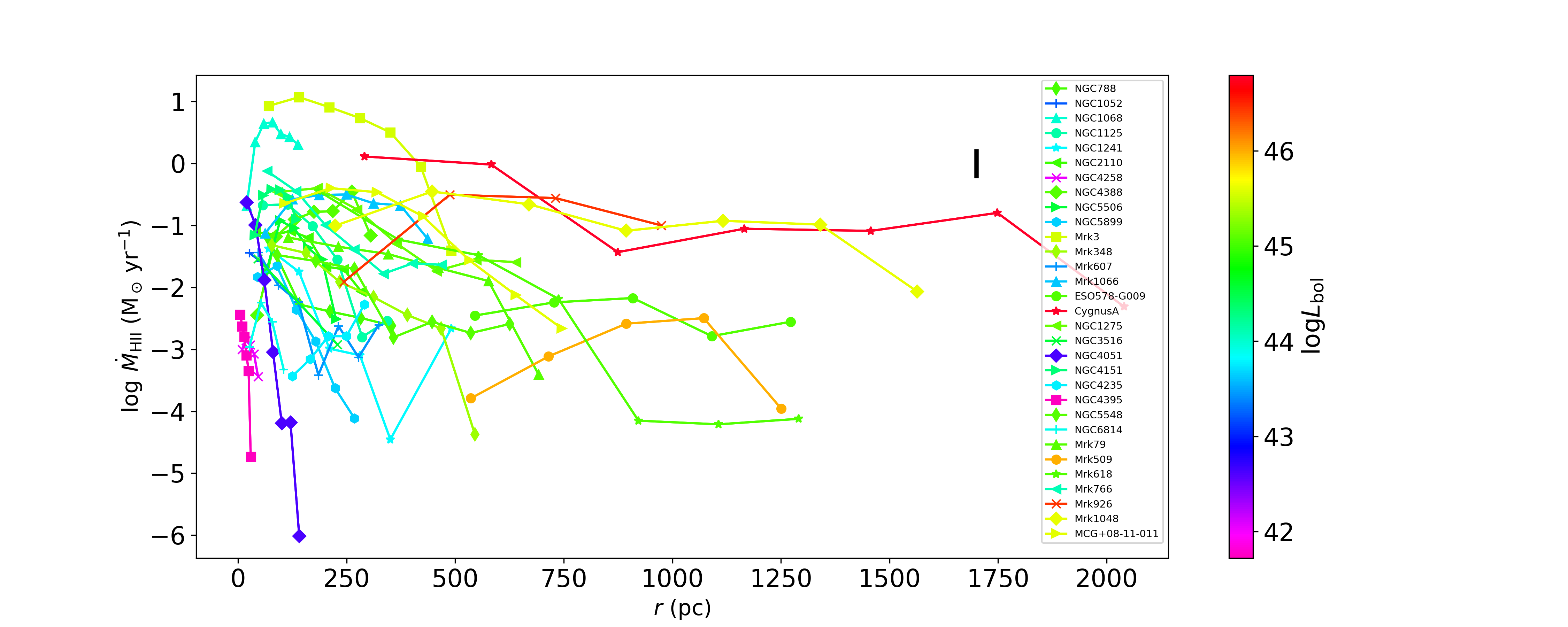}
    \includegraphics[width=1\textwidth]{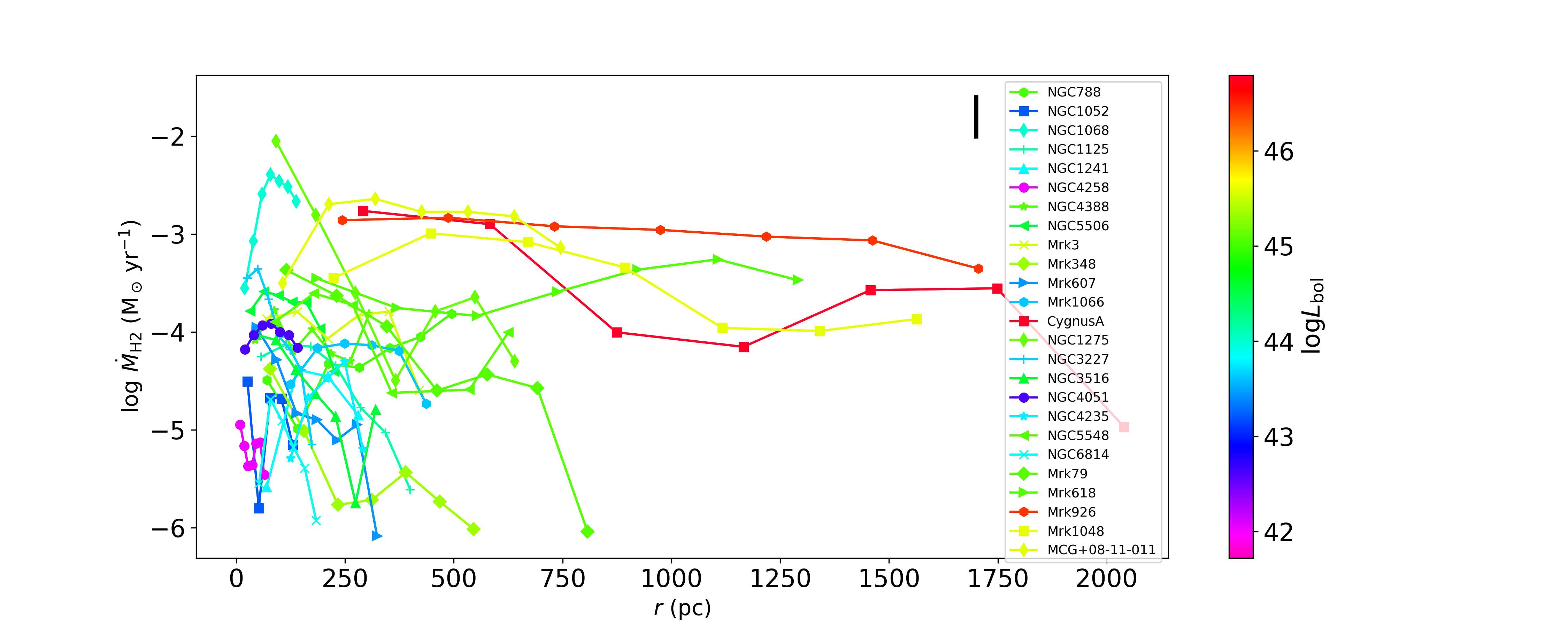}
    \includegraphics[width=1\textwidth]{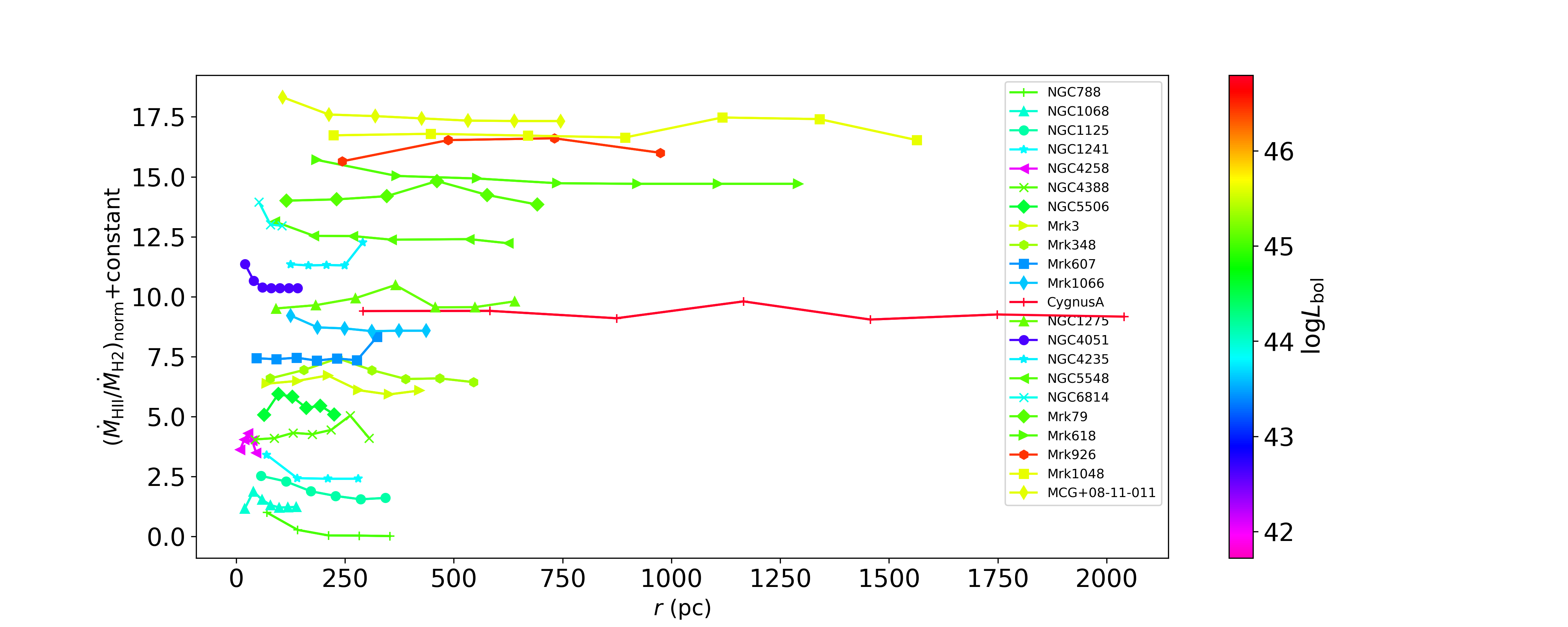}
\caption{\small Radial profiles of the mass-outflow rates in ionised (top) and hot molecular (middle) gas for the galaxies in our sample with detected outflows. The vertical bar in the top-right corner shows typical uncertainties in the mass-outflow rates. The bottom panel show radial profiles of the normalized ratio between the ionised and molecular outflow rates, plus an arbitrary constant, for objects with outflows in both phases.  All profiles are colour coded by the AGN bolometric luminosity, as indicated by the colour bars.}
    \label{fig:radial}
\end{figure*}

Many assumptions are needed to estimate the properties of outflows, which can lead to different results. These properties are affected mainly by the choice of geometries and densities \citep[e.g.][]{harrison18,baron19,lutz20,kakkad20,kakkad22,davies20,ruschel-dutra21,revalski21,revalski22}.  Here,  we estimate the mass of the gas in the outflow  ($M_{\rm out}$) by 
 \begin{equation}
         M_{\rm out}=\sum_i M_{\rm out}^i,  
 \end{equation}
     where the sum is done over all spaxels whose emission is dominated by the outflow as defined above and $M_{\rm out}^i$ is the mass of the outflow calculated for each spaxel $i$, using Eqs.~\ref{mh2} and \ref{mhii} for the molecular and ionised gas, respectively. Following \citet{ruschel-dutra21},  to compute the masses of the gas in the outflow, instead of using the total flux of the emission lines from the spaxels in the KDR ($F_{\rm KDR}$), we use only the fraction of the flux corresponding to absolute velocities larger than $W_{\rm 80}/2$ from the peak velocity. Then, we characterise the outflows in two different ways: (i)  adopting a spherical shell geometry and (ii) obtaining radial profiles of the properties and then using them
to estimate their peak values.
These outflow properties are calculated as follows:

\medskip
\noindent{\bf i. Spherical geometry: Global outflow properties }\\
In this method, we estimate the global or integrated outflow properties. 

 \begin{itemize}

 \item Velocity of the bulk of the outflow ($V_{\rm out}$), defined as 
 \begin{equation}
     V_{\rm out} = \frac{\langle W_{\rm 80\:KDR} F_{\rm KDR} \rangle}{\langle F_{\rm KDR} \rangle},
 \end{equation}
 which is an average velocity over the region dominated by outflows  where $W_{\rm 80\:KDR}$ and $F_{\rm KDR}$ are the $W_{\rm 80}$ values and fraction of the flux corresponding to absolute velocities larger than $W_{\rm 80}/2$ from the peak velocity of the corresponding emission lines for spaxels in the KDR.

 \item Radius of the bulk of the outflow ($R_{\rm out}$), defined as 
 \begin{equation}
     R_{\rm out} = \frac{\langle R_{\rm KDR} F_{\rm KDR} \rangle}{\langle F_{\rm KDR} \rangle},
 \end{equation}
 where $R_{\rm KDR}$ are the distances of outflow-dominated spaxels from the galaxy's nucleus.

\item Mass outflow rate computed by assuming a spherical geometry ($\dot{M}_{\rm out}^{\rm b}$), given by
    \begin{equation}
         \dot{M}_{\rm out}^{\rm b}=\frac{M_{\rm out} V_{\rm out}}{R_{\rm out}}.
     \end{equation}

         \item Kinetic power of the outflow for a spherical geometry ($\dot{E}_{\rm out}^{\rm b}$), given by
    \begin{equation}
         \dot{E}_{\rm out}^{\rm b}=\frac{1}{2}\dot{M}_{\rm out}^{\rm b} V_{\rm out}^2.
     \end{equation}
    
    \end{itemize}

\noindent{\bf ii. Radial profiles: peak outflow properties}

In this method, we calculate the properties as a function of distance from the nucleus and adopt as mass outflow rate and power their peak values. We compute the mass-outflow rates within circular apertures of 0\farcs25 width centred at the nucleus considering only spaxels whose line emission are dominated by the outflow component. For each shell, the mass rate [$\dot{M}_{\rm out}^{\rm sh}(r)$] and kinetic power [$\dot{E}_{\rm out}^{\rm sh}(r)$] of the outflow are computed by:
    
    \begin{equation}
         \dot{M}_{\rm out}^{\rm sh}(r)=\frac{M_{\rm out}^{\rm sh} V_{\rm out}^{\rm sh}}{\Delta\,R},
     \end{equation}
and  
       \begin{equation}
        \dot{E}_{\rm out}^{\rm sh}(r)=\frac{1}{2} \dot{M}_{\rm out}^{\rm sh}(r) (V_{\rm out}^{\rm sh})^2,
     \end{equation}
respectively. In these equations, $r$ corresponds to the distance of the centre of the shell from the nucleus,  $M_{\rm out}^{\rm sh}$ is the mass of the gas in the outflow in the shell obtained using Eqs. \ref{mh2} and \ref{mhii} and  $V_{\rm out}^{\rm sh}$ is the outflow velocity defined as the median of the $W_{\rm 80}$ values within the shell, and $\Delta\,R$ is the width of the shell (0\farcs25). Then, we define the outflow properties using the parameters below. 

 \begin{itemize}
     
 \item Radius corresponding to the peak of the outflow ($R_{\rm peak}$): The  $R_{\rm peak}$ is defined as the radius where the mass-outflow rate radial profile reaches its maximum value. In Fig.~\ref{fig:radial} we present the resulting radial profiles for the ionised and molecular gas mass outflow rate. 

 \item Maximum value of the mass-outflow rate ($\dot{M}_{\rm peak}$):  defined as the peak of the values computed within circular apertures of 0\farcs25 width (i.e. the maximum value of  $\dot{M}_{\rm out}^{\rm sh}$). 
 
  \item Maximum value of the kinetic power of the outflow  ($\dot{E}_{\rm peak}$): defined as the peak of the values computed circular apertures of 0\farcs25 width (i.e. the maximum value of    $\dot{E}_{\rm out}^{\rm sh}$). 
     
 \end{itemize}
 
 In Figure~\ref{fig:radial} we present the radial profiles of the mass-outflow rates in ionised (top panel) and hot molecular (bottom panel) gas. For most galaxies, the radial distribution of the mass outflow rates in both molecular and ionised gas shows an increase with radius from the nucleus until reaching a maximum value  at $R_{\rm peak}$, then decreasing with radius. A similar behaviour was obtained by \citet{revalski21} for six luminous Seyfert galaxies (including NGC\,1068, NGC\,4151 and Mrk\,3) based on observations of the [O\,{\sc iii}] emission line using long slit spectra obtained with the Space Telescope Imaging Spectrograph (STIS) and accurate determinations of radial density profiles.

\subsection{Uncertainties}\label{sec:uncert}

The uncertainties in the properties of AGN outflows are usually high because of the number of assumptions that have to be made to estimate them, such as the geometry, the electron density and velocity of the outflow.  The  electron density represents one of the major source of uncertainties in computing the mass-outflow rates in ionised gas due to different assumptions or tracers used to measure it. Depending on tracer of the electron density used, uncertainties of approximately one order of magnitude are expected for the derived mass outflow rates, as extensively discussed in recent works \citep[e.g.][]{baron19,davies20,revalski22}. 

The electron density in the AGN narrow line region is a strong function of the distance to the nucleus and calculating the masses of ionized gas require multicomponent photoionization models to reproduce the observed emission-line intensities, as done by  \citet{revalski22} for a sample of nearby Seyfert galaxies using HST STIS spectra. These authors found that using a constant density value of 10$^{2}$ cm$^{-3}$ overestimates the mass of ionised gas, while adopting a value of 10$^{3}$ cm$^{-3}$ results in an agreement within $\pm$1\,dex between the masses estimated from H recombination lines and those obtained from photoionization models. As mentioned above,  in this work we adopt $N_e=1\,000\,{\rm cm^{-3}}$, which is a typical value for AGN hosts using the [S\,{\sc ii}] doublet \citep[e.g.][]{dors14,dors20,perna17,Freitas18,kakkad18}, and thus the expected uncertainty regarding the density choice is $\sim$1\,dex in the outflow properties.

The uncertainty associated to the geometry of the outflow is smaller, with distinct geometries (e.g. conical and shells) resulting in overall similar values of the mass-outflow rates, and with typical standard deviations of the differences of $<$0.5 dex between the estimates using distinct geometries \citep[e.g.][]{kakkad22}.

Another source of uncertainty in the estimate of outflow properties using spatially resolved observations is associated to the selection criteria of the outflow dominated spaxels and the emission-line fluxes used to compute the mass of gas in the outflow. In order to estimate the effect of different assumptions, we estimate the global mass outflow rates using three sets of assumptions. The assumptions are the following:  {\it Method 1} -- we assume that spaxels with $W_{\rm 80}$ larger than the values presented in Tab.~\ref{tab:w80cut} are associated to outflows as in the calculation described in Sec.~\ref{sec:outflow} (which we will refer to as the  {\it adopted method} for comparison purposes), but we use the total line flux of the spaxel instead of only the fluxes of its wings as done in Sec.~\ref{sec:outflow}. A fraction of the emission-line flux may be due to the emission of the gas in the disc (at lower velocities), resulting in an overestimation of the gas mass in the outflow. Thus, {\it method 1} likely results in upper limits for the outflow properties.   {\it Method 2} -- the spaxels corresponding to outflows are selected using a single $W_{\rm 80}$ threshold of 500\,km\,s$^{-1}$ as defined by \citet{wylezalek20}, and the total flux of the line in each spaxel is used to compute the mass of gas. {\it Method 3} -- the same $W_{\rm 80}$ threshold  of {\it method 2} is used, and the mass of ionized gas is calculated using the flux corresponding to absolute velocities larger than $W_{\rm 80}/2$ from the peak velocity of the corresponding emission line. These assumptions will likely result in a lower limit of the mass outflow rate, as it does not include lower velocity outflows. 

\begin{figure*}
    \centering
    \includegraphics[width=1.1\textwidth]{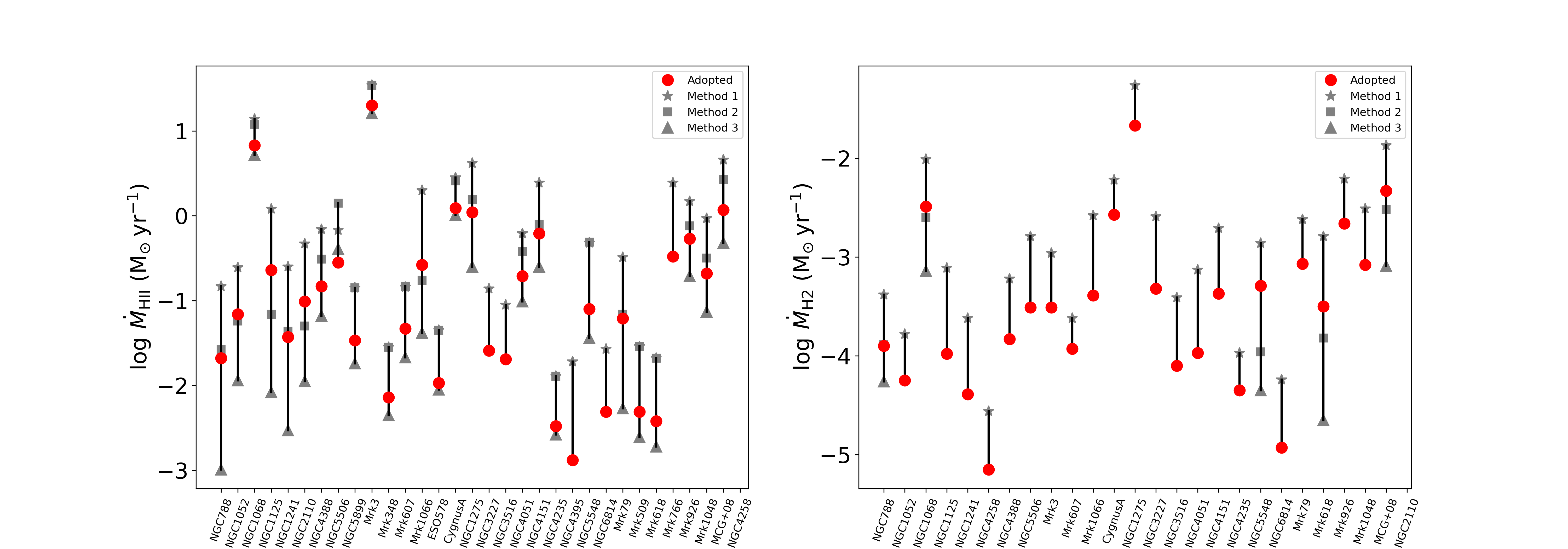}
\caption{\small Comparison of the global mass outflow rates in ionized (left) and hot molecular (right) gas obtained using different assumptions (see Sec.~\ref{fig:uncert}). }
    \label{fig:uncert}
\end{figure*}

In Figure~\ref{fig:uncert}, we present the comparison among the mass-outflow rates in ionized (left) and hot molecular (right) gas derived using the different set of assumptions, for each object. The mean differences between the maximum and minimum values are 1.0$\pm$0.5 dex for the ionized gas and  0.7$\pm$0.4 dex for the hot molecular gas. The highest discrepancies are of about two orders of magnitude for the ionized gas and one order of magnitude for the molecular gas, with the {\it adopted method} resulting in values between the maximum and minimum estimates for most objects. 

With these caveats in mind, we summarise the outflow properties and compare with values available in the literature, most of which share the same sources of uncertainty in measurements as ours.

\subsection{Summary of derived outflow properties and comparison with the literature}

\begin{figure*}
    \centering
    \includegraphics[width=1\textwidth]{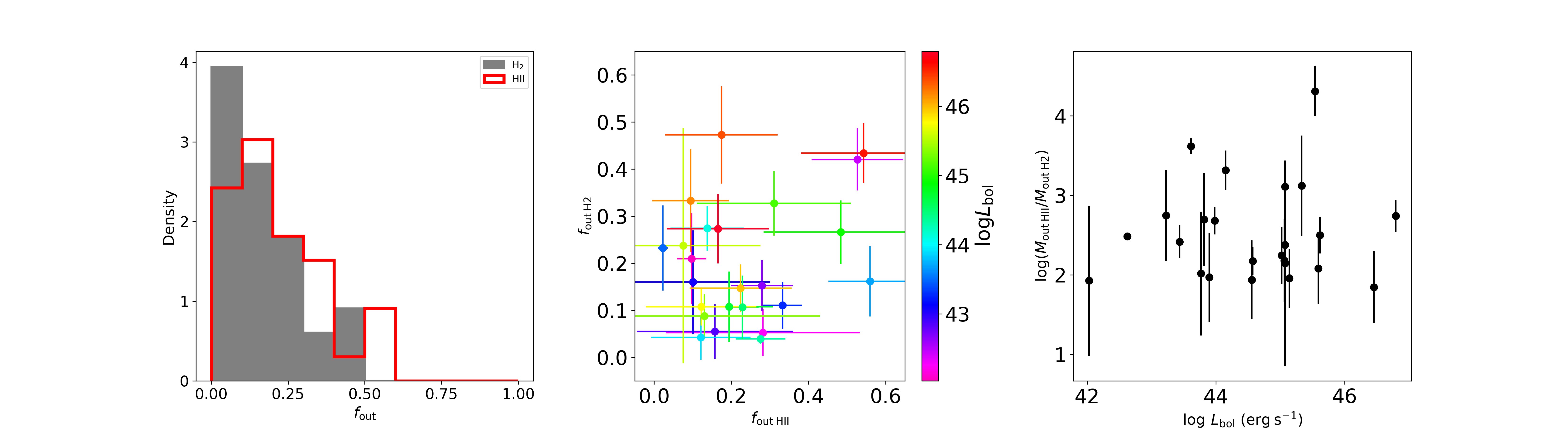}
\caption{\small Left: Distribution of the fraction of gas in the outflow ($f_{\rm out}=\frac{M_{\rm out}}{M_{\rm total}}$), compared to the total masses of ionised (in red) and molecular (in grey) gas. Middle: Plot of the $f_{\rm out}$ observed in H$_2$ (y-axis) vs. those seen in H\,{\sc ii} (x-axis), colour coded by the AGN bolometric luminosity. Right: Plot of the ratio of mass of ionised and molecular gas in the outflow versus the  AGN bolometric luminosity. }
    \label{fig:fraction}
\end{figure*}

In Tables~\ref{tab:ion_outflow} and \ref{tab:molec_outflow} we present the derived properties of the ionised and molecular outflows, respectively. The uncertainties in the outflow parameters quoted in the table are estimated by propagating the uncertainties in the fluxes of the \hml\ and \brg\ emission lines, the uncertainties in the radius (estimated as the the standard error of the radii of individual spaxels in the KDR)  and velocity of the outflow (estimated as  the standard error of $W_{\rm 80}$ values in the KDR). This uncertainties can be considered as lower limits, as systematic errors regarding the assumptions (e.g. densities and geometry) used to calculate the outflow properties may be the dominant source of uncertainties in deriving outflow properties, as discussed in Sec.~\ref{sec:discussion}.  The masses of ionised gas in the outflow are in the range 10$^3$--10$^7$~M$_\odot$, while the molecular outflows show masses in the range 10$^1$--10$^4$ M$_\odot$.   Figure~\ref{fig:fraction} shows the distributions of the mass fraction of the gas in the outflow relative to the total mass of molecular and ionised gas ($f_{\rm out}=M_{\rm out}/M_{\rm gas}$), obtained using Eq.~\ref{mh2} and \ref{mhii}. The mass of the gas in the outflow is estimated by considering only spaxels from the KDR, while the total gas masses are obtained by summing the contributions of all spaxels within the observed field-of-view with detected emission. For galaxies with no detected outflows, we assume $f_{\rm out}=0$. The $f_{\rm out}$ in ionised and molecular gas are listed in Tabs.~\ref{tab:ion_outflow} and \ref{tab:molec_outflow}, respectively. For most galaxies, the amount of outflowing gas corresponds to less than 30 per cent of the total gas reservoir in the central region of the galaxies, both in ionised and molecular gas. 

\begin{figure*}
    \centering
    \includegraphics[width=0.98\textwidth]{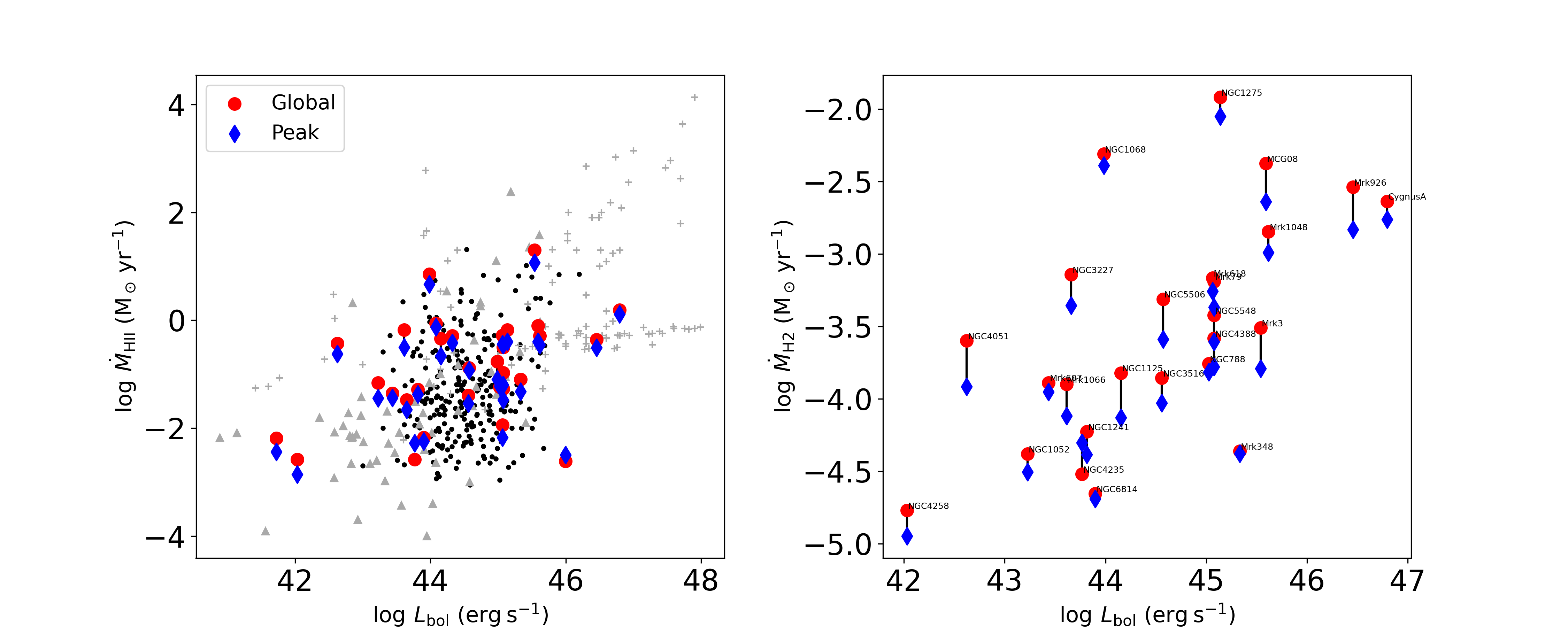}
        \includegraphics[width=0.98\textwidth]{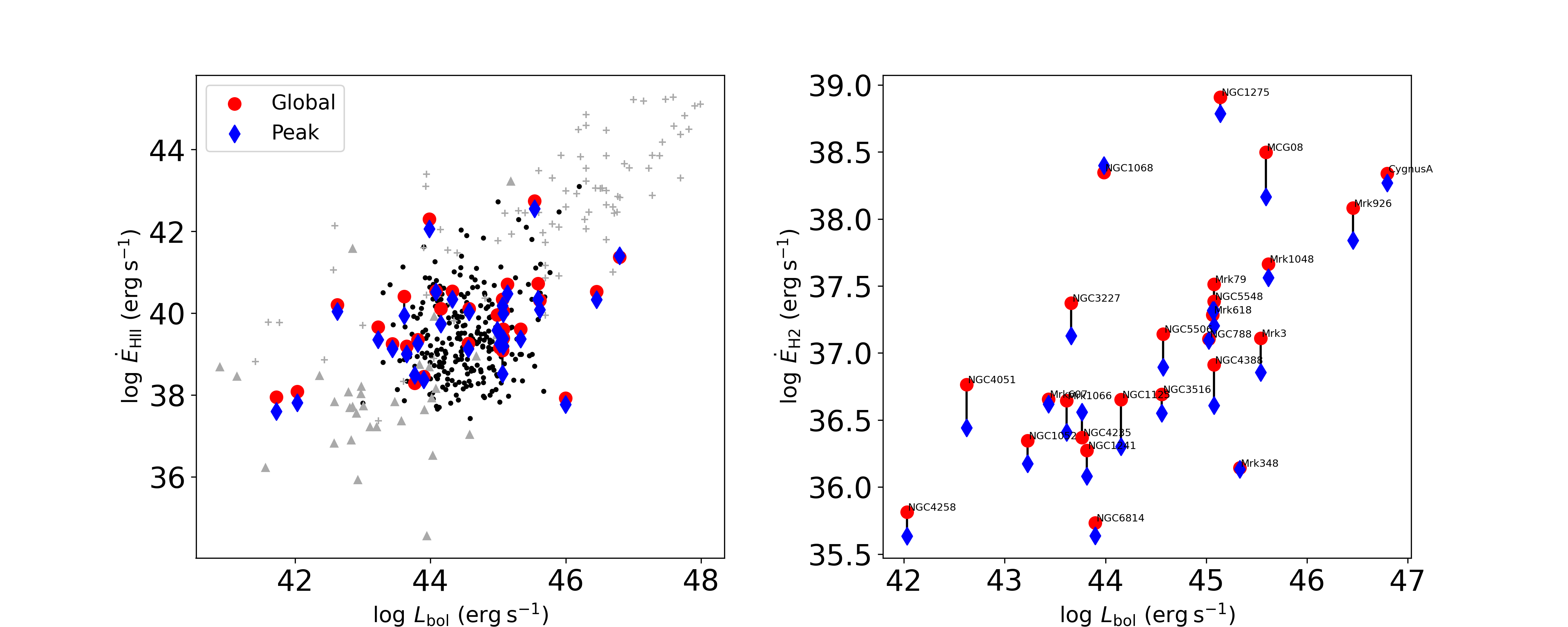}

\caption{\small  Top panels: plot of mass-outflow rates vs. AGN bolometric luminosity. Mass-outflow rates for the ionised (left) and hot molecular gas (right) obtained by assuming a spherical shell geometry (red circles) and peak from radial profiles (blue diamonds) are shown for each object. The grey and black points are a compilation of values from the literature, as described in the text. 
Bottom panels: same as top panels, but for the kinetic power of the outflows. Typical uncertainties in both parameters are 0.2 dex for both gas phases, as shown in Tabs.~\ref{tab:ion_outflow} and \ref{tab:molec_outflow}.}
    \label{fig:mout}
\end{figure*}

In the top panels of Figure~\ref{fig:mout} we show the plots of the mass outflow rates obtained for the ionised (left panel) and molecular gas against the AGN bolometric luminosity. 
For each object, we present estimates using the two approaches described above, for the global properties of the outflow (red circles) and peak value of the radial profile (blue diamond). The black points represent a compilation of measurements available in the literature for ionised outflows for nearby AGN, adopting various values of electron density for the outflow. These points include estimates for low luminosity AGN based on SDSS-III spectra using densities based on the ionisation parameter \citep{baron19}, luminous Seyferts based on nuclear spectra using densities measurements from auroral and transauroral lines \citep{davies20} and on long-slit HST spectra and photoionisation models \citep{revalski21}, and nearby QSOs based on HST spectra and photoionisation models \citep{falcao21}.  

As discussed in recent works, the choice of the method used to estimate the density of ionised outflows is one of the main sources of uncertainty to estimate their mass-outflow rates \citep{baron19,davies20,revalski22}, resulting in values that can differ by approximately one order of magnitude. For example, the most common method used in the optical to estimate the electron density, based on the [S\,{\sc ii}]$\lambda\lambda$6717,6731 doublet, provides values significantly lower than the real densities of ionised ouflows \citep{davies20}.  In Fig.~\ref{fig:mout}, we include mass-outflow rates from \citet{ruschel-dutra21}, \citet{deconto-machado22} and \citet[][ for their estimates using a circular 3 arcsec diameter aperture]{kakkad22}, estimated  using densities based on the  [S\,{\sc ii}] lines, measured from spatially resolved spectra.   These estimates are shown as grey triangles. 

\begin{landscape}
\begin{table}
    \caption{Properties of the ionised gas Outflows.  (1) Name of the galaxy; (2) Adopted distance: for most galaxies, the distances are estimated from the redshift, except for except for those with accurate distance determinations: NGC\,3227  \citep{tonry01} , NGC\,4051 \citep{yuan21},  NGC\,4151 \citep{yuan20},  NGC\,4258  \citep{reid19}, NGC\,4395  \citep{thim04} and NGC\,6814  \citep{bentz19}.  (3) AGN bolometric luminosity; (4) Radius of the bulk of the outflow (spherical geometry); (5) Radius of the peak of the outflow (spherical shells geometry); (6) Total mass of the outflow considering only spaxels in the KDR; (7) Mass fraction of the gas in the outflow; (8--12) Properties of the outflows estimated using the two methods described in the text.}
    \begin{tabular}{lccccccccccc}
    \hline
    (1) & (2) & (3) & (4) & (5) & (6) & (7) & (8) & (9) & (10) & (11) & (12)\\
 &   &  & & & & &  \multicolumn{3}{c}{Global/Bulk outflow} &  \multicolumn{2}{c}{Radial profiles}\\
    	  \cline{8-10} \cline{11-12}\\

   Galaxy & $D$ & $\log L_{\rm bol}$&$R_{\rm out}$ & $R_{\rm peak}$ & $\log M_{\rm out}$ & $f_{\rm out}$ & $V_{\rm out}$ &  $\log \dot{M}_{\rm out}^{\rm b}$ & $\log \dot{E}_{\rm out}^{\rm b}$ & $\log \dot{M}_{\rm peak}$ & $\log \dot{E}_{\rm peak}$ \\
   
    & [Mpc] & [${\rm erg\: s}^{-1}$] & [pc] & [pc] &  [M$_\odot$]   &&  [${\rm km\: s}^{-1}$]  & [M$_\odot{\rm yr^{-1}}$] & [${\rm erg\: s}^{-1}$] & [M$_\odot{\rm yr^{-1}}$] & [${\rm erg\: s}^{-1}$] \\
    
	     	 \hline
	     	 \multicolumn{12}{c}{Type 2}\\
 NGC788 & 58.3  & 45.02 &  165$\pm$39 &   70$\pm$30 & 3.93$\pm$0.49 & 0.04$\pm$0.04 & 284$\pm$49 &  -1.68$\pm$0.50  & 38.73$\pm$0.50 &  -1.66$\pm$0.50 & 38.82$\pm$0.50  \\    
NGC1052 & 21.4  & 43.23 &   97$\pm$34 &   25$\pm$34 & 3.54$\pm$0.46 & 0.28$\pm$0.25 & 459$\pm$37 &  -1.16$\pm$0.47  & 39.66$\pm$0.47 &  -1.45$\pm$0.52 & 39.36$\pm$0.52  \\    
NGC1068 & 16.3  & 43.98 &  414$\pm$29 &   78$\pm$24 & 5.74$\pm$0.19 & 0.50$\pm$0.13 & 922$\pm$42 &  0.83$\pm$0.19  & 42.26$\pm$0.19 &  0.61$\pm$0.21 & 41.97$\pm$0.22  \\    
NGC1125 & 47.1  & 44.15 &  168$\pm$104 &  114$\pm$102 & 4.83$\pm$0.14 & 0.14$\pm$0.07 & 302$\pm$32 &  -0.64$\pm$0.24  & 39.82$\pm$0.24 &  -0.95$\pm$0.33 & 39.43$\pm$0.34  \\    
NGC1241 & 57.9  & 43.82 &  149$\pm$40 &   70$\pm$31 & 4.00$\pm$0.68 & 0.12$\pm$0.20 & 371$\pm$45 &  -1.43$\pm$0.68  & 39.21$\pm$0.68 &  -1.51$\pm$0.68 & 39.12$\pm$0.68  \\    
NGC2110 & 33.4  & 44.99 &  283$\pm$41 &  121$\pm$33 & 4.39$\pm$0.32 & 0.18$\pm$0.09 & 424$\pm$37 &  -1.01$\pm$0.32  & 39.75$\pm$0.32 &  -1.36$\pm$0.33 & 39.32$\pm$0.33  \\    
NGC4258 & 7.6   & 42.03 &        --      &        --      &        --      &        --      &        --      &         --       &        --      &        --      &        --      \\    
NGC4388 & 36.0  & 45.08 &  495$\pm$80 &  261$\pm$41 & 4.90$\pm$0.28 & 0.16$\pm$0.07 & 366$\pm$39 &  -0.83$\pm$0.29  & 39.80$\pm$0.29 &  -0.69$\pm$0.29 & 39.76$\pm$0.29  \\    
NGC5506 & 26.6  & 44.57 &  334$\pm$46 &   96$\pm$40 & 4.70$\pm$0.08 & 0.05$\pm$0.01 & 552$\pm$42 &  -0.55$\pm$0.10  & 40.44$\pm$0.10 &  -0.57$\pm$0.17 & 40.38$\pm$0.18  \\    
NGC5899 & 36.9  & 43.65 &  153$\pm$33 &   89$\pm$29 & 3.74$\pm$0.43 & 0.16$\pm$0.13 & 381$\pm$42 &  -1.47$\pm$0.44  & 39.19$\pm$0.44 &  -1.65$\pm$0.44 & 39.00$\pm$0.44  \\    
   Mrk3 & 57.9  & 45.54 &  239$\pm$31 &  140$\pm$29 & 6.51$\pm$0.15 & 0.56$\pm$0.11 & 939$\pm$48 &  1.30$\pm$0.16  & 42.74$\pm$0.16 &  1.07$\pm$0.17 & 42.56$\pm$0.18  \\    
 Mrk348 & 64.3  & 45.33 &  346$\pm$60 &  233$\pm$40 & 3.57$\pm$0.42 & 0.02$\pm$0.04 & 474$\pm$50 &  -2.14$\pm$0.43  & 38.71$\pm$0.43 &  -2.15$\pm$0.43 & 38.59$\pm$0.43  \\    
 Mrk607 & 38.1  & 43.44 &  146$\pm$35 &   46$\pm$33 & 3.93$\pm$0.18 & 0.13$\pm$0.09 & 352$\pm$53 &  -1.33$\pm$0.20  & 39.27$\pm$0.21 &  -1.44$\pm$0.29 & 39.13$\pm$0.29  \\    
Mrk1066 & 51.4  & 43.61 &  365$\pm$40 &  249$\pm$33 & 5.16$\pm$0.15 & 0.11$\pm$0.06 & 374$\pm$31 &  -0.58$\pm$0.16  & 40.06$\pm$0.16 &  -0.85$\pm$0.17 & 39.62$\pm$0.17  \\    
ESO578-G009& 150.0  & 45.06 &  576$\pm$178 &  909$\pm$78 & 4.26$\pm$0.53 & 0.04$\pm$0.05 & 564$\pm$143 &  -1.97$\pm$0.54  & 39.03$\pm$0.54 &  -2.17$\pm$0.54 & 38.52$\pm$0.54  \\    
CygnusA & 240.4  & 46.80 &  268$\pm$41 &  291$\pm$39 & 6.19$\pm$0.25 & 0.28$\pm$0.10 & 574$\pm$45 &  0.09$\pm$0.25  & 41.11$\pm$0.25 &  -0.01$\pm$0.26 & 41.28$\pm$0.26  \\    
	     	 \hline
	     	 \multicolumn{12}{c}{Type 1}\\
	     	 \hline 
NGC1275 & 75.4  & 45.14 &  168$\pm$50 &   91$\pm$49 & 5.52$\pm$0.35 & 0.28$\pm$0.16 & 480$\pm$46 &  0.04$\pm$0.36  & 40.90$\pm$0.36 &  -0.13$\pm$0.38 & 40.67$\pm$0.38  \\    
NGC3227 & 20.5  & 43.66 &  201$\pm$38 &   49$\pm$27 & 3.47$\pm$0.33 & 0.03$\pm$0.05 & 402$\pm$46 &  -1.59$\pm$0.34  & 39.12$\pm$0.34 &  -1.60$\pm$0.36 & 39.09$\pm$0.36  \\    
NGC3516 & 37.7  & 44.56 &   67$\pm$35 &   45$\pm$35 & 3.21$\pm$0.76 & 0.21$\pm$0.20 & 381$\pm$46 &  -1.69$\pm$0.76  & 38.97$\pm$0.76 &  -1.89$\pm$0.76 & 38.78$\pm$0.76  \\    
NGC4051 & 16.6  & 42.62 &  109$\pm$43 &   20$\pm$40 & 4.04$\pm$0.81 & 0.17$\pm$0.20 & 371$\pm$50 &  -0.71$\pm$0.42  & 39.93$\pm$0.31 &  -0.90$\pm$0.37 & 39.77$\pm$0.28  \\    
NGC4151 & 15.8  & 44.32 &  393$\pm$44 &   76$\pm$39 & 5.00$\pm$0.17 & 0.28$\pm$0.06 & 430$\pm$30 &  -0.21$\pm$0.18  & 40.56$\pm$0.18 &  -0.48$\pm$0.24 & 40.21$\pm$0.24  \\    
NGC4235 & 34.3  & 43.77 &  603$\pm$84 &  290$\pm$30 & 3.21$\pm$0.65 & 0.17$\pm$0.28 & 472$\pm$80 &  -2.48$\pm$0.65  & 38.37$\pm$0.65 &  -2.21$\pm$0.65 & 38.55$\pm$0.65  \\    
NGC4395 & 4.0  & 41.72 &  195$\pm$48 &    4$\pm$44 & 1.70$\pm$0.40 & 0.07$\pm$0.05 & 233$\pm$31 &  -2.88$\pm$0.41  & 37.36$\pm$0.41 &  -3.15$\pm$1.02 & 36.89$\pm$1.02  \\    
NGC5548 & 73.7  & 45.08 &  215$\pm$49 &   89$\pm$45 & 4.57$\pm$1.13 & 0.11$\pm$0.20 & 383$\pm$37 &  -1.10$\pm$1.13  & 39.56$\pm$1.13 &  -1.32$\pm$1.13 & 39.36$\pm$1.13  \\    
NGC6814 & 21.6  & 43.90 &  188$\pm$44 &   52$\pm$40 & 2.78$\pm$0.64 & 0.09$\pm$0.14 & 364$\pm$44 &  -2.31$\pm$0.64  & 38.31$\pm$0.64 &  -2.38$\pm$0.65 & 38.24$\pm$0.65  \\    
  Mrk79 & 95.1  & 45.08 &  205$\pm$58 &  115$\pm$55 & 4.59$\pm$0.50 & 0.13$\pm$0.13 & 347$\pm$38 &  -1.21$\pm$0.50  & 39.37$\pm$0.50 &  -1.43$\pm$0.51 & 39.14$\pm$0.51  \\    
 Mrk509 & 147.4  & 46.00 &  505$\pm$67 & 1072$\pm$31 & 4.10$\pm$0.68 & 0.01$\pm$0.06 & 318$\pm$50 &  -2.31$\pm$0.68  & 38.19$\pm$0.68 &  -2.22$\pm$0.68 & 38.05$\pm$0.68  \\    
 Mrk618 & 152.1  & 45.06 &  336$\pm$68 &  553$\pm$40 & 3.78$\pm$0.77 & 0.00$\pm$0.02 & 354$\pm$58 &  -2.42$\pm$0.77  & 38.18$\pm$0.77 &  -2.15$\pm$0.77 & 38.47$\pm$0.77  \\    
 Mrk766 & 55.3  & 44.08 &  222$\pm$57 &   67$\pm$42 & 5.11$\pm$0.44 & 0.15$\pm$0.13 & 359$\pm$44 &  -0.48$\pm$0.44  & 40.13$\pm$0.44 &  -0.49$\pm$0.46 & 40.15$\pm$0.46  \\    
 Mrk926 & 201.0  & 46.45 &  225$\pm$42 &  487$\pm$38 & 5.73$\pm$0.39 & 0.21$\pm$0.14 & 493$\pm$38 &  -0.27$\pm$0.39  & 40.61$\pm$0.39 &  -0.42$\pm$0.39 & 40.42$\pm$0.39  \\    
Mrk1044 & 70.7  & 44.01 &        --      &        --      &        --      &        --      &        --      &         --       &        --      &        --      &        --      \\    
Mrk1048 & 184.3  & 45.62 &  287$\pm$56 &  446$\pm$48 & 5.51$\pm$0.47 & 0.21$\pm$0.19 & 374$\pm$39 &  -0.68$\pm$0.47  & 39.96$\pm$0.47 &  -0.90$\pm$0.47 & 39.64$\pm$0.47  \\    
MCG+08-11-011& 87.9  & 45.59 &  206$\pm$45 &  212$\pm$43 & 5.73$\pm$0.42 & 0.25$\pm$0.19 & 439$\pm$36 &  0.07$\pm$0.42  & 40.86$\pm$0.43 &  -0.26$\pm$0.43 & 40.45$\pm$0.43  \\       
  \hline

    \end{tabular}
    \label{tab:ion_outflow}
\end{table}
\end{landscape}

\begin{landscape}
\begin{table}
    \caption{Properties of the molecular outflows.  (1) Name of the galaxy; (2) Adopted distance: for most galaxies, the distances are estimated from the redshift, except for except for those with accurate distance determinations: NGC\,3227  \citep{tonry01} , NGC\,4051 \citep{yuan21},  NGC\,4151 \citep{yuan20},  NGC\,4258  \citep{reid19}, NGC\,4395  \citep{thim04} and NGC\,6814  \citep{bentz19}.   (3) AGN bolometric luminosity; (4) Radius of the bulk of the outflow (spherical geometry); (5) Radius of the peak of the outflow (spherical shells geometry); (6) Total mass of the outflow considering only spaxels in the KDR; (7) Mass fraction of the gas in the outflow; (8--12) Properties of the outflows estimated using the two methods described in the text.}
    \begin{tabular}{lccccccccccc}
    \hline
    (1) & (2) & (3) & (4) & (5) & (6) & (7) & (8) & (9) & (10) & (11) & (12)\\
 &   &  & & & & &  \multicolumn{3}{c}{Global/Bulk outflow} &  \multicolumn{2}{c}{Radial profiles}\\
    	  \cline{8-10} \cline{11-12}\\

   Galaxy & $D$ & $\log L_{\rm bol}$&$R_{\rm out}$ & $R_{\rm peak}$ & $\log M_{\rm out}$ & $f_{\rm out}$ & $V_{\rm out}$ &  $\log \dot{M}_{\rm out}^{\rm b}$ & $\log \dot{E}_{\rm out}^{\rm b}$ & $\log \dot{M}_{\rm peak}$ & $\log \dot{E}_{\rm peak}$ \\
   
    & [Mpc] & [${\rm erg\: s}^{-1}$] & [pc] & [pc] &  [M$_\odot$]   &&  [${\rm km\: s}^{-1}$]  & [M$_\odot{\rm yr^{-1}}$] & [${\rm erg\: s}^{-1}$] & [M$_\odot{\rm yr^{-1}}$] & [${\rm erg\: s}^{-1}$] \\

	     	 \hline
	     	 \multicolumn{12}{c}{Type 2}\\
\hline
 NGC788  &  58.3   &  45.02  &   557$\pm$54  &   494$\pm$29  &   1.95$\pm$0.42  &   0.15$\pm$0.11  &   516$\pm$47  &  -3.90$\pm$0.42  &  37.02$\pm$0.42  &   -3.88$\pm$0.43  &  37.03$\pm$0.43  \\    
NGC1052  &  21.4   &  43.23  &   257$\pm$41  &    25$\pm$36  &   0.92$\pm$0.39  &   0.07$\pm$0.05  &   408$\pm$44  &  -4.25$\pm$0.40  &  36.47$\pm$0.40  &   -4.37$\pm$0.49  &  36.31$\pm$0.49  \\    
NGC1068  &  16.3   &  43.98  &   530$\pm$29  &    78$\pm$24  &   2.92$\pm$0.08  &   0.29$\pm$0.08  &   375$\pm$27  &  -2.49$\pm$0.08  &  38.16$\pm$0.09  &   -2.56$\pm$0.14  &  38.18$\pm$0.15  \\    
NGC1125  &  47.1   &  44.15  &   251$\pm$104  &   114$\pm$102  &   1.65$\pm$0.32  &   0.11$\pm$0.05  &   309$\pm$28  &  -3.98$\pm$0.33  &  36.50$\pm$0.33  &   -4.29$\pm$0.39  &  36.16$\pm$0.39  \\    
NGC1241  &  57.9   &  43.82  &   230$\pm$33  &   140$\pm$31  &   1.28$\pm$0.32  &   0.04$\pm$0.06  &   317$\pm$30  &  -4.39$\pm$0.32  &  36.11$\pm$0.32  &   -4.56$\pm$0.34  &  35.91$\pm$0.34  \\    
NGC2110  &  33.4   &  44.99  &          --       &          --       &         --       &          --       &        --      &         --       &         --       &         --       &        --   \\    
NGC4258  &  7.6   &  42.03  &   412$\pm$73  &          --       &   -0.22$\pm$0.31  &   0.06$\pm$0.09  &   418$\pm$54  &  -5.15$\pm$0.32  &  35.59$\pm$0.32  &   -5.07$\pm$0.39  &  35.54$\pm$0.39  \\    
NGC4388  &  36.0   &  45.08  &   419$\pm$77  &   305$\pm$42  &   1.87$\pm$0.23  &   0.06$\pm$0.06  &   332$\pm$45  &  -3.83$\pm$0.24  &  36.71$\pm$0.25  &   -3.93$\pm$0.25  &  36.49$\pm$0.25  \\    
NGC5506  &  26.6   &  44.57  &   291$\pm$43  &    64$\pm$39  &   1.92$\pm$0.14  &   0.12$\pm$0.07  &   318$\pm$28  &  -3.51$\pm$0.16  &  36.99$\pm$0.16  &   -3.73$\pm$0.21  &  36.76$\pm$0.21  \\    
NGC5899  &  36.9   &  43.65  &          --       &          --       &         --       &          --       &        --      &         --       &         --       &         --       &        --  \\    
   Mrk3  &  57.9   &  45.54  &   294$\pm$40  &   350$\pm$30  &   2.21$\pm$0.30  &   0.16$\pm$0.08  &   363$\pm$31  &  -3.51$\pm$0.30  &  37.11$\pm$0.30  &   -3.79$\pm$0.31  &  36.86$\pm$0.31  \\    
 Mrk348  &  64.3   &  45.33  &          --       &          --       &         --       &          --       &        --      &         --       &         --       &         --       &        --   \\    
 Mrk607  &  38.1   &  43.44  &   178$\pm$36  &    46$\pm$32  &   1.44$\pm$0.17  &   0.22$\pm$0.05  &   333$\pm$40  &  -3.93$\pm$0.18  &  36.61$\pm$0.19  &   -4.01$\pm$0.27  &  36.56$\pm$0.28  \\    
Mrk1066  &  51.4   &  43.61  &   361$\pm$37  &   186$\pm$33  &   2.53$\pm$0.11  &   0.15$\pm$0.06  &   243$\pm$26  &  -3.39$\pm$0.12  &  36.88$\pm$0.13  &   -3.59$\pm$0.13  &  36.61$\pm$0.14  \\    
ESO578-G009&   150.0   &  45.06  &          --       &          --       &         --       &          --       &        --      &         --       &         --       &         --       &   -- \\    
CygnusA  &  240.4   &  46.80  &   218$\pm$42  &   291$\pm$39  &   3.48$\pm$0.14  &   0.14$\pm$0.07  &   527$\pm$44  &  -2.57$\pm$0.15  &  38.38$\pm$0.16  &   -2.70$\pm$0.16  &  38.33$\pm$0.17  \\ 
        \hline
	     	 \multicolumn{12}{c}{Type 1}\\
	     	 \hline   
NGC1275  &  75.4   &  45.14  &   165$\pm$51  &    91$\pm$49  &   3.88$\pm$0.12  &   0.33$\pm$0.05  &   402$\pm$45  &  -1.67$\pm$0.16  &  39.04$\pm$0.17  &   -1.81$\pm$0.23  &  38.94$\pm$0.24  \\    
NGC3227  &  20.5   &  43.66  &   210$\pm$29  &    49$\pm$27  &   1.85$\pm$0.13  &   0.09$\pm$0.04  &   330$\pm$32  &  -3.32$\pm$0.14  &  37.22$\pm$0.15  &   -3.56$\pm$0.23  &  36.99$\pm$0.23  \\    
NGC3516  &  37.7   &  44.56  &   255$\pm$49  &    45$\pm$34  &   1.40$\pm$0.30  &   0.15$\pm$0.07  &   336$\pm$35  &  -4.10$\pm$0.30  &  36.45$\pm$0.30  &   -4.27$\pm$0.35  &  36.30$\pm$0.35  \\    
NGC4051  &  16.6   &  42.62  &   404$\pm$42  &    80$\pm$39  &   1.46$\pm$0.13  &   0.14$\pm$0.07  &   278$\pm$25  &  -3.97$\pm$0.14  &  36.42$\pm$0.14  &   -4.29$\pm$0.48  &  36.07$\pm$0.48  \\    
NGC4151  &  15.8   &  44.32  &   396$\pm$42  &    76$\pm$39  &   1.98$\pm$0.17  &   0.20$\pm$0.05  &   310$\pm$25  &  -3.37$\pm$0.17  &  37.12$\pm$0.18  &   -3.66$\pm$0.23  &  36.70$\pm$0.24  \\    
NGC4235  &  34.3   &  43.77  &   512$\pm$45  &   249$\pm$29  &   1.25$\pm$0.23  &   0.13$\pm$0.04  &   492$\pm$36  &  -4.35$\pm$0.23  &  36.54$\pm$0.23  &   -4.15$\pm$0.24  &  36.71$\pm$0.24  \\    
NGC4395  &  4.0   &  41.72  &          --       &          --       &         --       &          --       &        --      &         --       &         --       &         --       &        --    \\    
NGC5548  &  73.7   &  45.08  &   229$\pm$48  &   178$\pm$44  &   2.33$\pm$0.69  &   0.27$\pm$0.25  &   445$\pm$31  &  -3.29$\pm$0.69  &  37.51$\pm$0.69  &   -3.48$\pm$0.70  &  37.19$\pm$0.70  \\    
NGC6814  &  21.6   &  43.90  &   500$\pm$77  &    78$\pm$38  &   0.68$\pm$0.40  &   0.06$\pm$0.04  &   289$\pm$38  &  -4.93$\pm$0.40  &  35.49$\pm$0.40  &   -4.99$\pm$0.43  &  35.34$\pm$0.43  \\    
  Mrk79  &  95.1   &  45.08  &   164$\pm$56  &   115$\pm$55  &   2.58$\pm$0.19  &   0.19$\pm$0.05  &   398$\pm$39  &  -3.07$\pm$0.22  &  37.63$\pm$0.22  &   -3.25$\pm$0.26  &  37.44$\pm$0.26  \\    
 Mrk509  &  147.4   &  46.00  &          --       &          --       &         --       &          --       &        --      &         --       &         --       &         --       &        --  \\    
 Mrk618  &  152.1   &  45.06  &   404$\pm$44  &   922$\pm$39  &   2.81$\pm$0.34  &   0.13$\pm$0.07  &   335$\pm$28  &  -3.50$\pm$0.34  &  37.05$\pm$0.34  &   -3.61$\pm$0.34  &  37.01$\pm$0.35  \\    
 Mrk766  &  55.3   &  44.08  &          --       &          --       &         --       &          --       &        --      &         --       &         --       &         --       &        --   \\    
 Mrk926  &  201.0   &  46.45  &   313$\pm$41  &   487$\pm$38  &   3.64$\pm$0.23  &   0.32$\pm$0.10  &   354$\pm$29  &  -2.66$\pm$0.23  &  37.94$\pm$0.23  &   -2.94$\pm$0.23  &  37.74$\pm$0.23  \\    
Mrk1044  &  70.7   &  44.01  &          --       &          --       &         --       &          --       &        --      &         --       &         --       &         --       &        --   \\    
Mrk1048  &  184.3   &  45.62  &   299$\pm$51  &   446$\pm$48  &   3.21$\pm$0.19  &   0.26$\pm$0.07  &   316$\pm$28  &  -3.08$\pm$0.20  &  37.42$\pm$0.20  &   -3.23$\pm$0.20  &  37.32$\pm$0.20  \\    
MCG+08-11-011&  87.9   &  45.59  &   366$\pm$47  &   425$\pm$42  &   3.58$\pm$0.20  &   0.35$\pm$0.10  &   439$\pm$28  &  -2.33$\pm$0.20  &  38.45$\pm$0.21  &   -2.60$\pm$0.21  &  38.11$\pm$0.22  \\       
        \hline

    \end{tabular}
    \label{tab:molec_outflow}
\end{table}
\end{landscape}

In addition, a wide range of densities --  mostly in the range 10$^2$--10$^4$ cm$^{-3}$ \citep[e.g.][]{liu13,diniz19,kakkad20} --  have been adopted to determine the properties of outflows over the last decade, in case it cannot be directly estimated from the data used. As can be seen in Eq.~\ref{mhii}, the mass of ionised gas is inversely proportional to the electron density and thus, we scale the mass outflow-rates from the literature to the adopted density in this work ($N_e=1\,000\,{\rm cm^{-3}}$) and show them as grey crosses in Fig.~\ref{fig:mout}. This compilation is available as supplementary material and includes the estimates presented in \citet{fiore17}, and based on IFS of nearby Seyfert galaxies \citep{rogemar_n7582,rogemar_m79,rogemar_n5929,rogemar_n5643,rogemar_cygA,rogemar_m1066kin,rogemar_m1157,muller-sanchez11,barbosa14,allan14,allan16,mingozzi19,Shimizu19,diniz19,couto20,avery21,bianchin22,kakkad22}, QSOs at $z\approx0.3$ \citep{oliveira21}, and $z=2-3$ \citep{kakkad20,vayner21}.

The ionised mass-outflow rates estimated for our sample span two orders of magnitude,  ranging from 10$^{-3}$ to 10$^{1}$  M$_\odot$\,yr$^{-1}$, in agreement with the values available in the literature (Fig.~\ref{fig:mout}). 
The top-right panel of Fig.~\ref{fig:mout} shows a plot of the mass-outflow rate in hot molecular gas vs. the bolometric luminosity for our sample. Hot molecular gas outflows are scarce in the literature and trace only a small fraction of the molecular gas reservoir in the central region of galaxies \citep[e.g.][]{dale05,mazzalay13}. We find outflow rates in the range from 10$^{-5}$ to 10$^{-2}$  M$_\odot$\,yr$^{-1}$, considering estimates using the two methods. These values are consistent with the  estimates for the hot molecular gas, available in the literature \citep[e.g.][]{Diniz15,rogemar_N1275,bianchin22}.

The bottom panels of Fig.~\ref{fig:mout} show plots of the kinetic power of the outflows for the ionised (left panel) and molecular (right panel) gas versus the AGN bolometric luminosity. We include estimates of the kinetic power from the literature, obtained from the same references used to compile the values of mass-outflow rates, described above. The kinetic powers of the ionised outflows in our sample are in the range $\sim$10$^{37}$--10$^{43}$ erg\,s$^{-1}$, in good agreement with values from the literature at the same range of bolometric luminosity. The kinetic powers of the hot molecular outflows are on average three orders of magnitude lower than those in ionised gas, with values ranging from 10$^{35}$ to 10$^{39}$ erg\,s$^{-1}$.

\section{Discussion}\label{sec:discussion}

\begin{figure}
    \centering
    \includegraphics[width=0.48\textwidth]{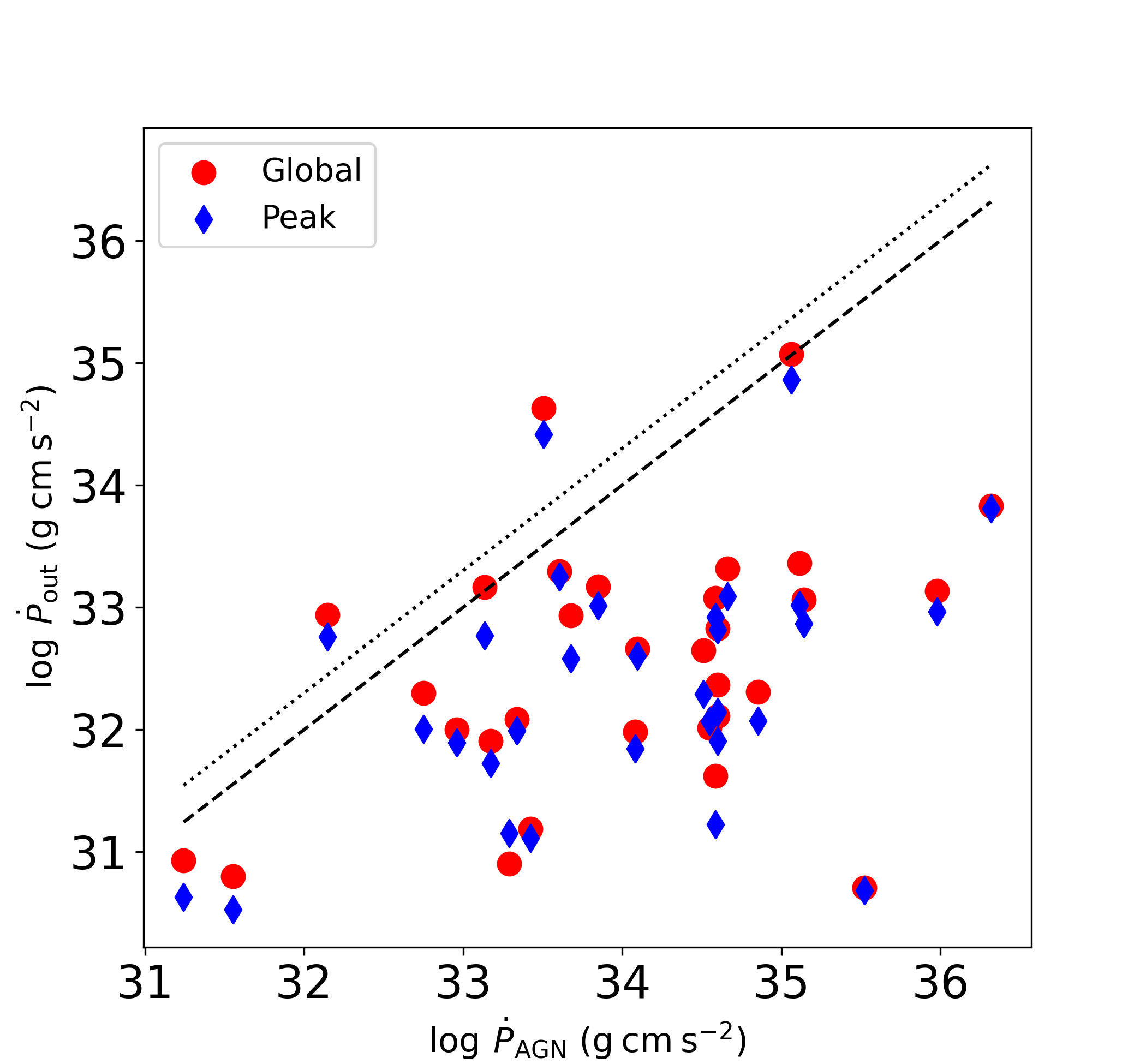}
    \caption{\small  Momentum flux of the ionised outflows versus the AGN photon momentum flux for our sample. Red circles represent estimates obtained by assuming a spherical shell geometry and blue diamonds represent peak estimates from radial profiles. The dotted and dashed lines correspond to constant ratios of 2:1 and 1:1, respectively,useful to investigate the origin of the outflows.}
    \label{fig:momentum}
\end{figure}

In luminous quasars, the gas emission is usually dominated by outflowing gas \citep[e.g.][]{kakkad20,vayner21}. Although ionised outflows are also frequently reported in lower luminosity AGN, a significant fraction of line emission in the inner kpc arises from gas with motions dominated by the gravitational potential of the galaxy \citep[e.g.][]{davies07,schonell19,ruschel-dutra21,fonseca-faria21,bianchin22}. We find that 79 per cent of our sample present ionised outflows. Recently, \citet{ruschel-dutra21} found outflows in 70 per cent of their sample composed of 30 AGN at $z\leq0.02$ using optical Gemini GMOS-IFU observations, while \citet{kakkad22} studied ionised outflows in a sample of 22 X-ray selected AGN at $z\leq0.1$ observed with the MUSE instrument at the VLT.   

We find that the ionised outflows correspond to a median of 15 per cent of the mass of ionised gas in our sample. As shown in Table~\ref{tab:ion_outflow} and Fig.~\ref{fig:fraction}, our sample spans a wide range in fraction of gas in the outflow relative to the total gas mass, which is consistent with previous observations \citep[e.g.][]{muller-sanchez11,rogemar_m79,rogemar_N1275,rogemar_cygA,may18,bianchin22}. As shown in Fig.~\ref{fig:fraction}, there is a positive trend between the fraction of the gas in the outflow in ionised and molecular gas phases (middle panel), but there is no clear relation bewteen the masses of the gas in the outflow in both phases with the luminosity (right panel).

The fraction of objects with hot molecular outflows in our sample is 76 per cent. The contribution of hot molecular outflows to the total mass of hot molecular gas in the inner region of the observed galaxies is 13 per cent, while if we consider only galaxies with outflows, this contribution slightly increases to 15 per cent. This is consistent with previous observations of nearby active galaxies, which indicate that the hot molecular hydrogen in the inner kpc of nearby AGN hosts is more restricted to the galaxy disc, while the ionised gas usually presents an ouflowing component traced by gas that extends to higher latitudes above the disc \citep[e.g.][]{rogemar_n5929,rogemar_sample,rogemar21_extended,ramos-almeida17,sb19,speranza22}.

 We can compare the properties of the outflows observed in both, hot molecular and ionised gas, phases. Wind scaling relations suggest that the molecular gas phases are relatively more important in lower luminosity objects, but similar to the ionised gas phase in higher luminosity AGN, as indicated by the higher slopes of the correlations of winds properties with the AGN bolometric luminosity observed in ionised gas, relative to those seen in cold molecular gas \citep{fiore17}.  As shown in Fig.~\ref{fig:fraction}, we do not find a clear relation between the ratio of the gas masses of ionised and hot molecular gas with the luminosity. 
In addition,  simulations aimed at investigating outflow properties as a function of radius for different gas phases, indicate that the molecular phase of the outflow is generated from the cooling of the gas trapped into the outflow, and so one would expect that the molecular phase is more important at larger radii, relative to the ionised phase \citep[e.g.][]{ferrara16,costa18,richings18b,richings21}. As shown in the bottom panel of Fig.~\ref{fig:radial}, we do not find a clear relation between the mass-outflow rates in ionised and hot molecular gas with the distance from the centre.  A possible explanation for the absence of relations between the relative outflow properties in both phases with the luminosity and radius, is that the hot molecular gas phase observed via the H$_2$ emission lines represents only the  heated surface of a much larger colder molecular gas reservoir \citep{dale05,mazzalay13}, considered in the simulations.

The mass-outflow rates and kinetic powers of the outflows estimated for our sample are in agreement with previous measurements for AGN of similar luminosity. However, as can be seen in Fig.~\ref{fig:mout}, the scatters of the relations of $\log\:\dot{M}_{\rm out}$ versus $\log L_{\rm bol}$ and  $\log\:\dot{E}_{\rm out}$ versus $\log L_{\rm bol}$ are high, if we consider all measurements available in the literature and used for comparison here.  The density of the outflow is a function of the radius, as well as the ionised gas density at a given radius is a function of the AGN luminosity \citep{davies20,revalski22}, and thus the use of a fixed density value to estimate the mass-outflow rate may reduce the scatter of the correlation between $\log\:\dot{M}_{\rm out}$ and $\log L_{\rm bol}$. In the luminosity range of our sample, the mass-outflow rates in ionised gas cover four orders of magnitude, while the kinetic powers spans three orders of magnitude.  The wide ranges of values observed for  $\log\:\dot{M}_{\rm out}$ and $\log\:\dot{E}_{\rm out}$ are in good agreement with previous works and are partially due to the assumptions made to calculate the outflow properties by distinct works, which include assumptions on the electron density estimates, geometry, and velocity of the outflows \citep[see ][ for a detailed discussion]{davies20}.
An important caveat is that outflow rates are defined as the amount of material passing through a common radius, so they may be estimated globally for the entire outflow, or outflow rates measured from individual spaxels, may be added azimuthally to produce radial outflow rate profiles (see Figure~\ref{fig:radial} and \citealp{revalski21}). However, outflow rates cannot be added radially. These instantaneous outflow rates are directly related to the spatial resolution of the data, such that higher spatial sampling yields larger instantaneous rates, because they account for material passing through multiple boundaries for a fixed evacuation time \citep{veilleux17,kakkad22}. This can explain why mass outflow rates and kinetic powers estimated by radially summing the individual spaxels are systematically larger by up to two orders of magnitude as compared to the global and radial estimates \citep[e.g.][]{kakkad22}

We can compare the  kinetic coupling efficiencies ($\dot{E}_{K}/L_{\rm bol}$) for observed ionised outflows in our sample with theoretical predictions. For AGN feedback to become efficient in suppressing star formation, the models require a minimum coupling efficiency ($\varepsilon_f$) in the range of 0.5--20 per cent \citep{dimatteo05,hopkins_elvis10,dubois_horizon_14,Schaye_eagle_15,weinberger17}. However, as discussed in \citet{harrison18}, it is unlikely that all the injected energy becomes kinetic power in the outflow and a direct comparison between observed $\dot{E}_{K}/L_{\rm bol}$  and predicted $\varepsilon_f$ is not straightforward. Indeed, recent numerical simulations indicate that the kinetic energy of the outflows represents less  than 20 per cent of the total  emitted outflow energy \citep{richings18b}.

With the above caveat in mind, we estimate the kinetic coupling efficiencies for the ionised gas in our sample. We find that none object in our sample present outflows with kinetic powers corresponding to more than 0.5 per cent of the AGN bolometric luminosity.  The median value of the kinetic coupling efficiency using the global kinetic power of the ionised outflows is $\dot{E}_{K}/L_{\rm bol}\approx1.8\times10^{-3}$.
The kinetic powers of the hot molecular outflows are about 2 orders of magnitude lower than those of the ionised gas, and thus they are also not powerful enough to suppress star formation in the galaxies.  However, besides the fact that $\dot{E}_{K}/L_{\rm bol}<\varepsilon_f$ mentioned above, the outflows in AGN are seen in multiple gas phases and the kinetic power of dense cold molecular outflows are expected to be larger. Thus, even if the ionised outflows seen here are not powerful enough to suppress star formation in the host galaxies, we cannot discard AGN feedback as an important mechanism in shaping the evolution of the galaxies in our sample.

In order to investigate the physical mechanism that drives the outflows observed in our sample, we compute the momentum flux of the outflow by $\dot{P}_{\rm out} = \dot{M} \times v_{\rm out}$, where $v_{\rm out}$ is the velocity of the outflow. In Figure~\ref{fig:momentum}, we present a plot of $\dot{P}_{\rm out}$ versus the photon momentum flux ($\dot{P}_{\rm AGN}=L_{\rm AGN}/c$, where $L_{\rm AGN}$ is the AGN bolometric luminosity), which can yield insights into the origin of winds \citep{murr05,thom15,costa18,Veilleux20,vayner21}. To estimate $\dot{P}_{\rm out}$, we use the same definition for the  velocity of the outflow previously used to calculate the mass-outflow rates.  The dotted and dashed lines show a 2:1 and 1:1 constant relations respectively, that can be used to investigate the the driving mechanism of the outflows. Theoretical studies suggest that $\dot{P}_{\rm out} \gtrsim 2 \dot{P}_{\rm AGN}$ on scales $\lesssim$1\,kpc are due to radiation pressure driven winds in a high-density, optically thick environment, where far-infrared photons are scattered multiple times \citep{thom15,costa18}. Values of $\dot{P}_{\rm out} \lesssim 1 \dot{P}_{\rm AGN}$ are usually attributed to radiation-pressure driven winds in low density environments or  shocked AGN winds \citep{fauc12b}. Most objects in our sample are below the 1:1 line in Fig.~\ref{fig:momentum}, indicating that the winds are driven by radiation-pressure in low-density environment, with possible contribution from shocks as suggested also in our previous studies \citep[e.g.][]{rogemar21_extended,rogemar21_exc}.

\section{Conclusions}\label{sec:conc}

We have studied the molecular and ionised gas kinematics of the inner 0.04--2\,kpc of a sample of 33 AGN hosts with $0.001\lesssim z\lesssim0.056$ and hard X-ray luminosities of $41\lesssim \log L_{\rm X}/({\rm erg\, s^{-1}}) \lesssim45$. The K-band observations were performed with the Gemini NIFS instrument, with a field of view covering from the inner 75$\times$75~pc$^2$ to 3.6$\times$3.6~kpc$^2$ at spatial resolutions of 6 to 250\,pc and velocity resolution of $\sigma_{\rm inst}\sim$20\,km\,s$^{-1}$. We use the $W_{\rm 80}$, $V_{\rm peak}$ and $V_{\rm cen}$ parameters for the \hml\ and \brg\ emission lines to identify regions where the gas motions are dominated by kinematic disturbances due to the AGN and regions where the gas motions are due the gravitational potential of galaxies. Our main conclusions are:

\begin{itemize}

    \item   We identify ionised gas  kinematically disturbed regions (KDRs) in 31 galaxies (94 per cent)  of our sample, while 25 objects (76 per cent) present KDRs in molecular gas. 
    
    \item   We attribute the KDR as being produced by AGN outflows and estimate their mass-outflow rates and kinetic powers in two ways: (i) by assuming an spherical geometry, resulting in {\it global outflow} properties and (ii) adopting the {\it peak} outflow properties, derived from their radial profiles. 
    
    \item  The masses of the outflowing gas are in the ranges 10$^2$--10$^7$ M$_\odot$ and 10$^0$--10$^4$ M$_\odot$ for the ionised and hot molecular gas, respectively. These values correspond to median fractions of the gas in the outflow relative to the total amount of gas of  about 15 per cent for both ionised and hot molecular gas (within a typical covered region of a few 100\,pc radius at the galaxies).
    
 \item   The mass-outflow rates in ionised gas are in the range 10$^{-3}$--10$^{1}$ M$_\odot$\,yr$^{-1}$. The kinetic powers of the ionised outflows are in the range $\sim$10$^{37}$--10$^{43}$ erg\,s$^{-1}$, being smaller than 0.5 per cent of the AGN bolometric luminosity for most objects, with a median  kinetic coupling efficiency in our sample is $\dot{E}_{K}/L_{\rm bol}\approx1.8\times10^{-3}$. The estimated mass-outflow rates and kinetic powers of the outflows are consistent with previous estimates for objects in the same luminosity range, but a large scatter of the wind scaling relations is seen in the lower luminosity range.
 
 \item   The mass-outflow rates in molecular gas range from 10$^{-5}$ to 10$^{-2}$  M$_\odot$\,yr$^{-1}$, and the kinetic power of the outflows are in the range  10$^{35}$--10$^{39}$ erg\,s$^{-1}$. Both mass outflow rates and powers present positive correlations with the AGN bolometric luminosity.
 
 \item   The momentum flux of the ionised outflows are lower than the photon momentum flux of the accretion disc in most objects, indicating that the observed outflows are consistent with radiation-pressure driven winds in low density environments including possible contribution of shocked AGN winds. 
\end{itemize}

In summary, our results support the presence of kinematic disturbances produced by the AGN in most sources, with a higher impact in the galaxy produced by the ionised gas outflows as compared to that of the hot molecular gas. This can be attributed mostly to the small mass in this latter gas phase, but its kinematics is also dominated by lower velocities than observed in the ionised gas. Observations in cold molecular gas should be made to investigate the presence of outflows in this gas phase.

\section*{Acknowledgements}
We thank to an anonymous referee for the suggestions which helped us to improve this paper. 
RAR acknowledges financial support from Conselho Nacional de Desenvolvimento Cient\'ifico e Tecnol\'ogico and Funda\c c\~ao de Amparo \`a pesquisa do Estado do Rio Grande do Sul. RR thanks to CNPq (grant 311223/2020-6,  304927/2017-1 and  400352/2016-8) and FAPERGS (grant 16/2551-0000251-7 and 19/1750-2).  MB thanks the financial support from Coordena\c c\~ao de Aperfei\c coamento de Pessoal de N\'ivel Superior - Brasil (CAPES) - Finance Code 001. NLZ is supported by the J. Robert Oppenheimer Visiting Professorship and the Bershadsky Fund. NZD acknowledges support from the Agencia Estatal de Investigaci\'on del Ministerio de Ciencia e Innovaci\'on (AEI-MCINN) under project with reference PID2019-107010GB100. 
ARA acknowledges CNPq (grant 312036/2019-1) for partial support to this work
Based on observations obtained at the Gemini Observatory, which is operated by the Association of Universities for Research in Astronomy, Inc., under a cooperative agreement with the NSF on behalf of the Gemini partnership: the National Science Foundation (United States), National Research Council (Canada), CONICYT (Chile), Ministerio de Ciencia, Tecnolog\'{i}a e Innovaci\'{o}n Productiva (Argentina), Minist\'{e}rio da Ci\^{e}ncia, Tecnologia e Inova\c{c}\~{a}o (Brazil), and Korea Astronomy and Space Science Institute (Republic of Korea).  This research has made use of NASA's Astrophysics Data System Bibliographic Services. This research has made use of the NASA/IPAC Extragalactic Database (NED), which is operated by the Jet Propulsion Laboratory, California Institute of Technology, under contract with the National Aeronautics and Space Administration.

\section*{Data availability}
The data used in this paper is available in the Gemini Science Archive at https://archive.gemini.edu/searchform. Processed datacubes used will be shared on reasonable request to the corresponding author.




\bibliographystyle{mnras}
\bibliography{paper_arxiv} 




\appendix
\include{appendix}
{\centering \bf Supplementary Materials}\\

\section{Results for individual galaxies}

\begin{figure*}
    \centering
    \includegraphics[width=0.98\textwidth]{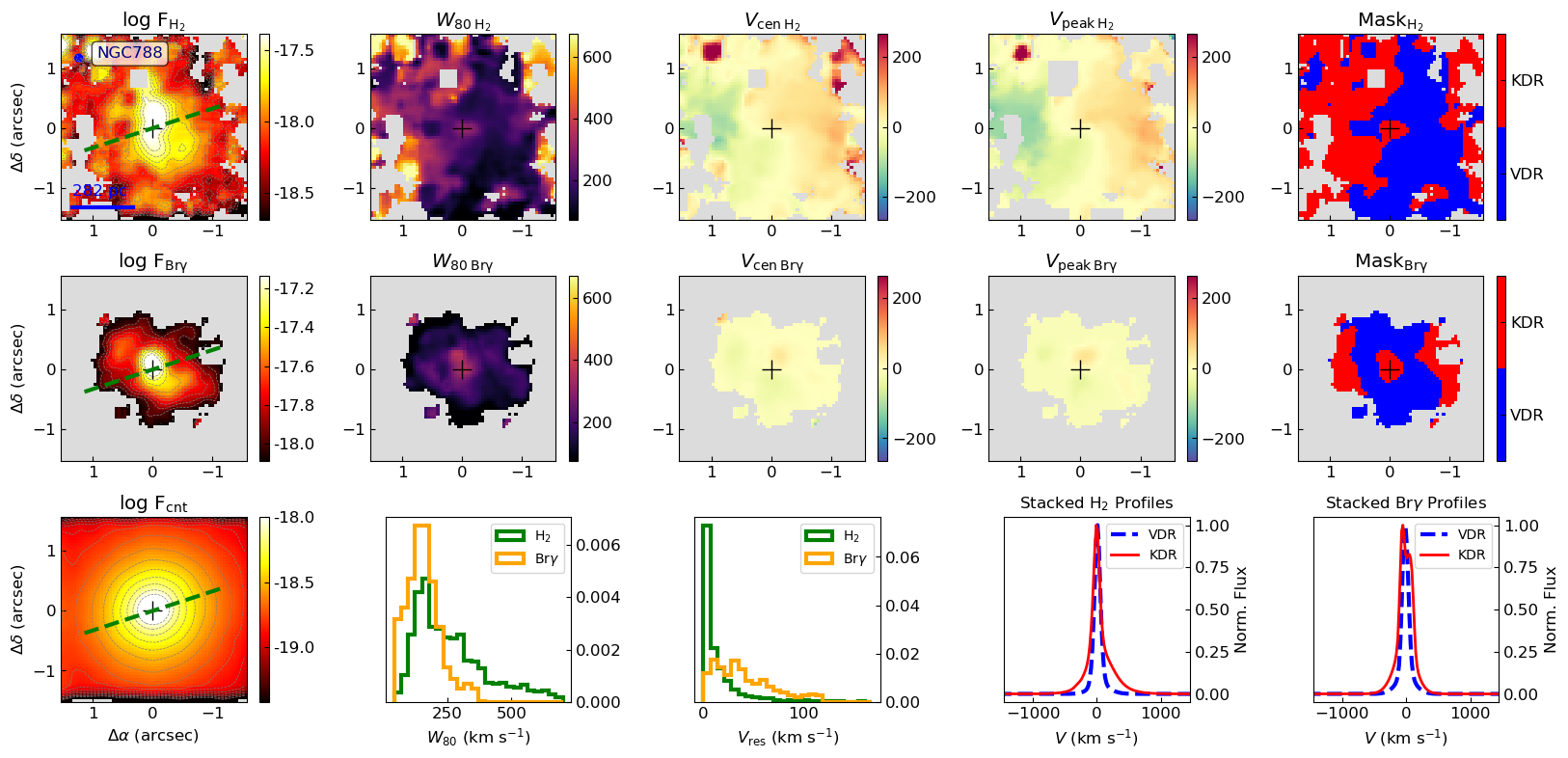}
  \caption{\small Results for NGC788. The first row  shows the results for the \hml\, and second row  show results for the the \brg\ emission line. From left to right: emission line flux distribution, $W_{\rm 80}$, $V_{\rm cen}$,  $V_{\rm peak}$, and a ``kinematic map'' identifying the kinematically disturbed region (KDR) in red and the virially-dominated region (VDR) in blue. The colour bars show fluxes in erg\,s$^{-1}$\,cm$^{-2}$\,spaxel$^{-1}$ and velocities in km\,s$^{-1}$.  The grey areas identify locations where the emission-line amplitude is below 3 times the continuum noise amplitude (3$\sigma$). The bottom rows show a K-band continuum image in erg\,s$^{-1}$\,cm$^{-2}$\,\AA$^{-1}$\,spaxel$^{-1}$, the density distributions of $W_{\rm 80}$ and $V_{\rm res}=|V_{\rm cen}-V_{\rm peak}|$ and stacked profiles of the H$_2$ and Br$\gamma$ emission lines from the VDR and KDR. Stacked profiles for the KDR are presented only if it corresponds to at least 10 per cent of the spaxels with detected emission. The green dashed lines in the leftmost panels show the orientation of the major axis of the large-scale disk, as presented in \citet{rogemar21_extended}. In all maps, North is up and East is to the left.}
    \label{fig:n788}
\end{figure*}

\begin{figure*}
    \centering
    \includegraphics[width=0.98\textwidth]{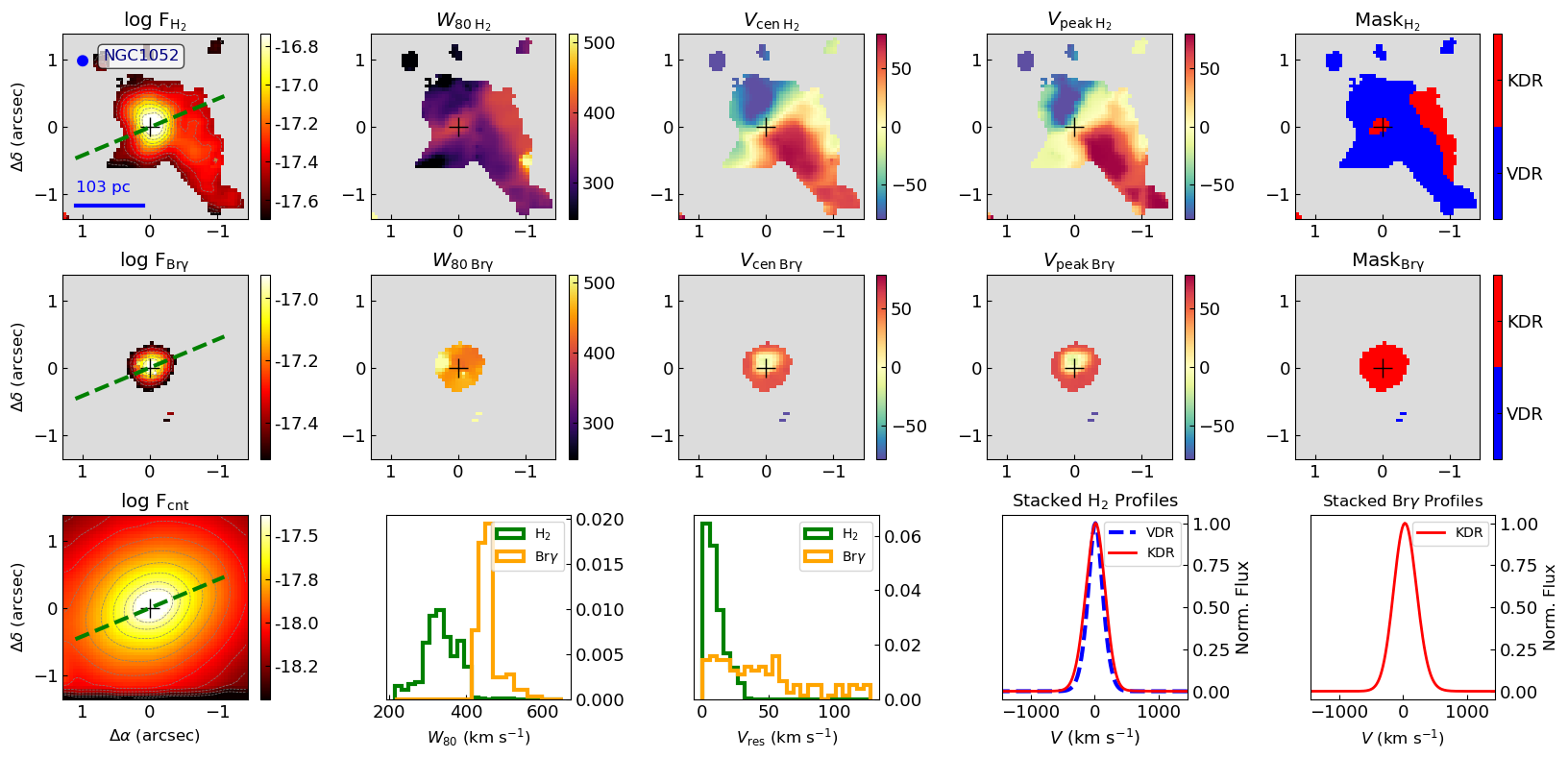}
  \caption{\small Same as Fig.~\ref{fig:n788}, but for NGC\,1052.}
    \label{fig:n1052}
\end{figure*}

\begin{figure*}
    \centering
    \includegraphics[width=0.98\textwidth]{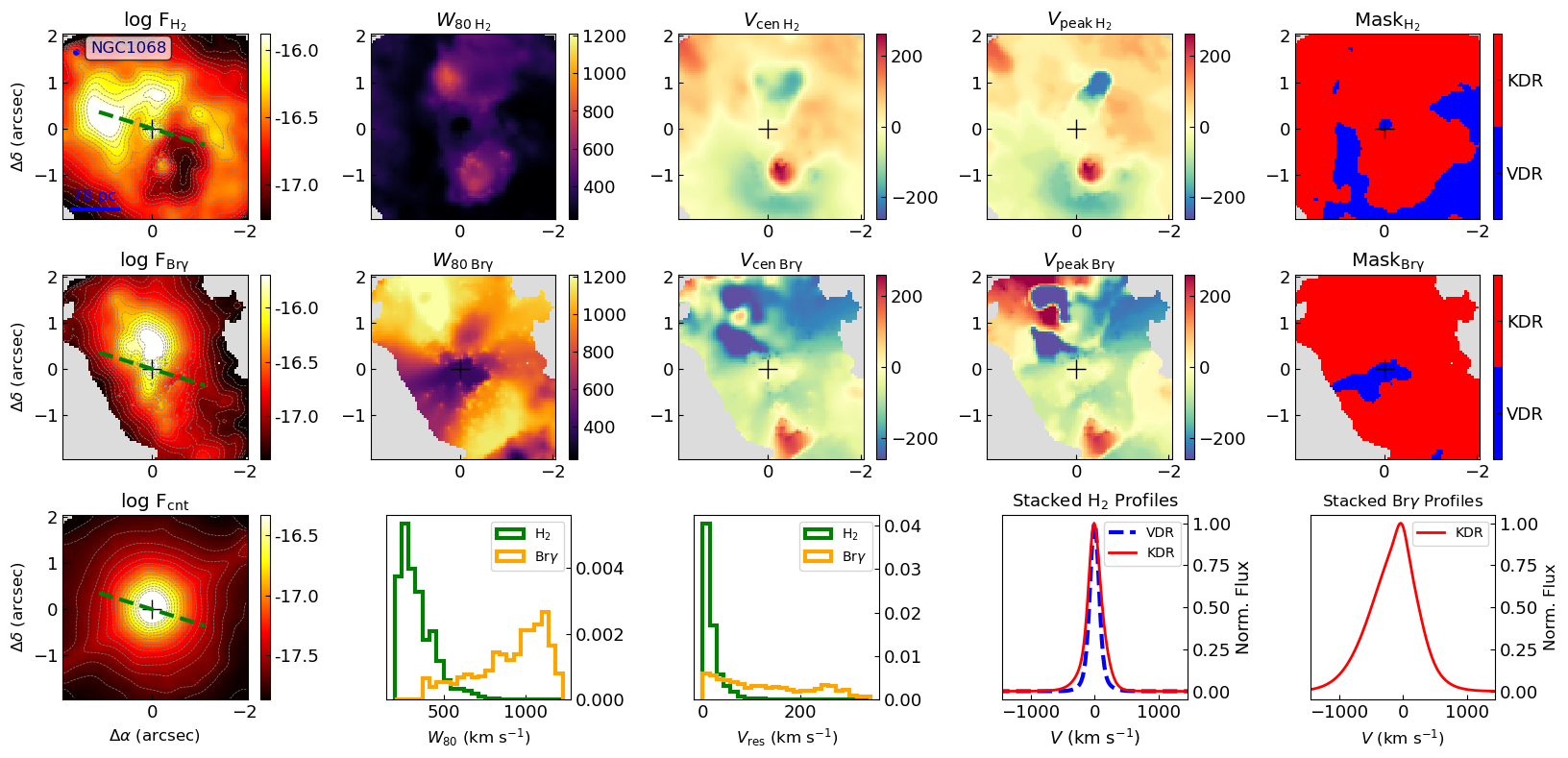}
 
\caption{\small Same as Fig.~\ref{fig:n788}, but for NGC\,1068.}
    \label{fig:n1068}
\end{figure*}

\begin{figure*}
    \centering
    \includegraphics[width=0.98\textwidth]{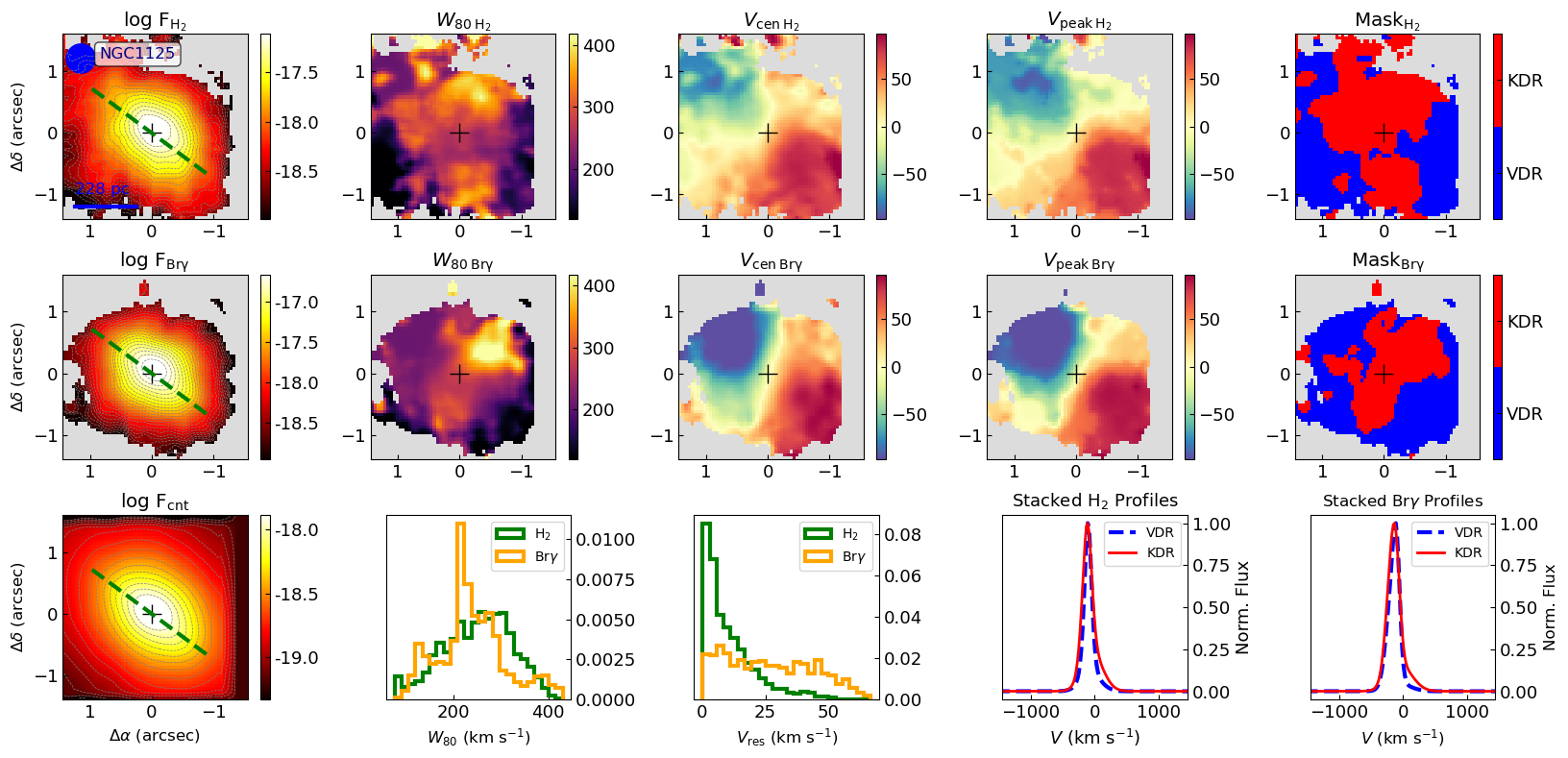}
 
\caption{\small Same as Fig.~\ref{fig:n788}, but for NGC\,1125.}
    \label{fig:n1125}
\end{figure*}

\begin{figure*}
    \centering
    \includegraphics[width=0.98\textwidth]{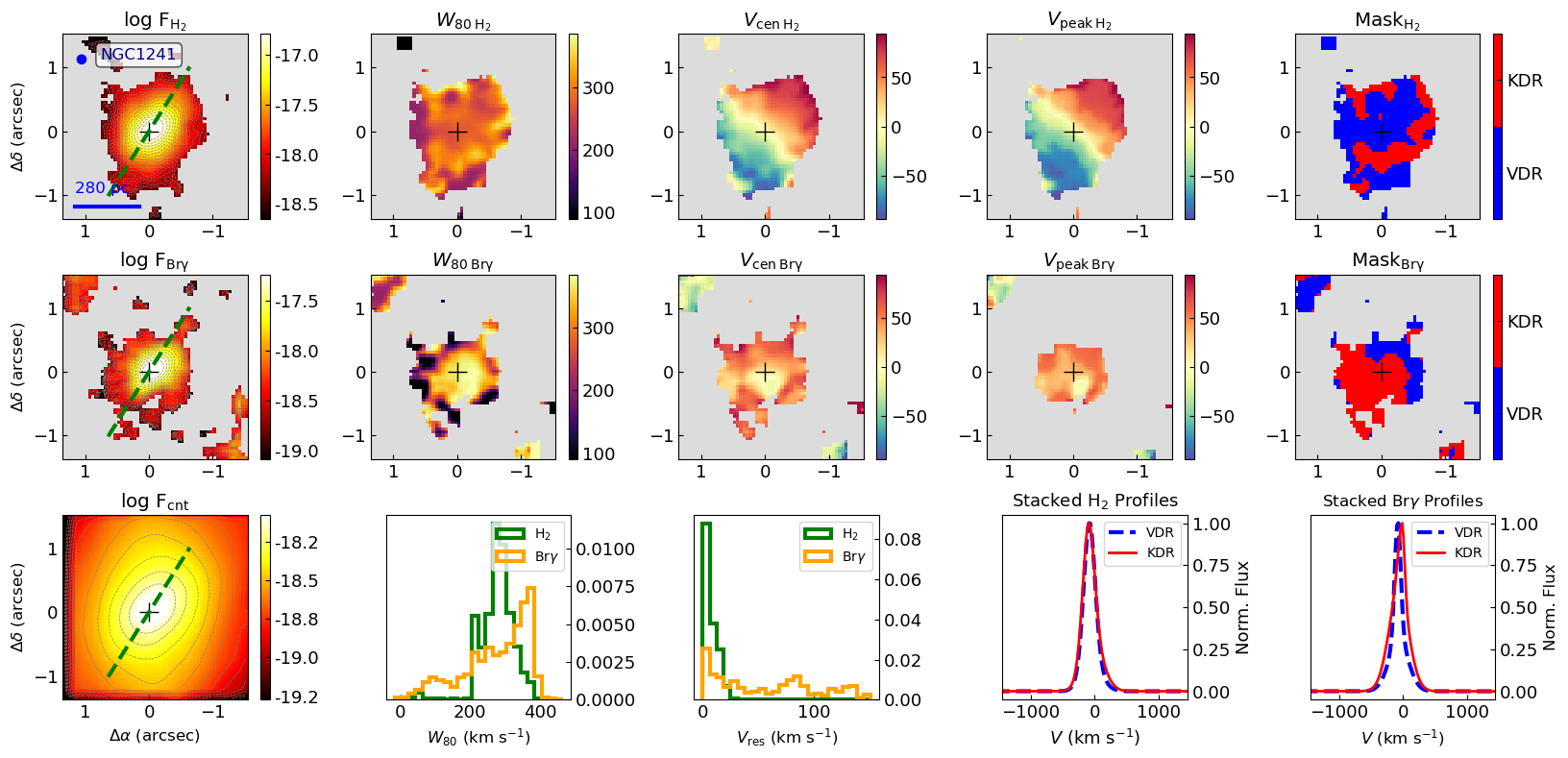}
 
\caption{\small Same as Fig.~\ref{fig:n788}, but for NGC\,1241.}
    \label{fig:n1241}
\end{figure*}

\begin{figure*}
    \centering
    \includegraphics[width=0.98\textwidth]{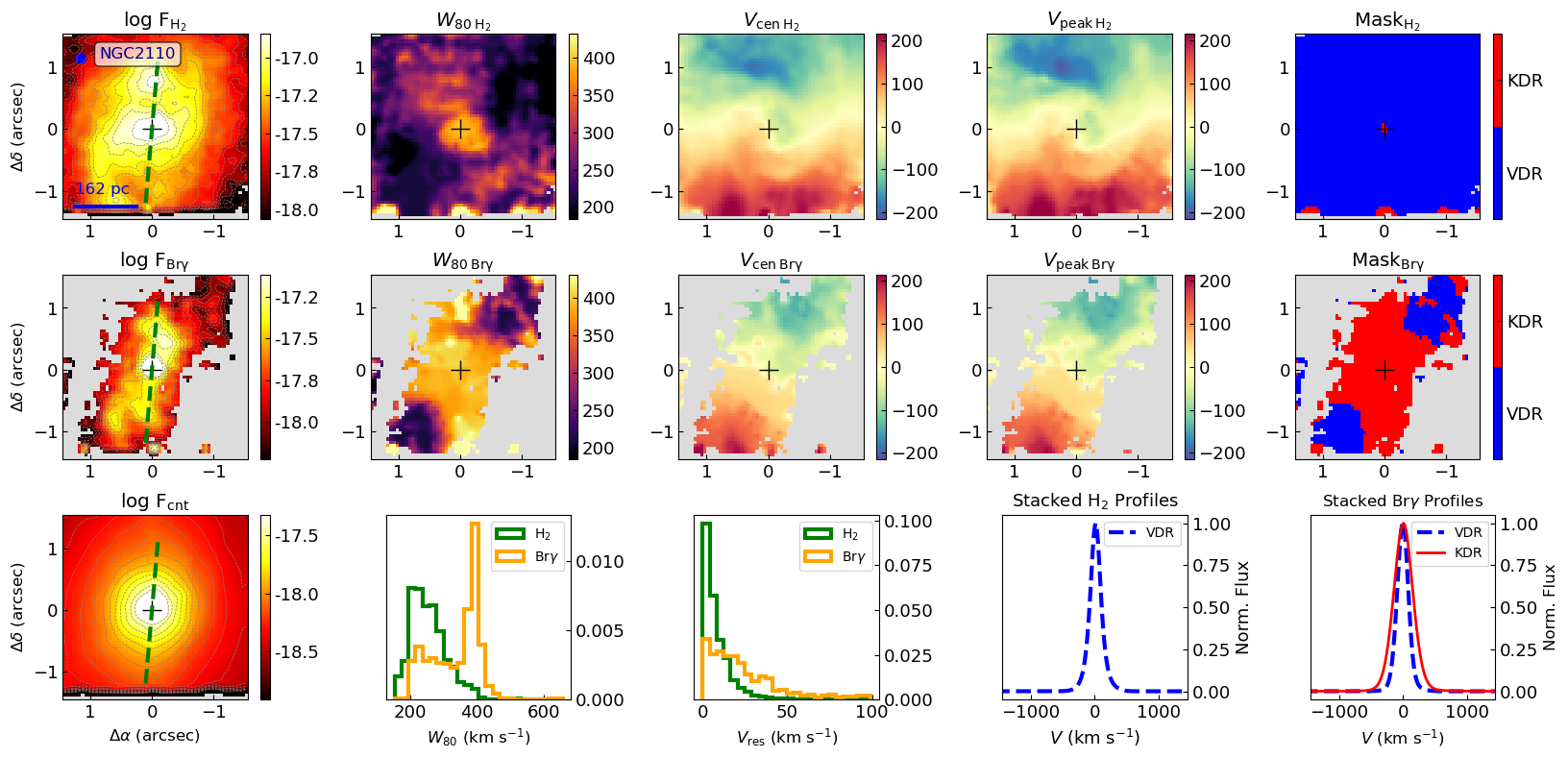}
 
\caption{\small Same as Fig.~\ref{fig:n788}, but for NGC\,2110.}
    \label{fig:n2110}
\end{figure*}

\begin{figure*}
    \centering
    \includegraphics[width=0.98\textwidth]{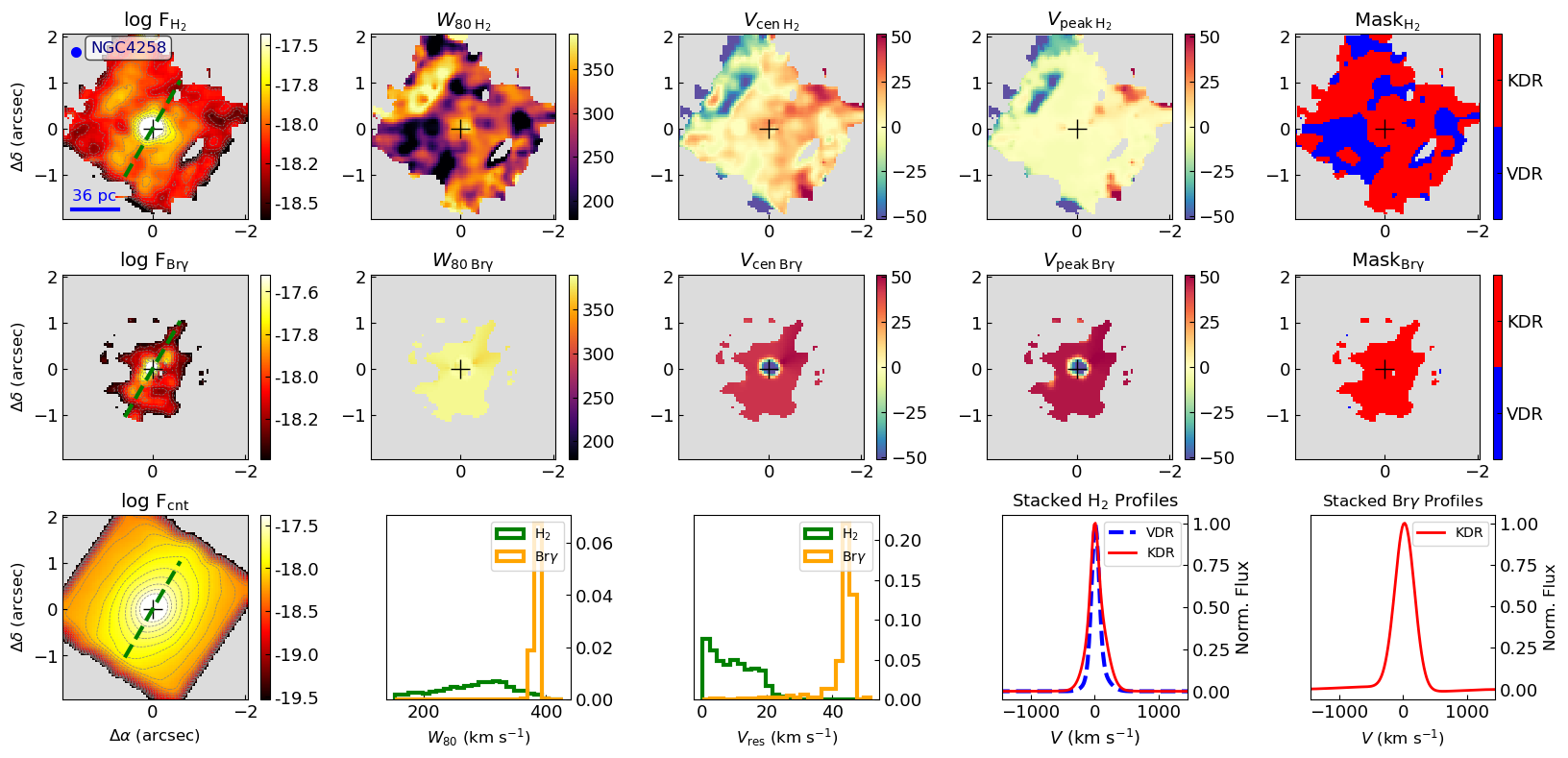}
\caption{\small Same as Fig.~\ref{fig:n788}, but for NGC\,4258.}
    \label{fig:n4258}
\end{figure*}

\begin{figure*}
    \centering
    \includegraphics[width=0.98\textwidth]{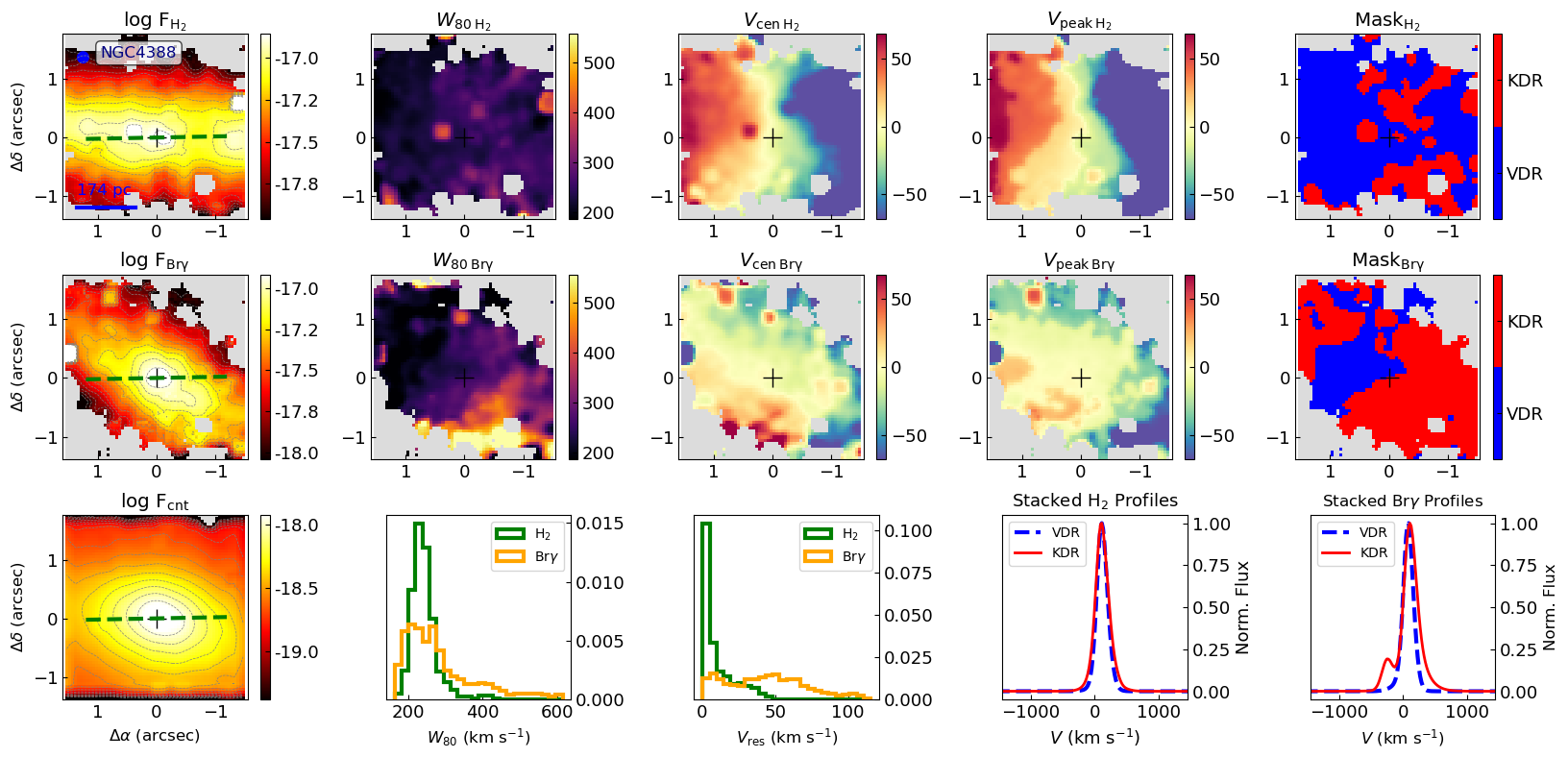}
 
\caption{\small Same as Fig.~\ref{fig:n788}, but for NGC\,4388.}
    \label{fig:n4388}
\end{figure*}

\begin{figure*}
    \centering
    \includegraphics[width=0.98\textwidth]{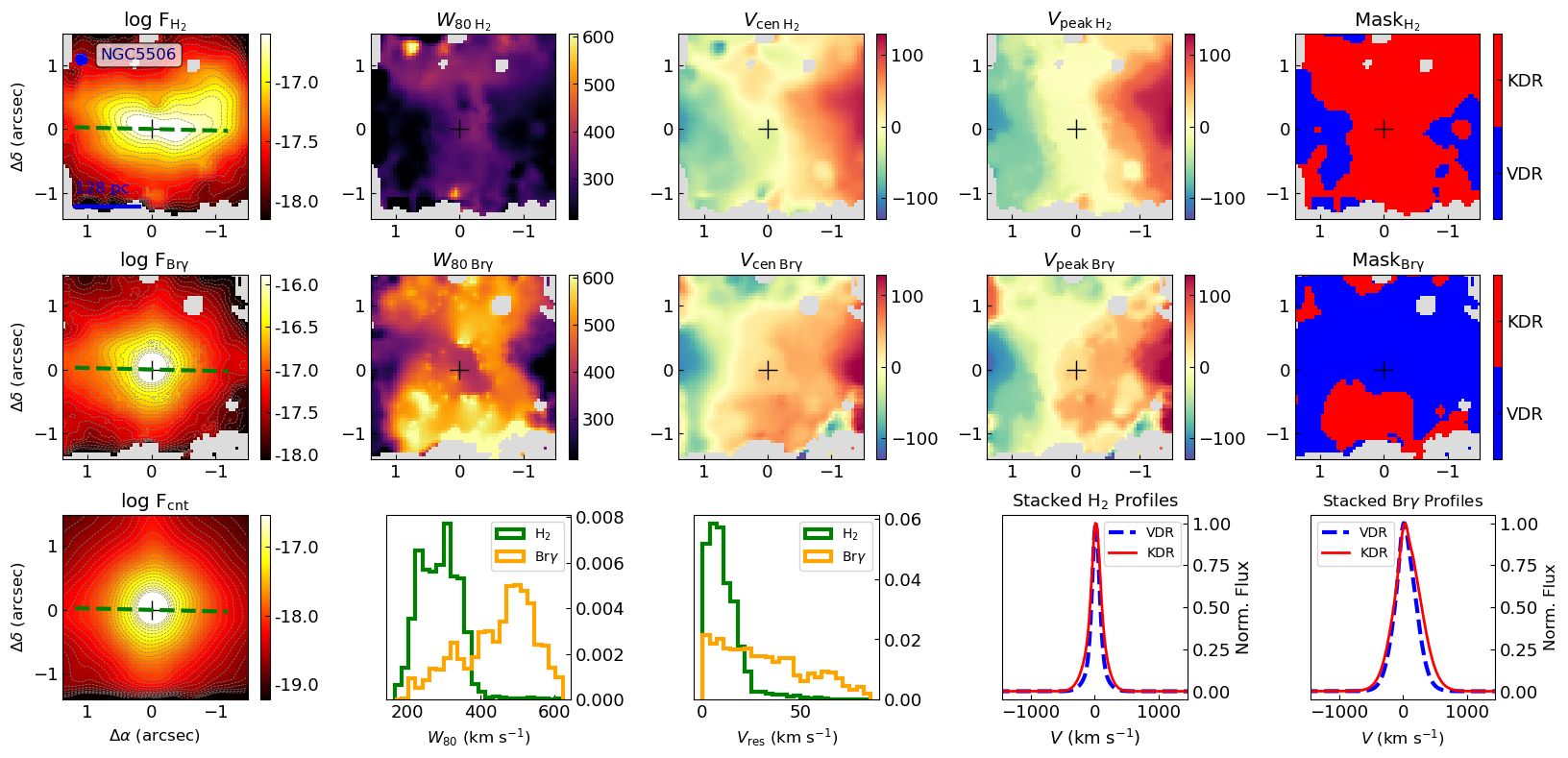}
 \caption{\small Same as Fig.~\ref{fig:n788}, but for NGC\,5506.}
    \label{fig:n5506}
\end{figure*}

\begin{figure*}
    \centering
    \includegraphics[width=0.98\textwidth]{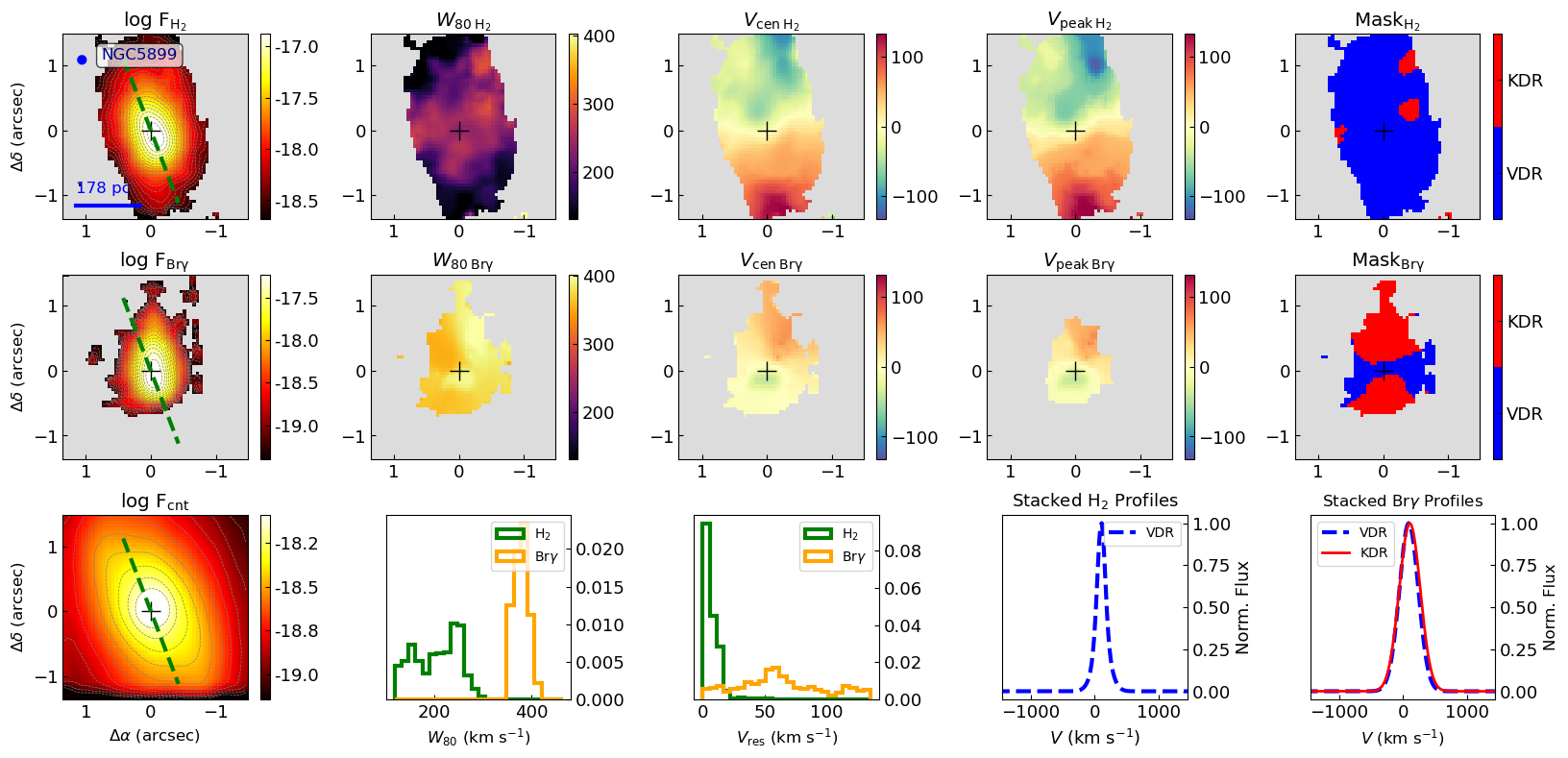}

\caption{\small Same as Fig.~\ref{fig:n788}, but for NGC\,5899.}
    \label{fig:n5899}
\end{figure*}

\begin{figure*}
    \centering
    \includegraphics[width=0.98\textwidth]{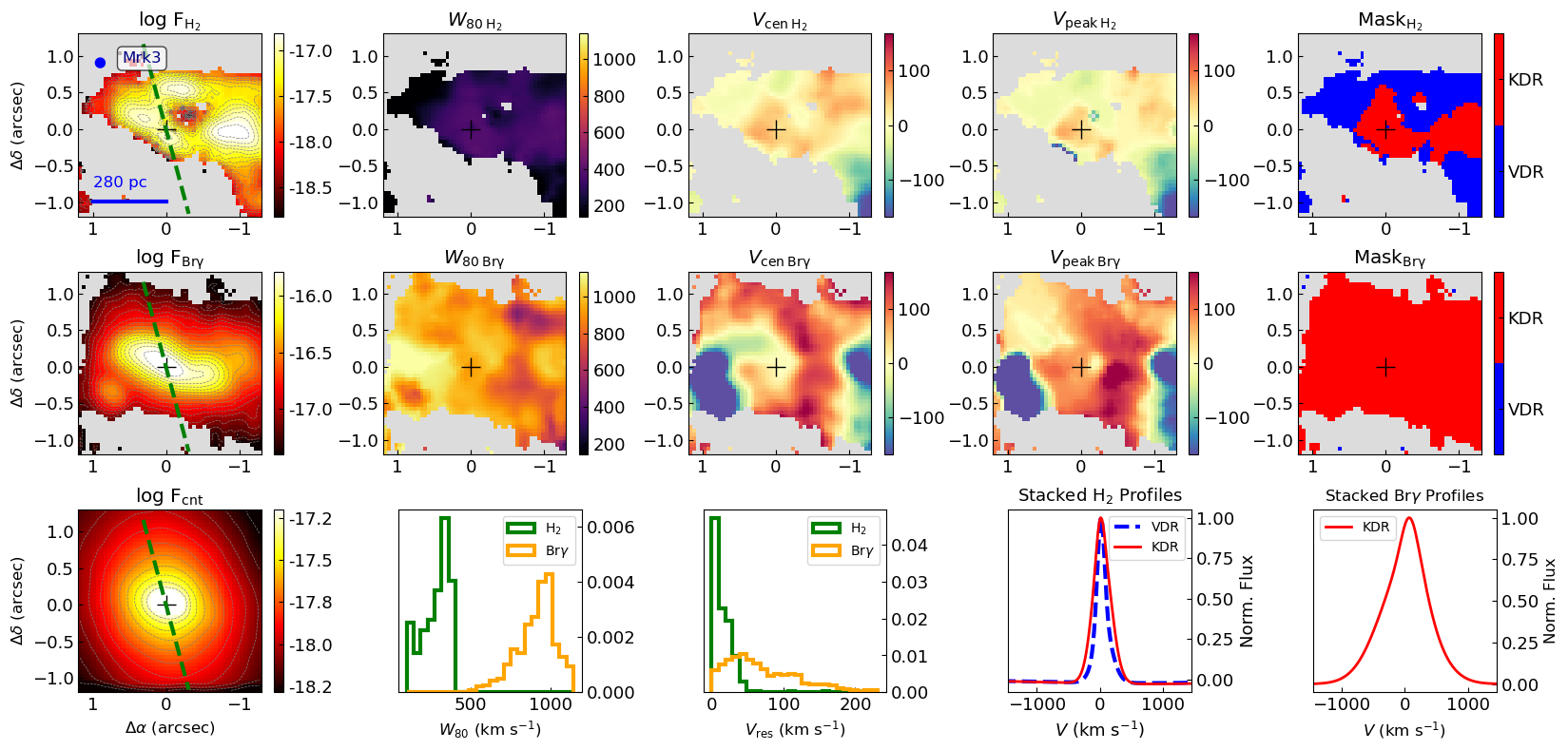}

\caption{\small Same as Fig.~\ref{fig:n788}, but for Mrk\,3.}
    \label{fig:mrk3}
\end{figure*}

\begin{figure*}
    \centering
    \includegraphics[width=0.98\textwidth]{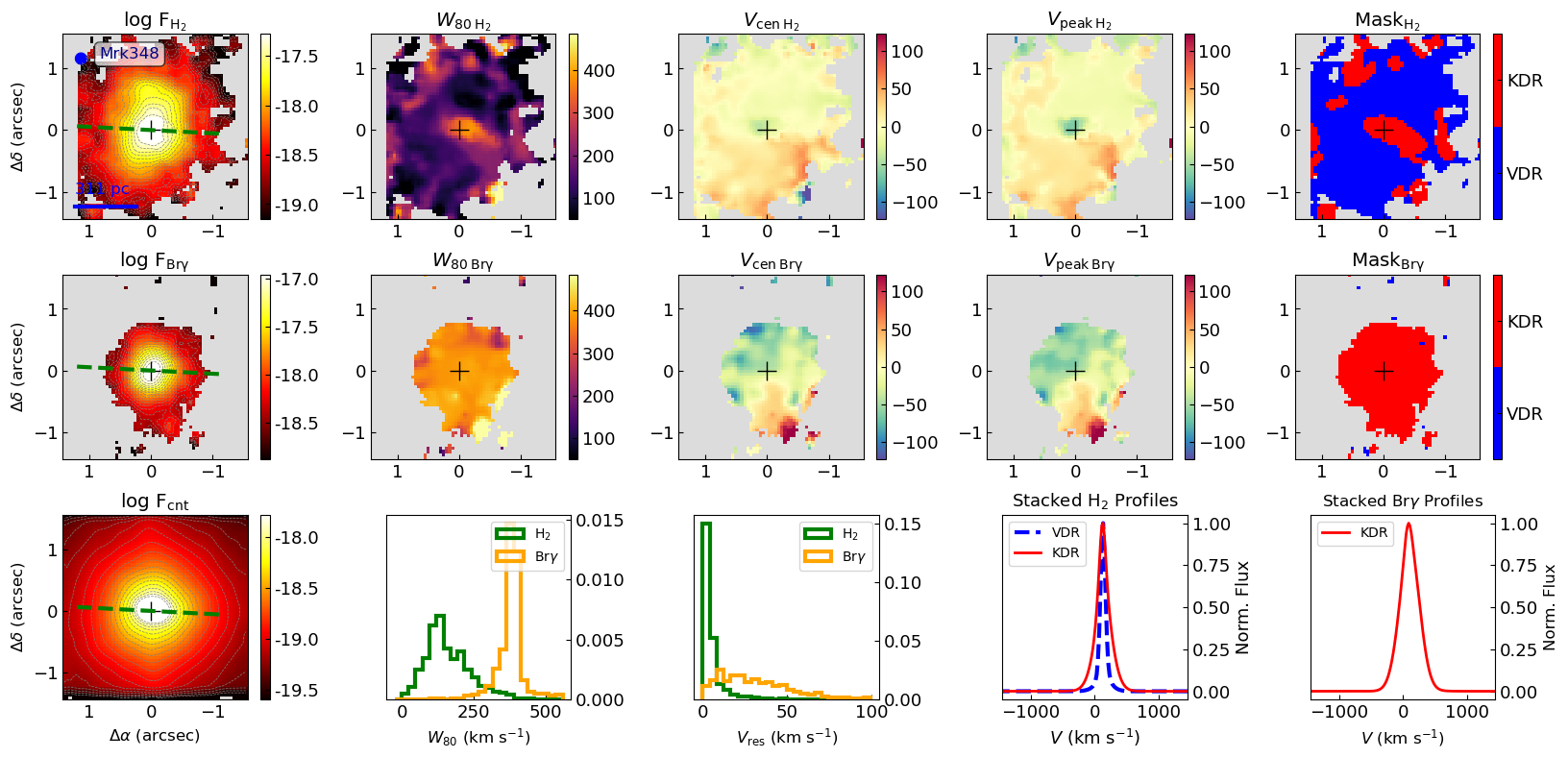}

\caption{\small Same as Fig.~\ref{fig:n788}, but for Mrk\,348.}
    \label{fig:mrk348}
\end{figure*}

\begin{figure*}
    \centering
    \includegraphics[width=0.98\textwidth]{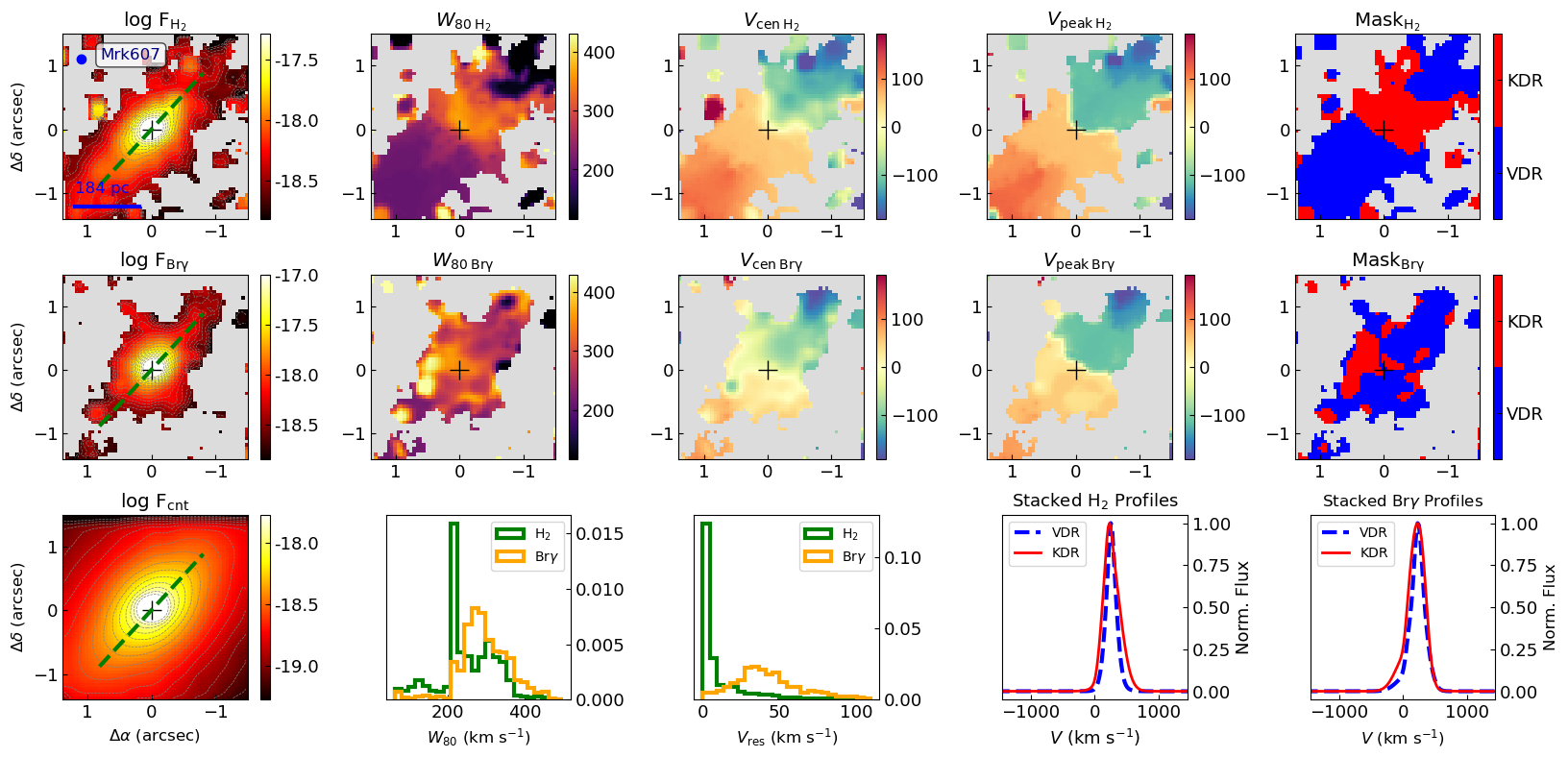}

\caption{\small Same as Fig.~\ref{fig:n788}, but for Mrk\,607.}
    \label{fig:mrk607}
\end{figure*}

\begin{figure*}
    \centering
    \includegraphics[width=0.98\textwidth]{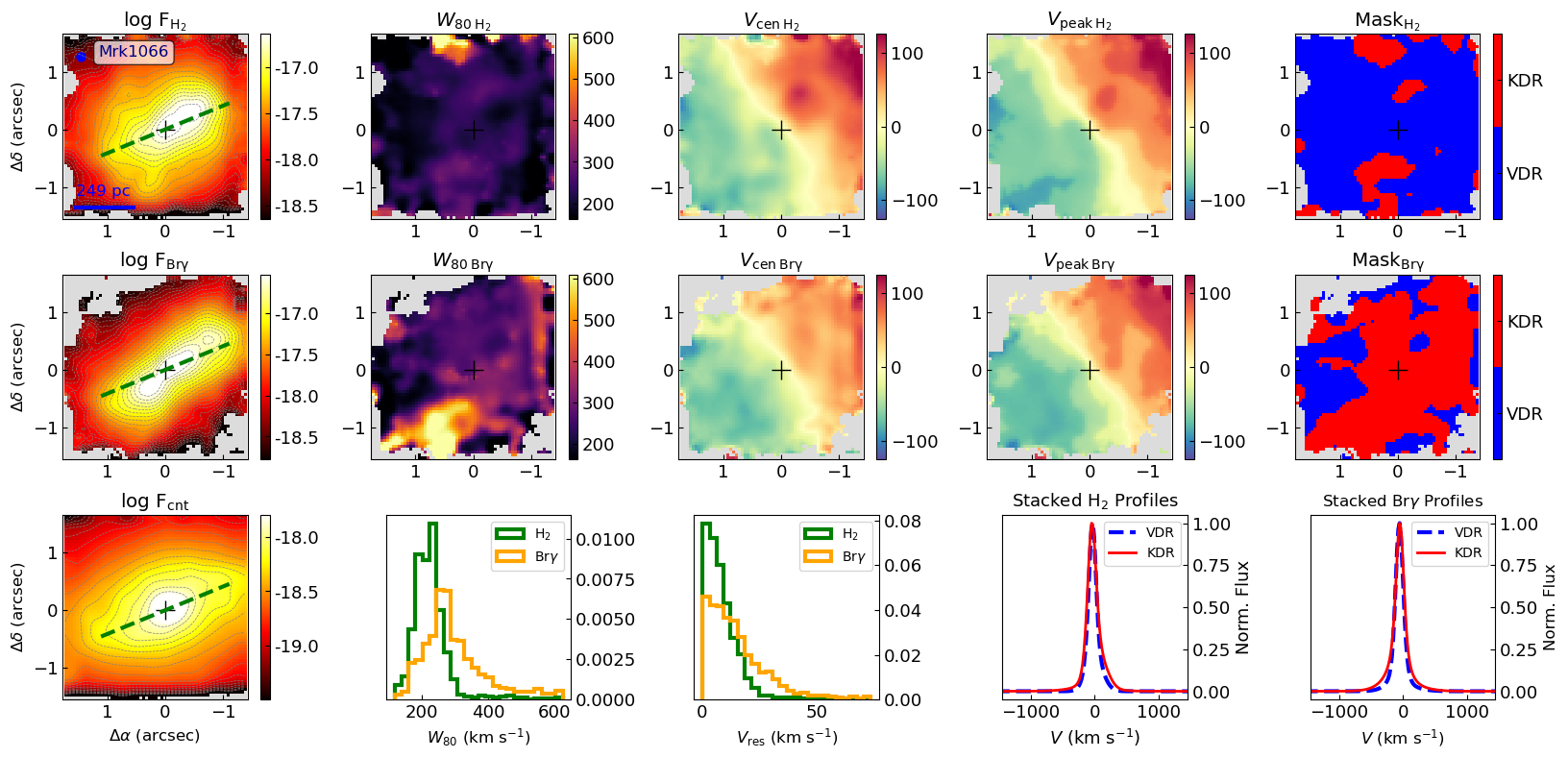}

\caption{\small Same as Fig.~\ref{fig:n788}, but for Mrk\,1066.}
    \label{fig:mrk1066}
\end{figure*}

\begin{figure*}
    \centering
    \includegraphics[width=0.98\textwidth]{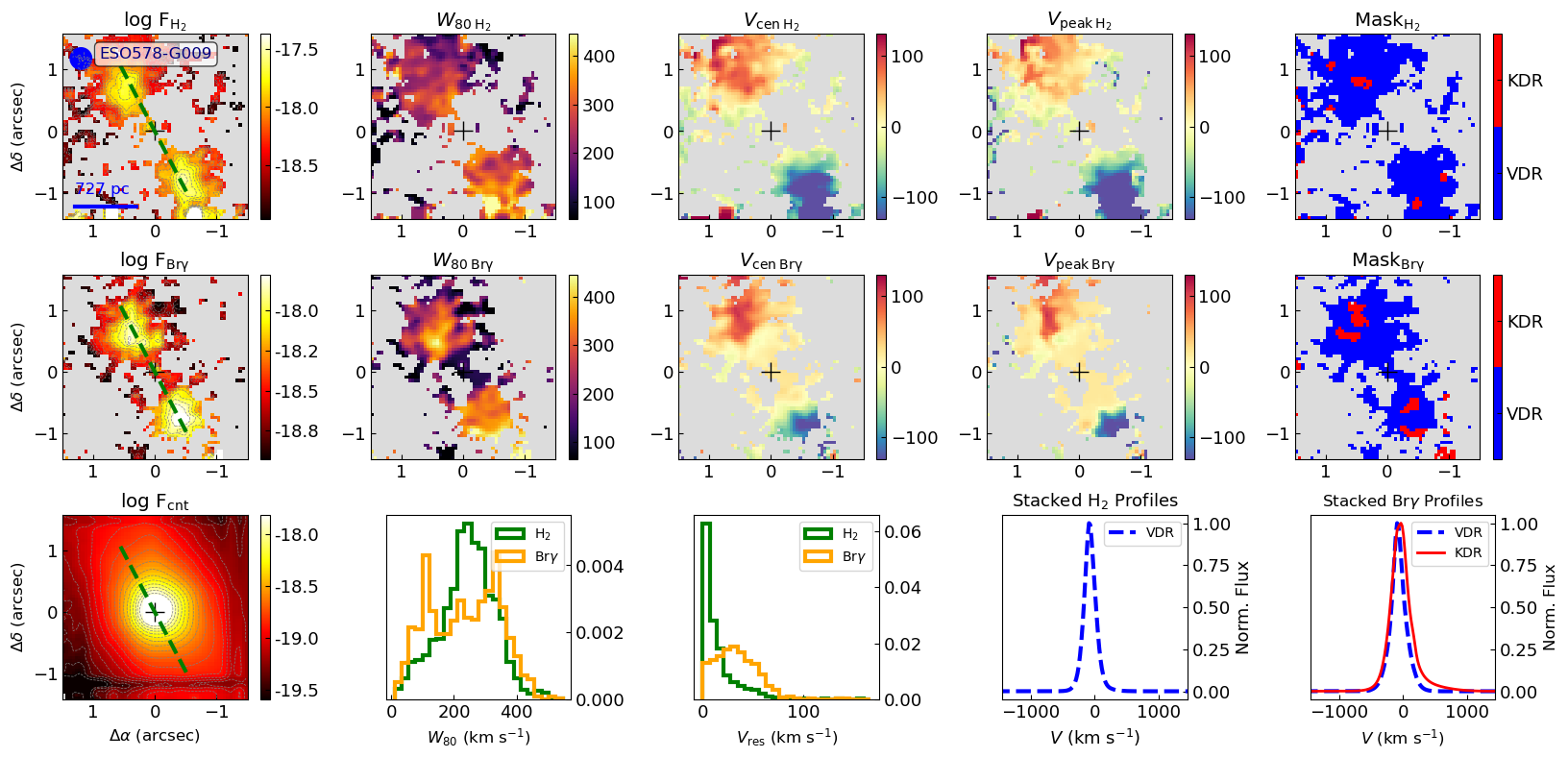}

\caption{\small Same as Fig.~\ref{fig:n788}, but for ESO\:578-G009.}
    \label{fig:eso578}
\end{figure*}

\begin{figure*}
    \centering
    \includegraphics[width=0.98\textwidth]{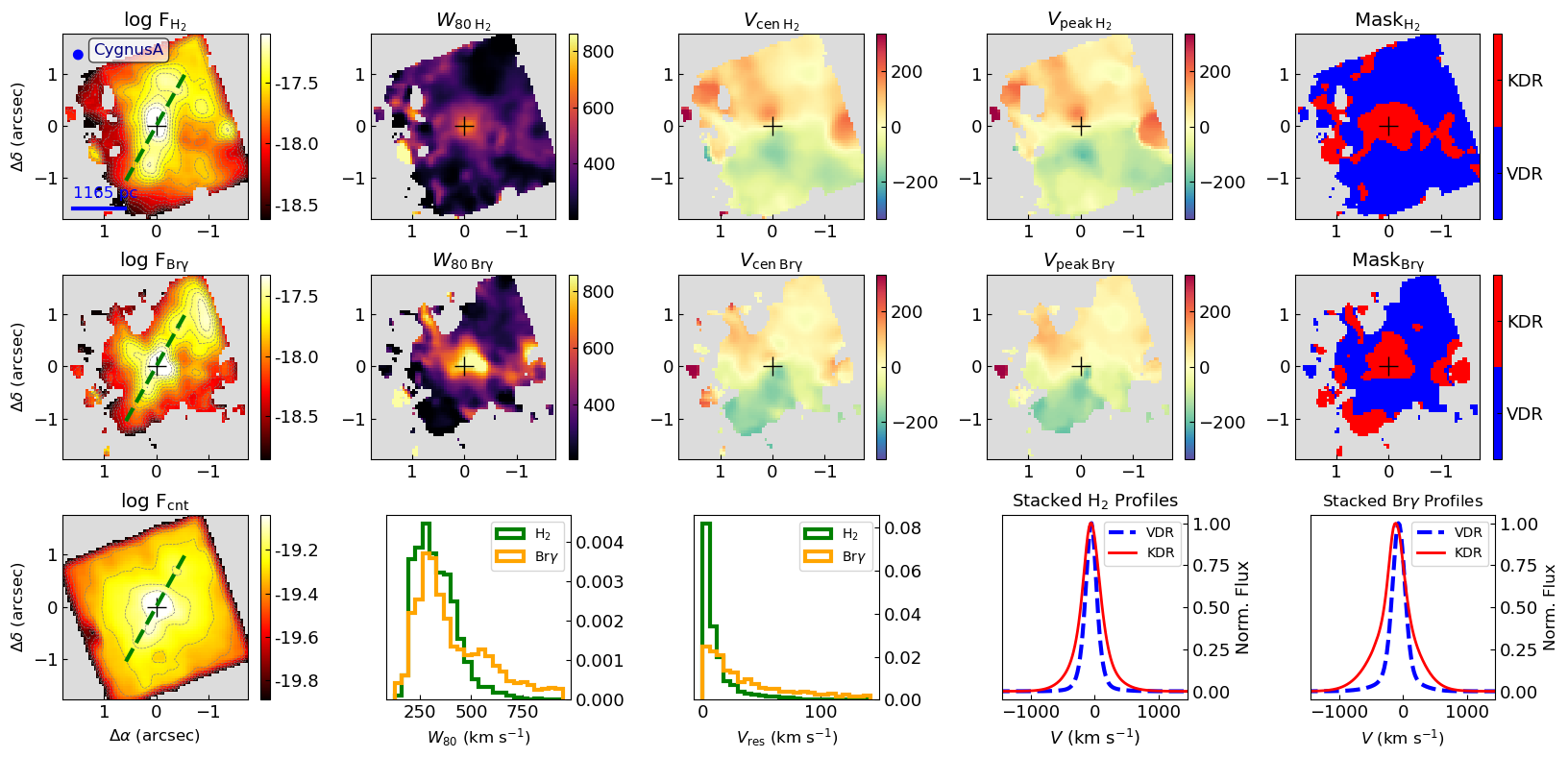}

\caption{\small Same as Fig.~\ref{fig:n788}, but for Cygnus\:A.}
    \label{fig:cygnusA}
\end{figure*}

\begin{figure*}
    \centering
    \includegraphics[width=0.98\textwidth]{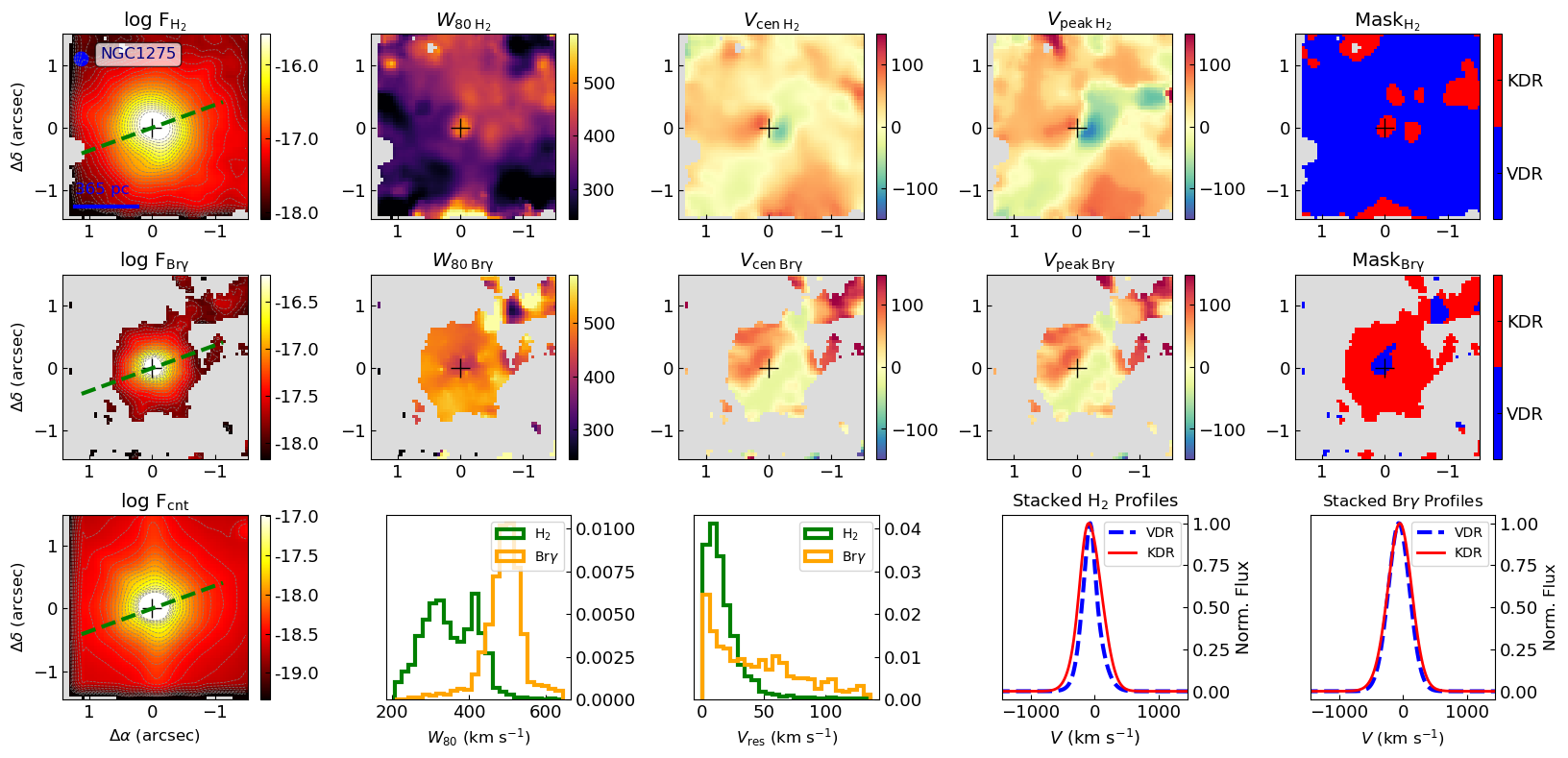}

\caption{\small Same as Fig.~\ref{fig:n788}, but for NGC\,1275.}
    \label{fig:n1275}
\end{figure*}

\begin{figure*}
    \centering
    \includegraphics[width=0.98\textwidth]{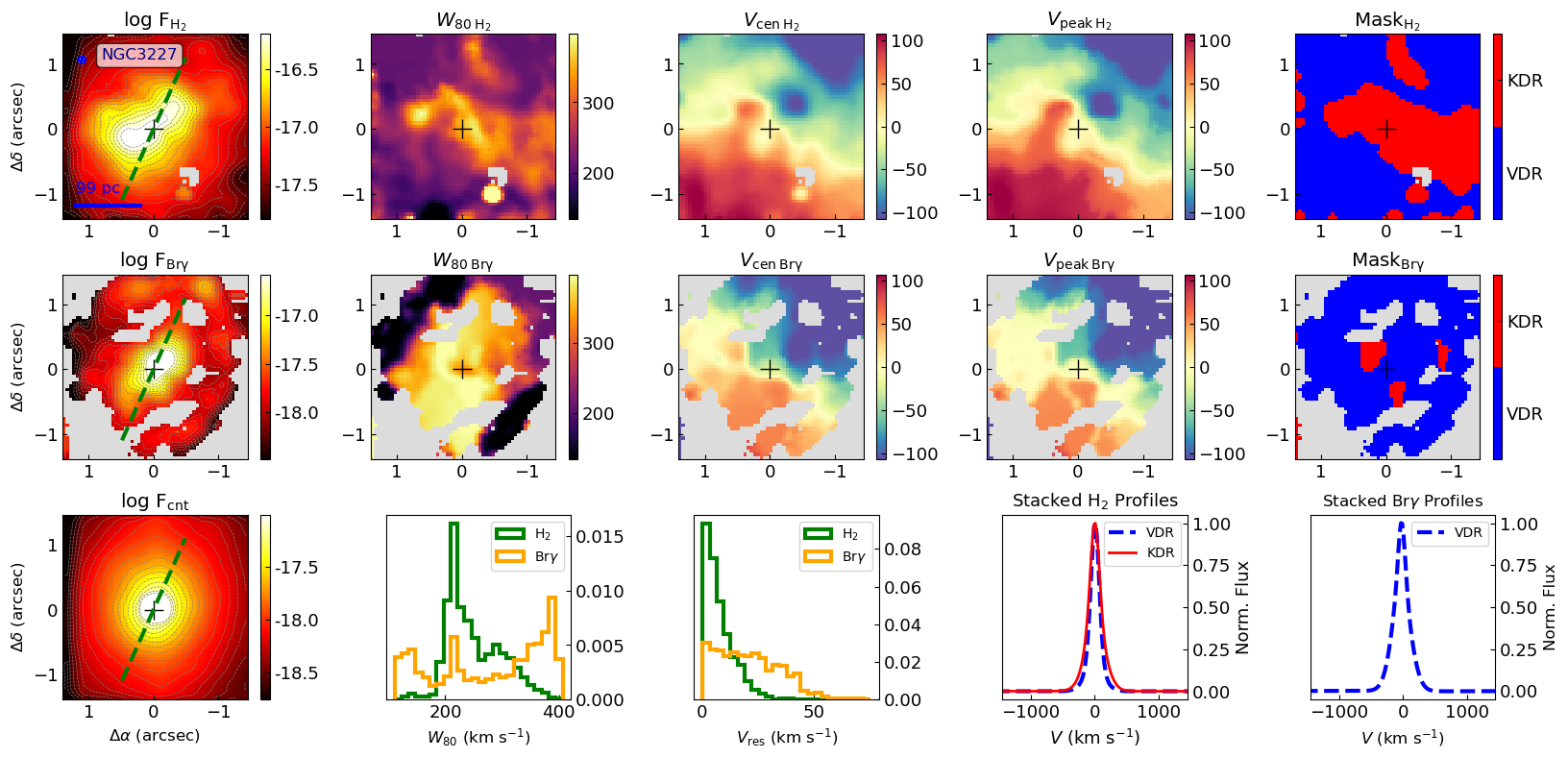}

\caption{\small Same as Fig.~\ref{fig:n788}, but for NGC\,3227.}
    \label{fig:n3227}
\end{figure*}

\begin{figure*}
    \centering
    \includegraphics[width=0.98\textwidth]{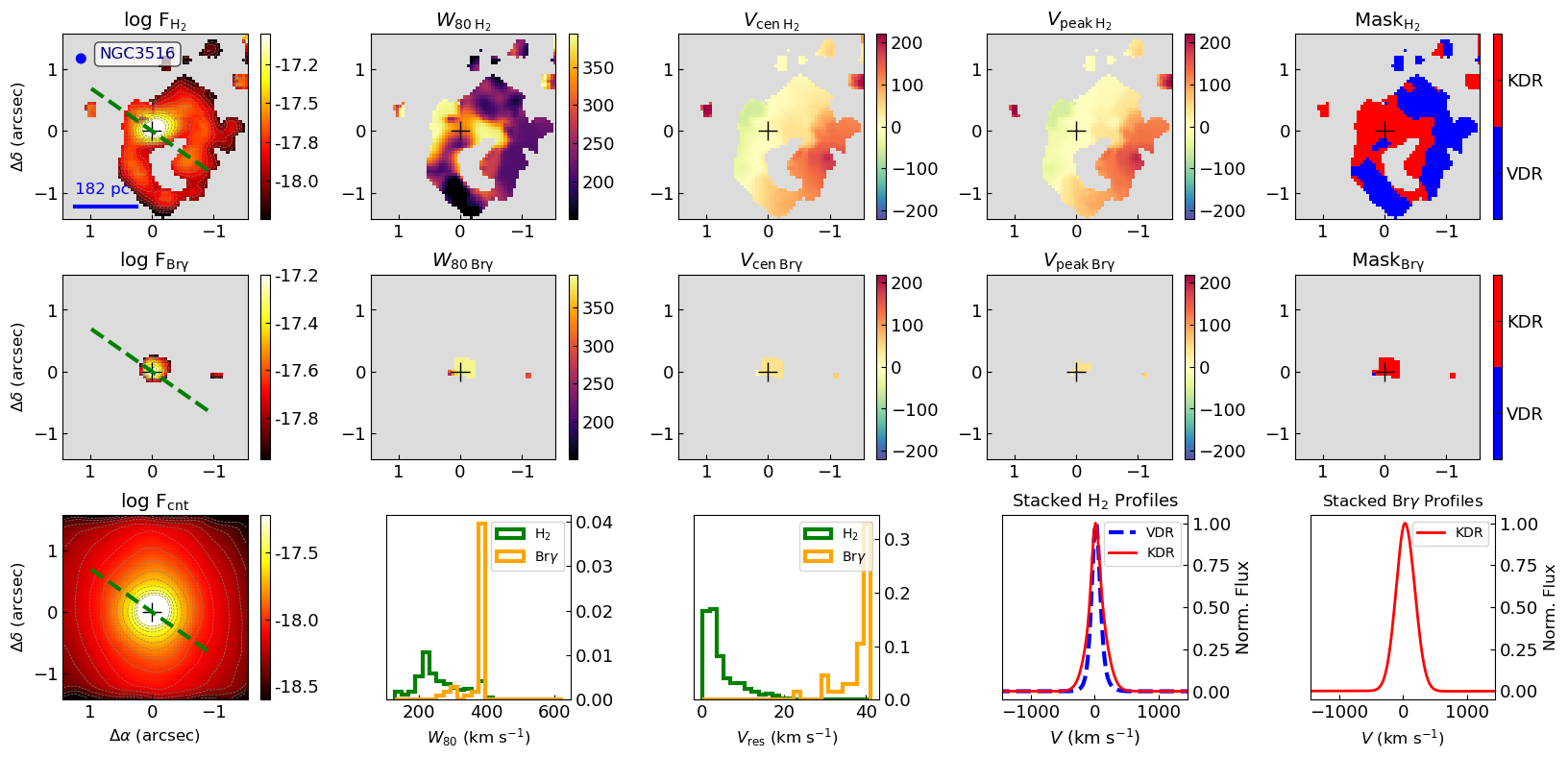}
\caption{\small Same as Fig.~\ref{fig:n788}, but for NGC\,3516.}
    \label{fig:n3516}
\end{figure*}

\begin{figure*}
    \centering
    \includegraphics[width=0.98\textwidth]{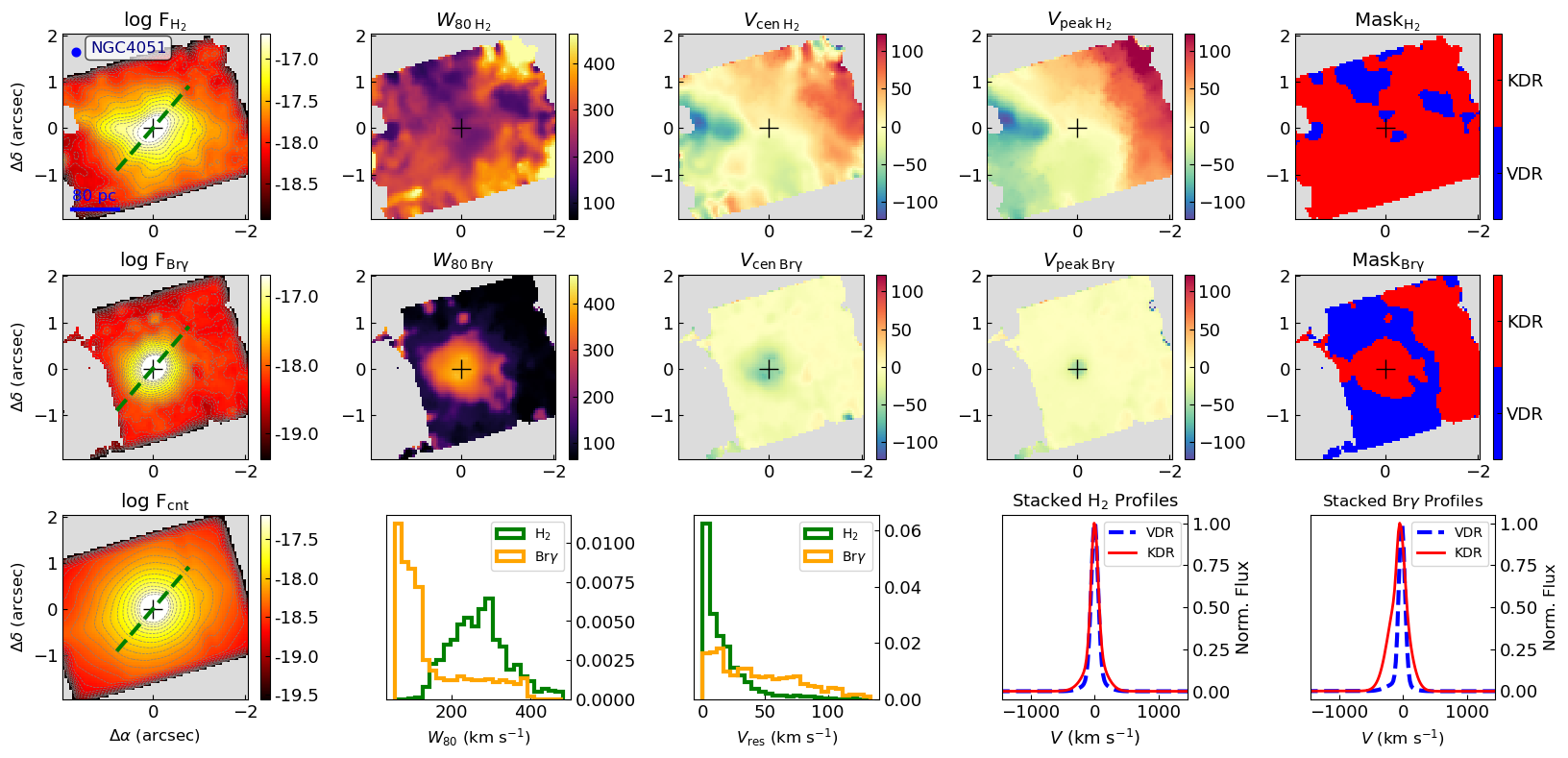}

\caption{\small Same as Fig.~\ref{fig:n788}, but for NGC\,4051.}
    \label{fig:n4051}
\end{figure*}

\begin{figure*}
    \centering
    \includegraphics[width=0.98\textwidth]{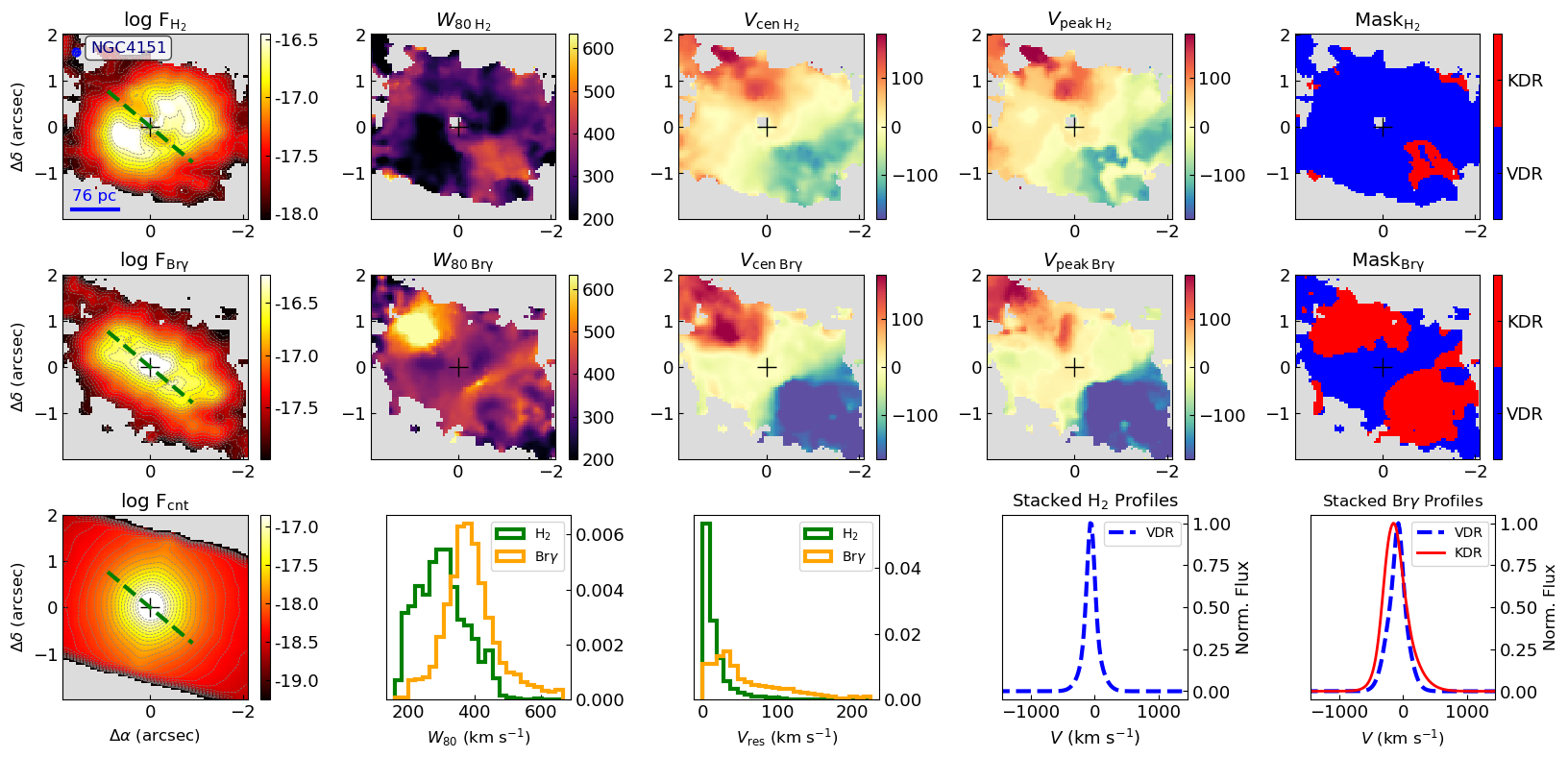}
\caption{\small Same as Fig.~\ref{fig:n788}, but for NGC\,4151.}
    \label{fig:n4151}
\end{figure*}

\begin{figure*}
    \centering
    \includegraphics[width=0.98\textwidth]{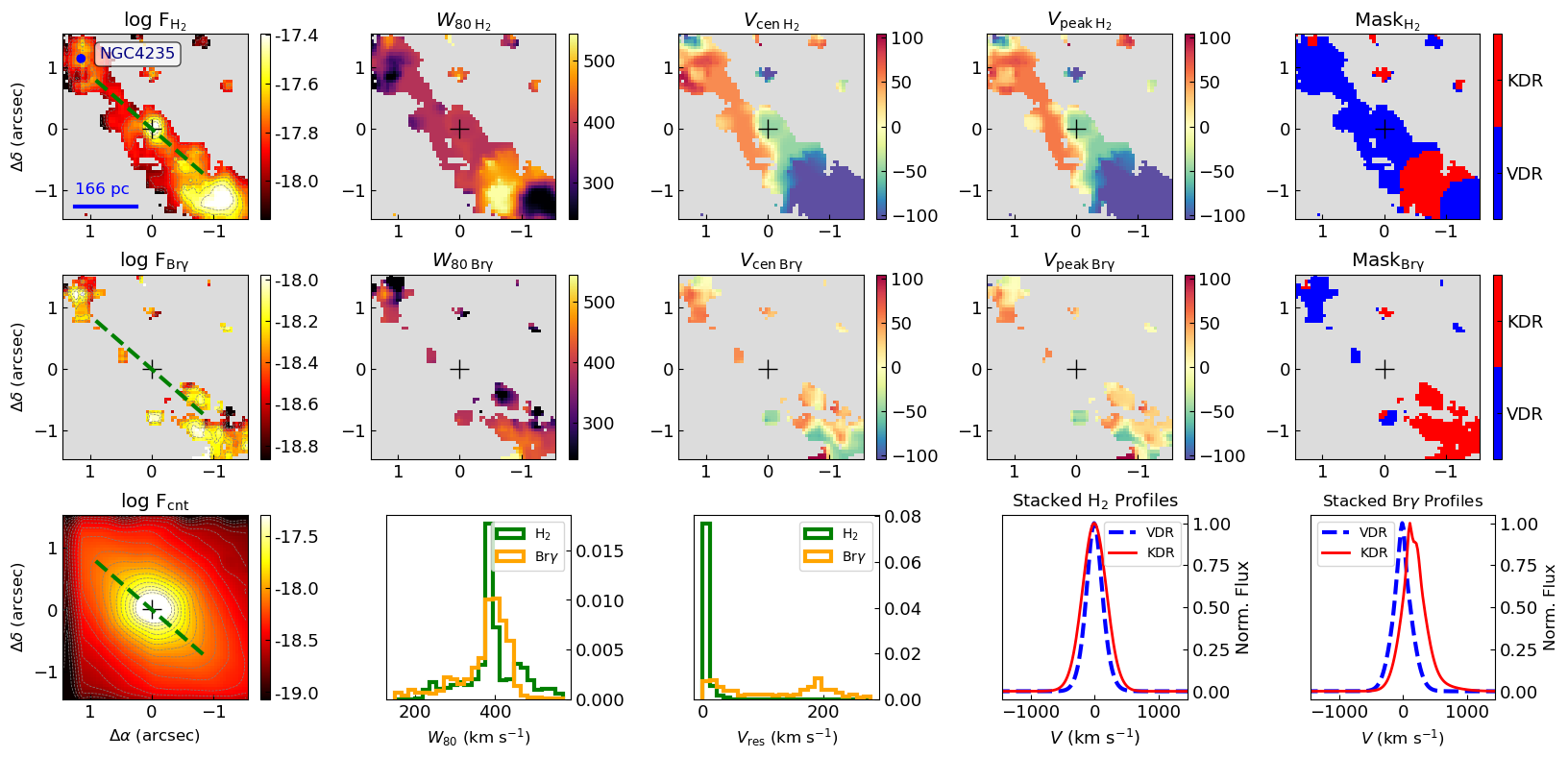}

\caption{\small Same as Fig.~\ref{fig:n788}, but for NGC\,4325.}
    \label{fig:n4235}
\end{figure*}

\begin{figure*}
    \centering
    \includegraphics[width=0.98\textwidth]{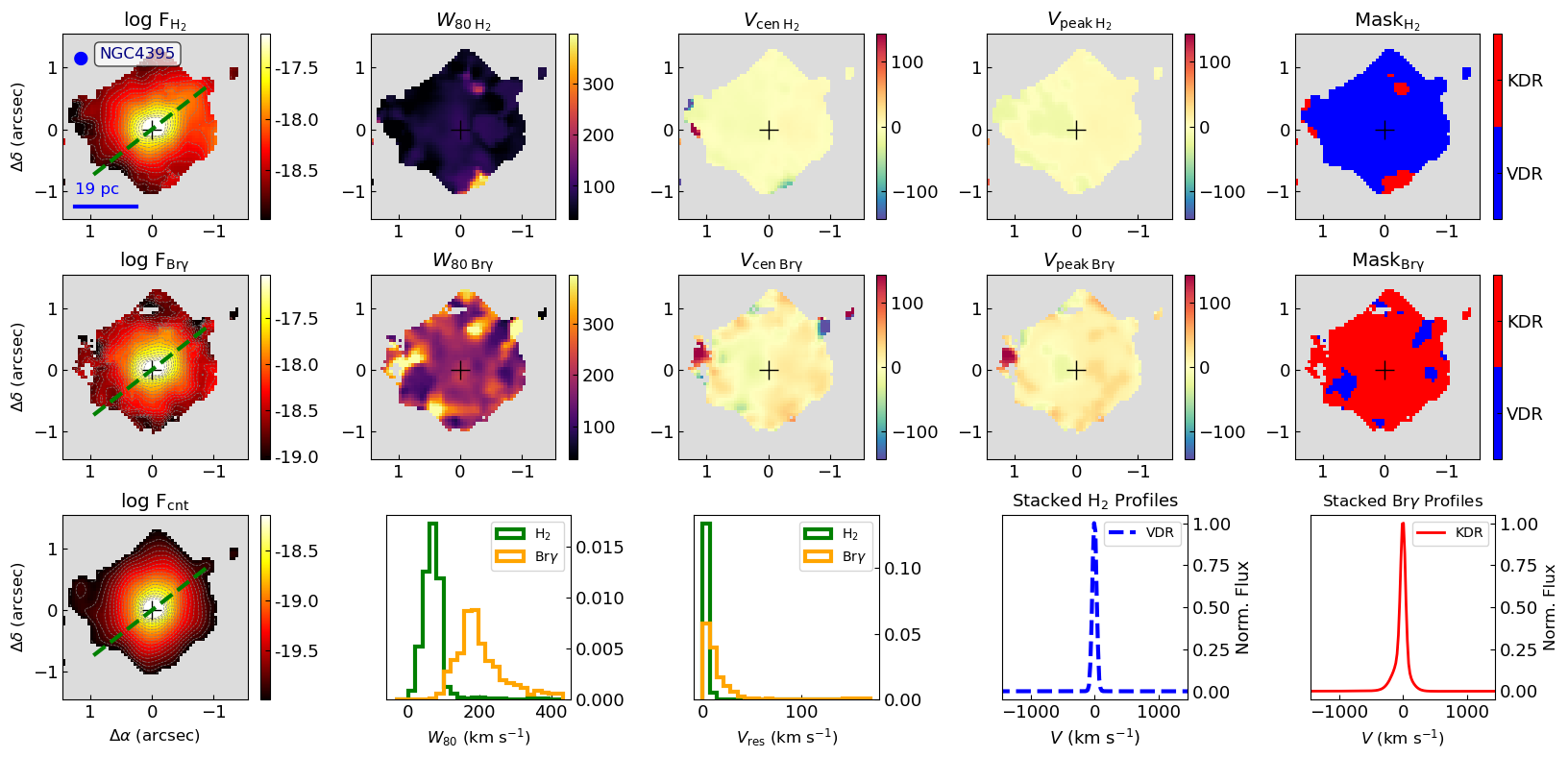}

\caption{\small Same as Fig.~\ref{fig:n788}, but for NGC\,4395.}
    \label{fig:n4395}
\end{figure*}

\begin{figure*}
    \centering
    \includegraphics[width=0.98\textwidth]{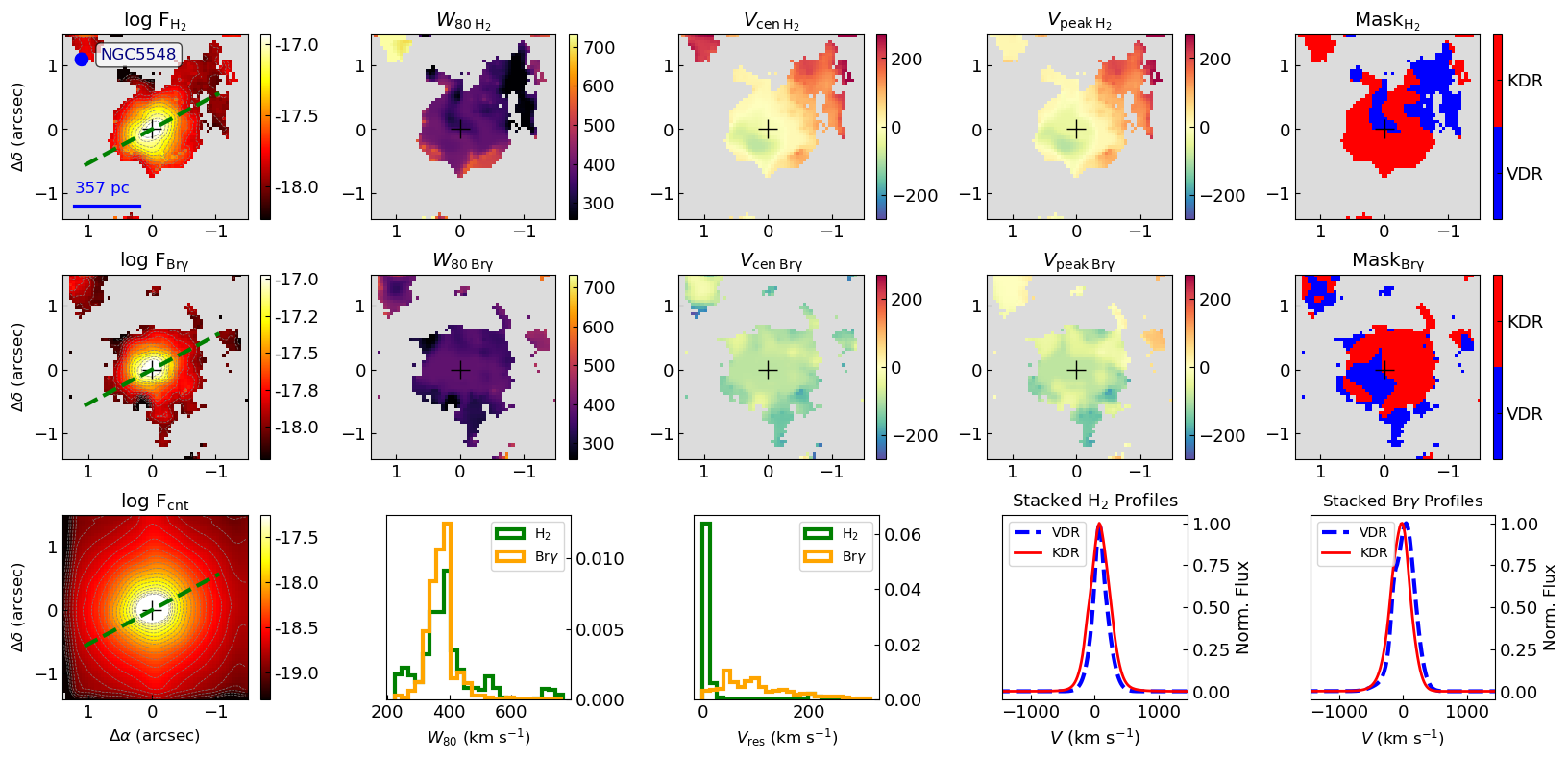}

\caption{\small Same as Fig.~\ref{fig:n788}, but for NGC\,5548.}
    \label{fig:n5548}
\end{figure*}

\begin{figure*}
    \centering
    \includegraphics[width=0.98\textwidth]{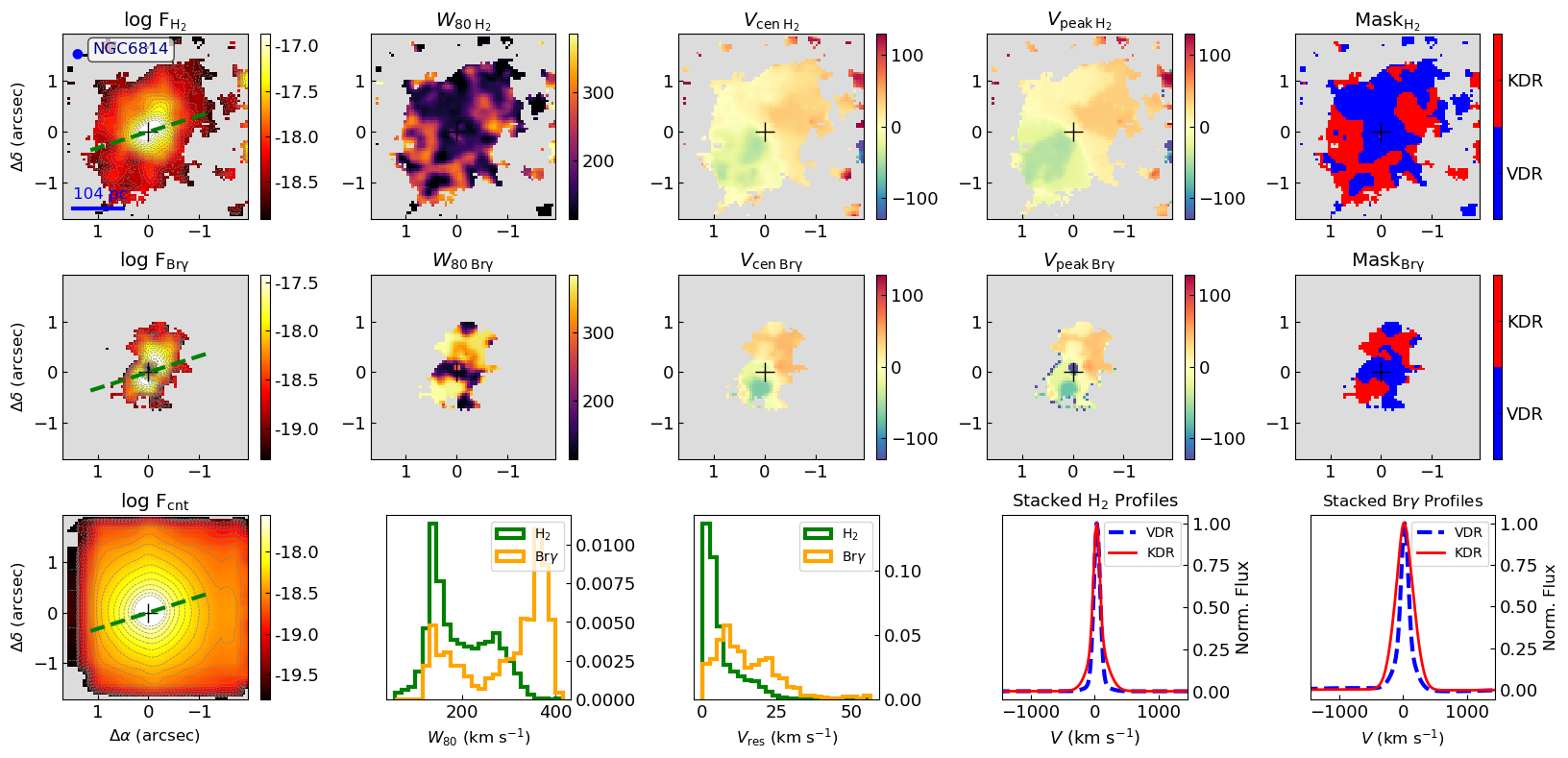}

\caption{\small Same as Fig.~\ref{fig:n788}, but for NGC\,6814.}
    \label{fig:n6814}
\end{figure*}

\begin{figure*}
    \centering
    \includegraphics[width=0.98\textwidth]{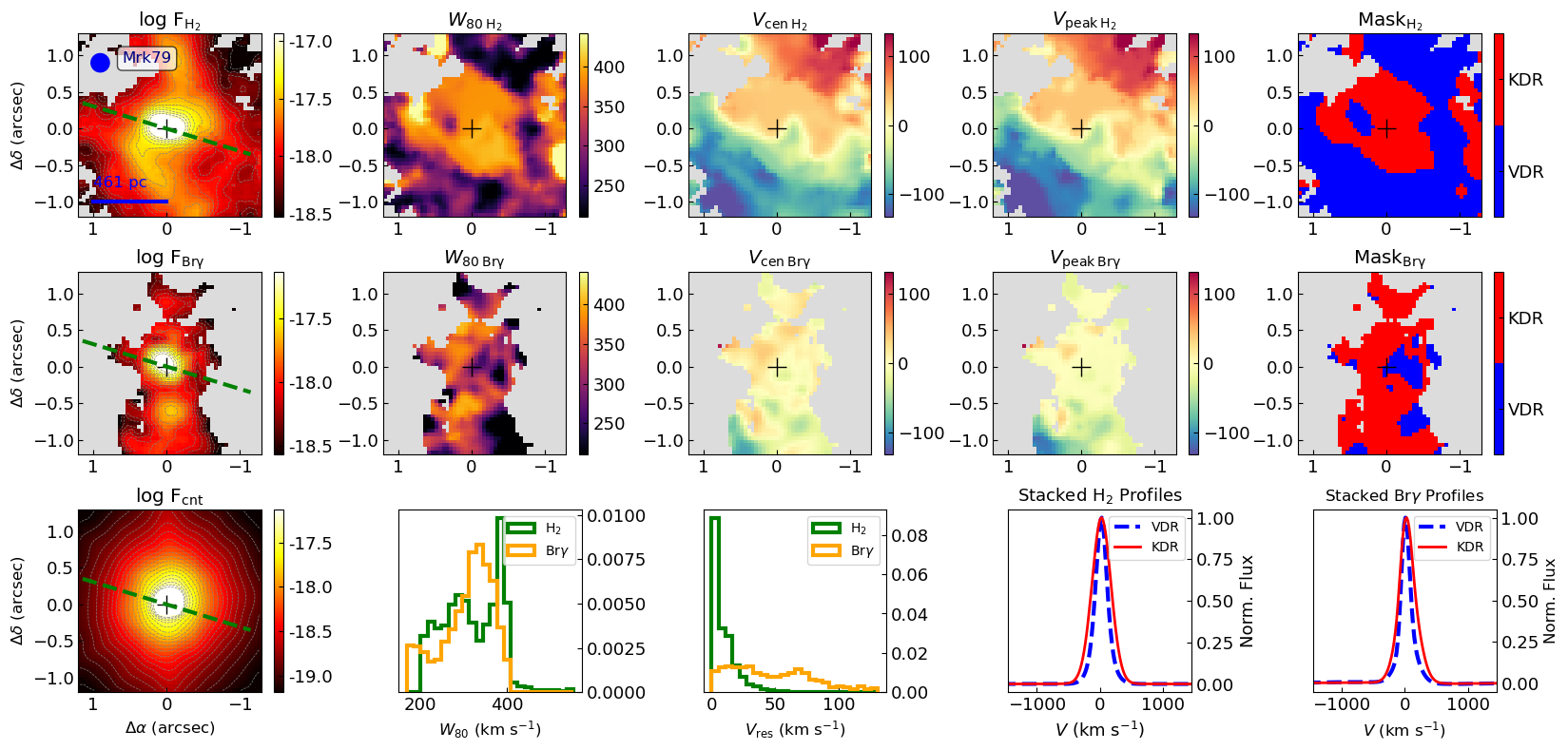}

\caption{\small Same as Fig.~\ref{fig:n788}, but for Mrk\,79.}
    \label{fig:mrk79}
\end{figure*}

\begin{figure*}
    \centering
    \includegraphics[width=0.98\textwidth]{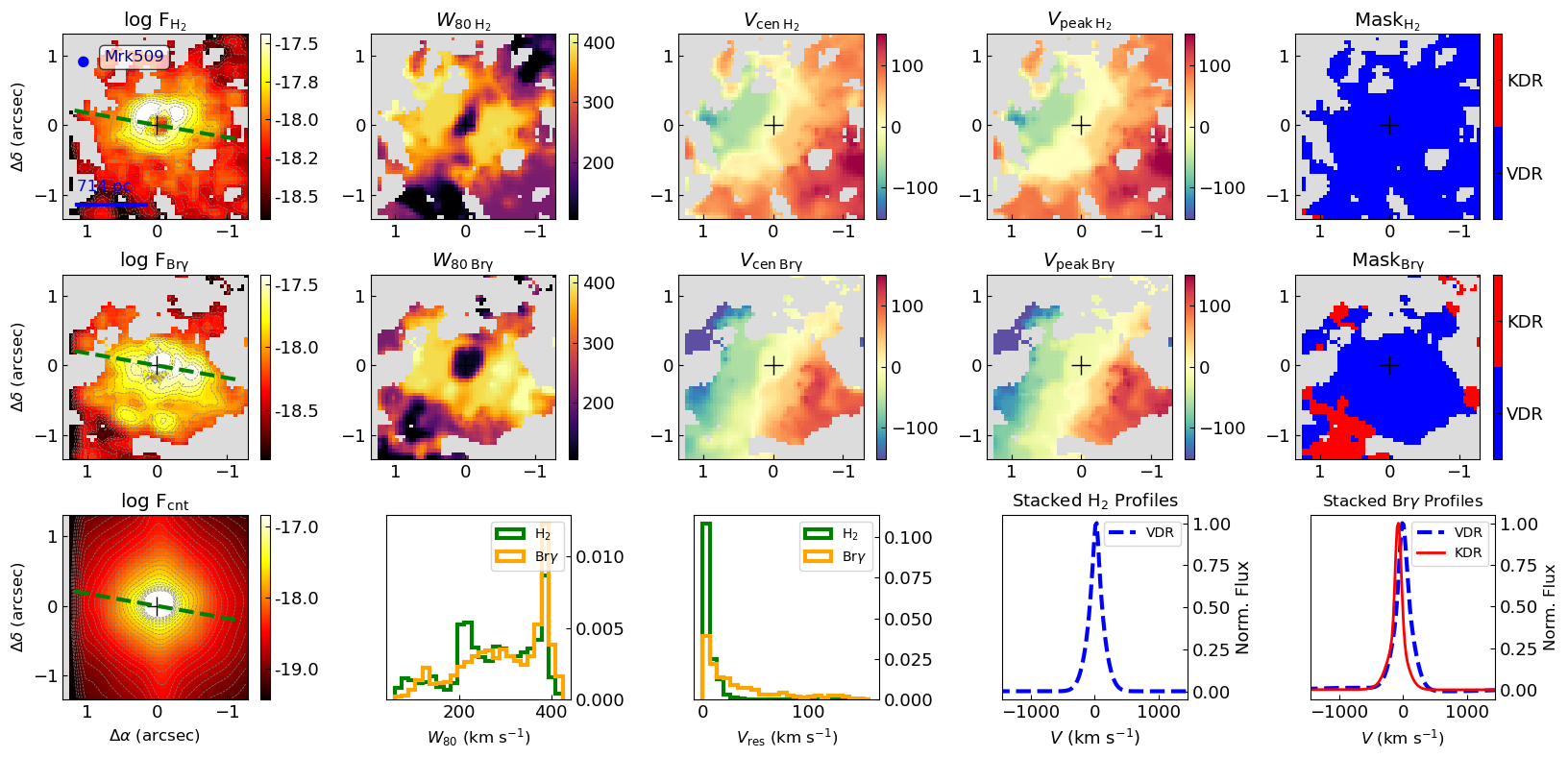}

\caption{\small Same as Fig.~\ref{fig:n788}, but for Mrk\,509.}
    \label{fig:mrk509}
\end{figure*}

\begin{figure*}
    \centering
    \includegraphics[width=0.98\textwidth]{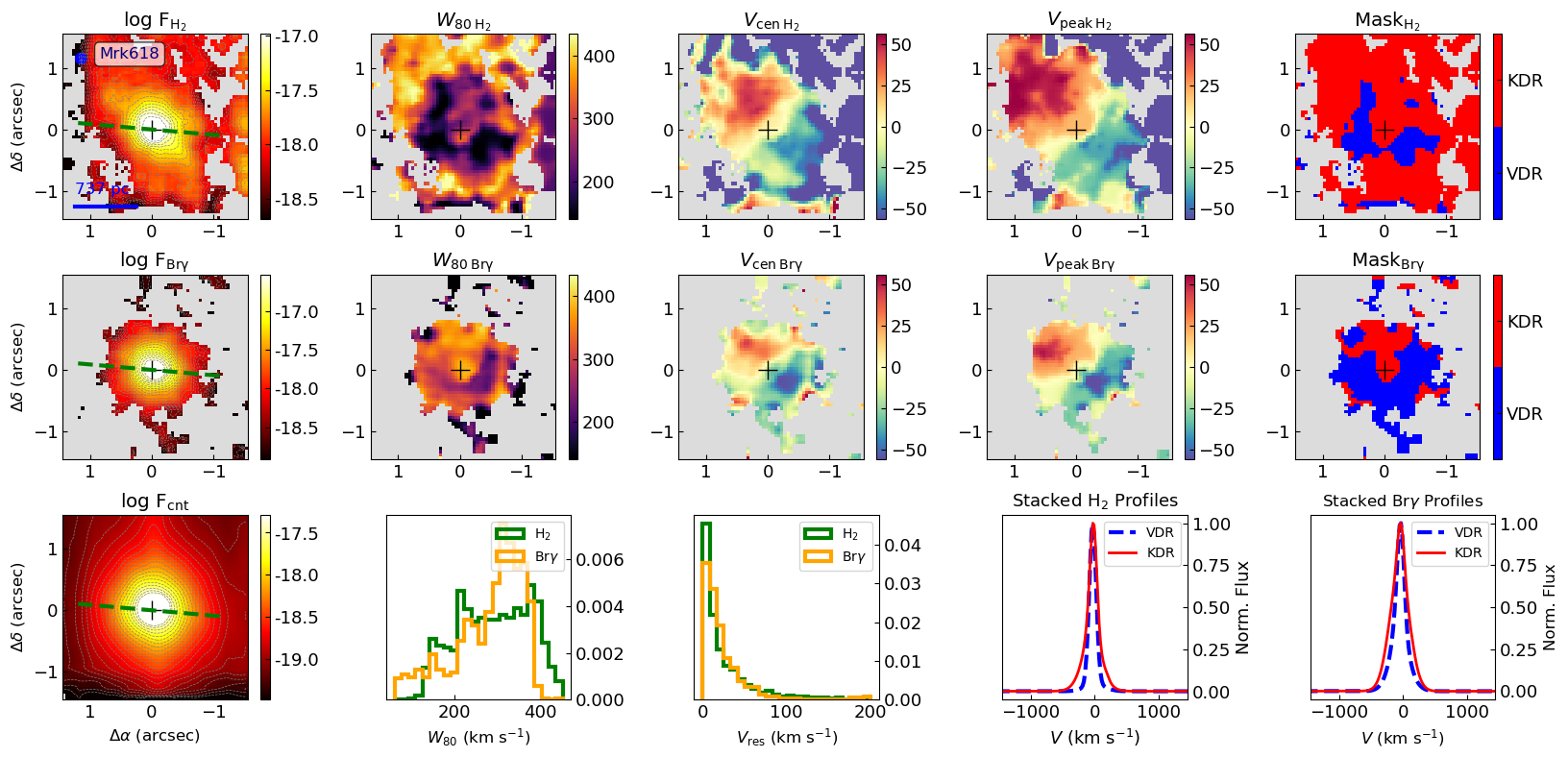}
\caption{\small Same as Fig.~\ref{fig:n788}, but for Mrk\,618.}
    \label{fig:mrk618}
\end{figure*}

\begin{figure*}
    \centering
    \includegraphics[width=0.98\textwidth]{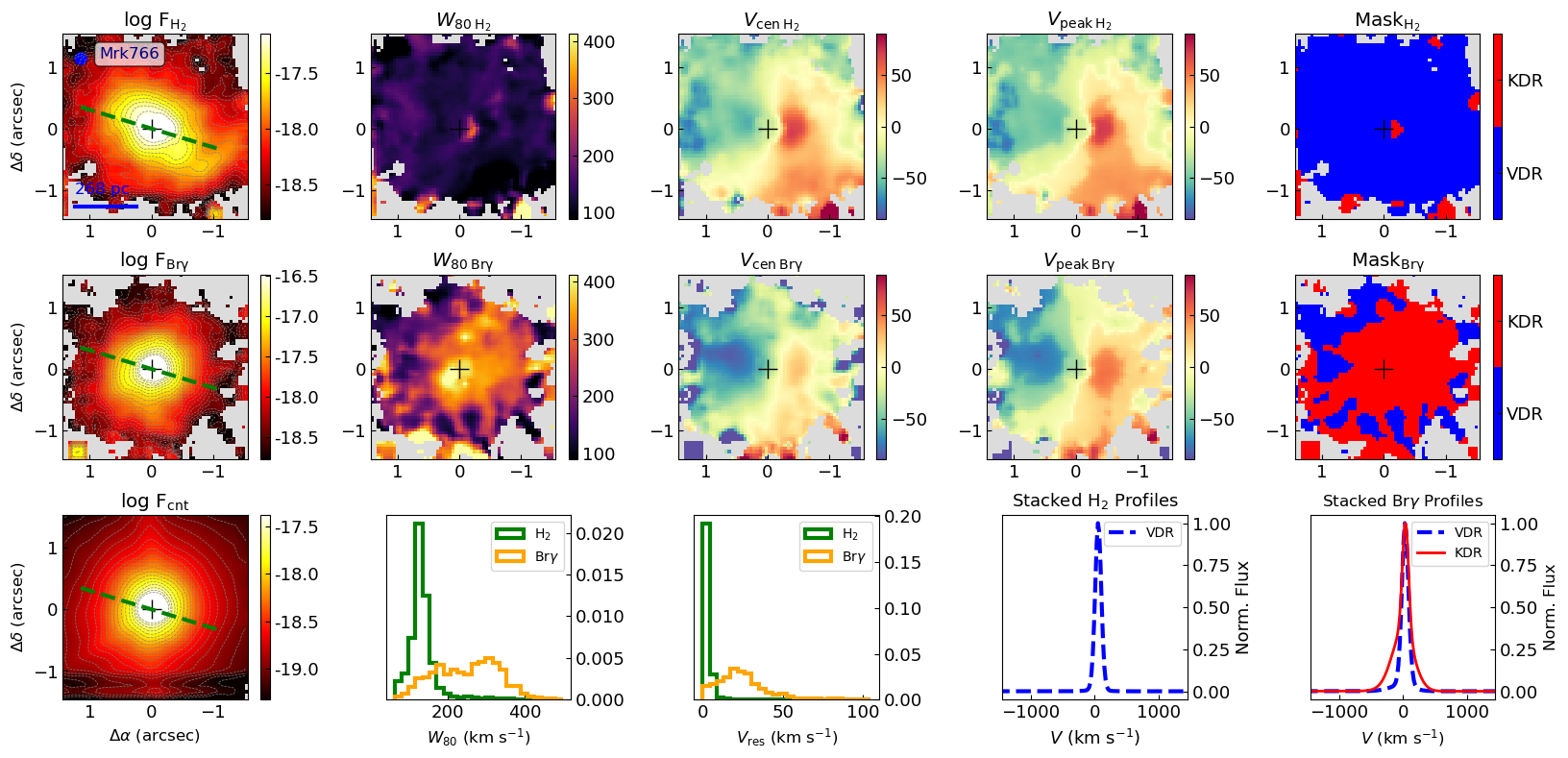}

\caption{\small Same as Fig.~\ref{fig:n788}, but for Mrk\,766.}
    \label{fig:mrk766}
\end{figure*}

\begin{figure*}
    \centering
    \includegraphics[width=0.98\textwidth]{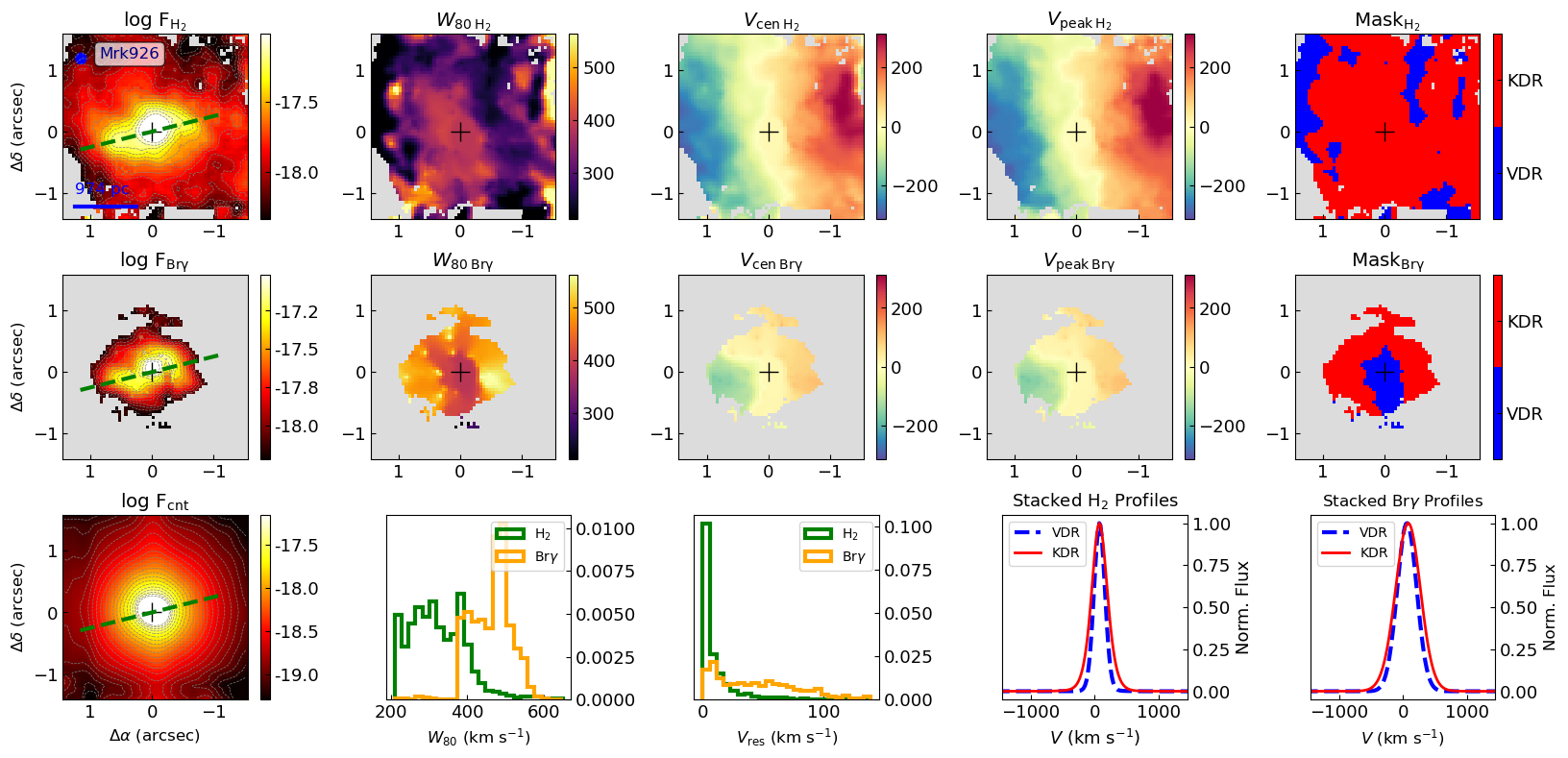}

\caption{\small Same as Fig.~\ref{fig:n788}, but for Mrk\,926.}
    \label{fig:mrk926}
\end{figure*}

\begin{figure*}
    \centering
    \includegraphics[width=0.98\textwidth]{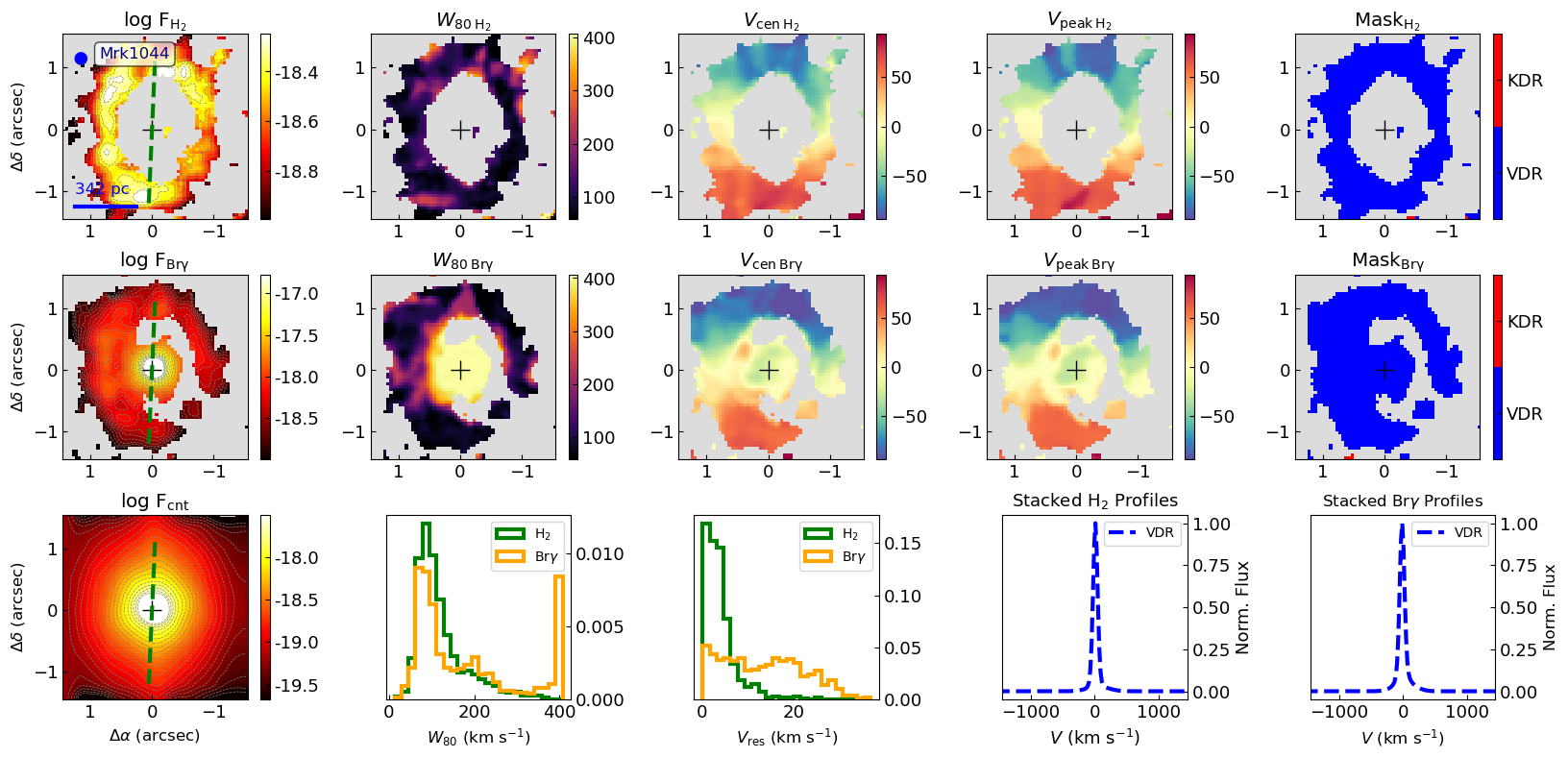}

\caption{\small Same as Fig.~\ref{fig:n788}, but for Mrk\,1044.}
    \label{fig:mrk4044}
\end{figure*}

\begin{figure*}
    \centering
    \includegraphics[width=0.98\textwidth]{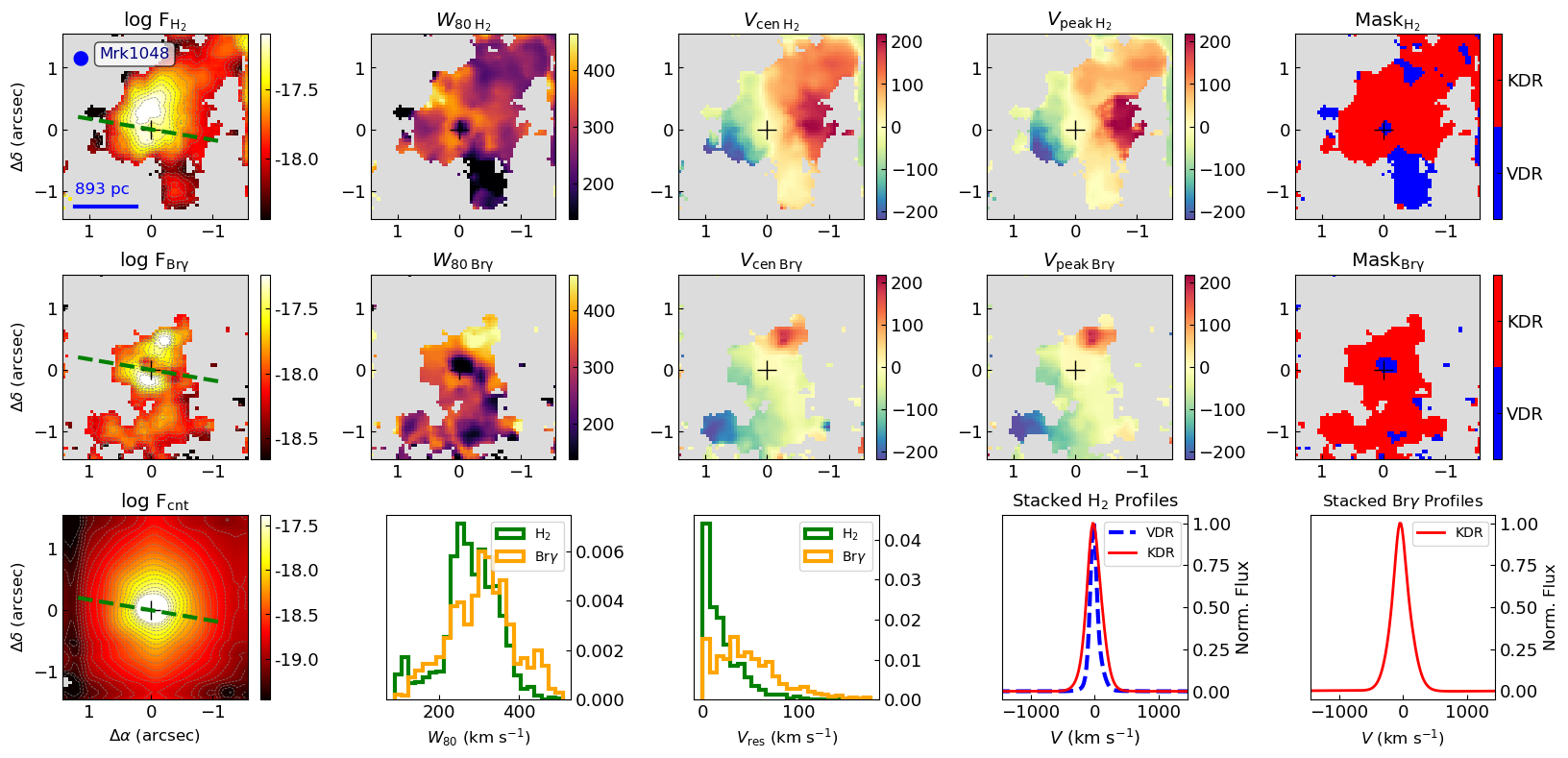}

\caption{\small Same as Fig.~\ref{fig:n788}, but for Mrk\,1048.}
    \label{fig:mrk1048}
\end{figure*}

\begin{figure*}
    \centering
    \includegraphics[width=0.98\textwidth]{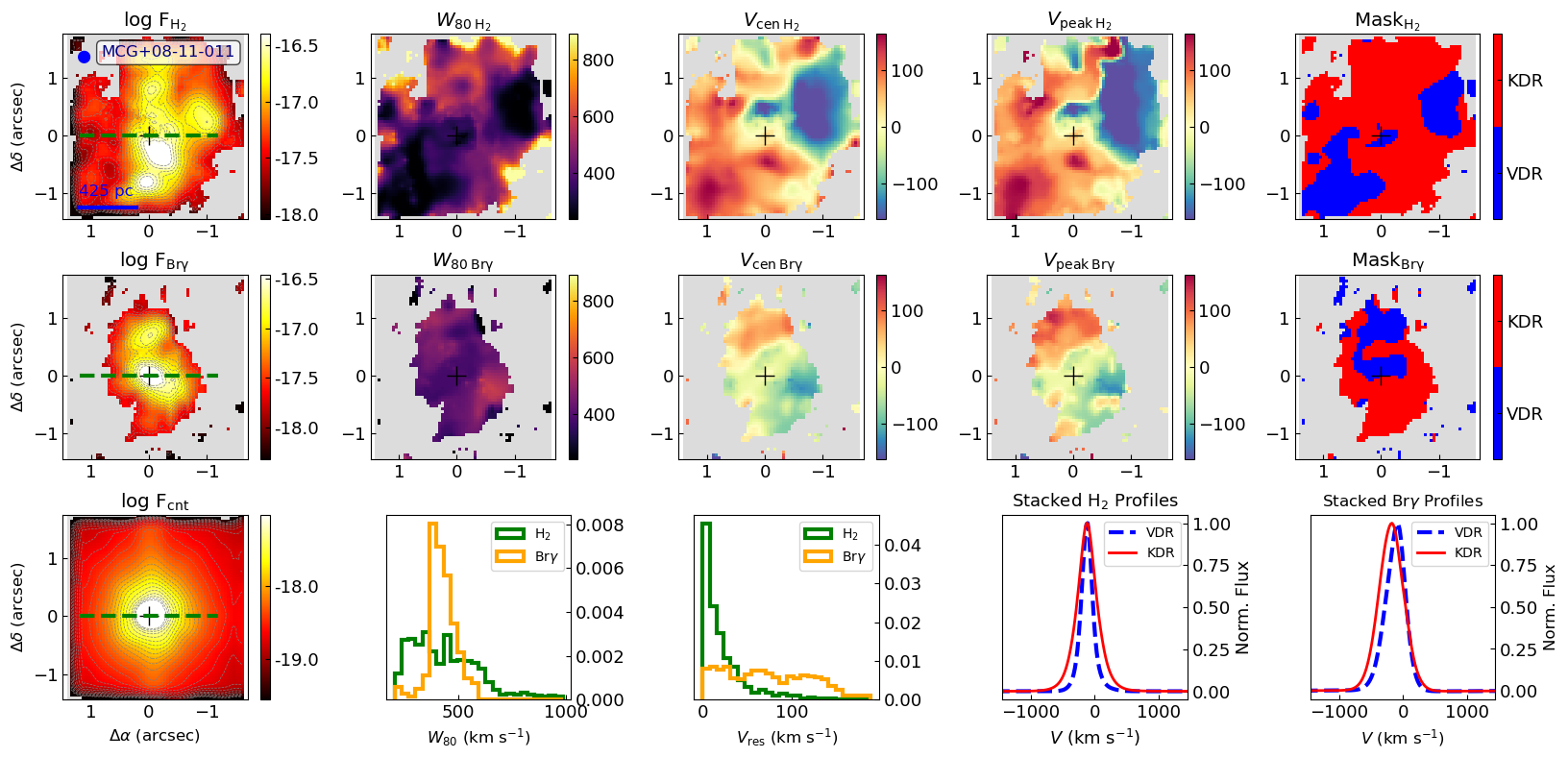}

\caption{\small Same as Fig.~\ref{fig:n788}, but for MCG+08-11-011.}
    \label{fig:mcg08}
\end{figure*}

In figures \ref{fig:n788} to \ref{fig:mcg08}, we present the maps, distributions and stacked profiles for all galaxies in our sample. 

\bsp	
\label{lastpage}
\end{document}